\documentclass[iop]{emulateapj}
\pdfoutput=1

\usepackage{amssymb}
\usepackage{amsmath}
\usepackage{wrapfig}
\usepackage{rotating}
\usepackage{hyperref}
\usepackage[utf8]{inputenc}
\usepackage{boondox-cal}
\usepackage{calrsfs}
\DeclareMathAlphabet{\pazocal}{OMS}{zplm}{m}{n}

\newcommand\uiks{$u^*i^\prime K_s$}
\newcommand\ks{$K_s$}
\newcommand\ip{$i^\prime$}
\newcommand\ust{$u^*$}
\newcommand\gp{$g^\prime$}
\newcommand\rp{$r^\prime$}

\graphicspath{{./}{figures/}}

\received{November 16, 2018}
\revised{February 19, 2019}
\accepted{March 15, 2019}
\shorttitle{Spectroscopy of NGC 4258 GCCs}
\shortauthors{Gonz\'alez-L\'opezlira et al.}

\begin{document}

\title{Spectroscopy of NGC~4258 globular cluster candidates: membership confirmation and kinematics.}

%\correspondingauthor{Rosa A.\ Gonz\'alez-L\'opezlira}
\email{r.gonzalez@irya.unam.mx}

%\author[0000-0003-1557-4931]{Rosa A.\ Gonz\'alez-L\'opezlira}
\author{Rosa A.\ Gonz\'alez-L\'opezlira}
\affiliation{Instituto de Radioastronomia y Astrofisica, UNAM, Campus Morelia,
     Michoacan, Mexico, C.P.\ 58089}

\author{Y.\ D.\ Mayya}
\affiliation{Instituto Nacional de Astrof\'{\i}sica, \'Optica y Electr\'onica, Luis Enrique Erro 1, Tonantzintla, Puebla, C.P. 72840, Mexico}

\author{Laurent Loinard}
\affiliation{Instituto de Radioastronomia y Astrofisica, UNAM, Campus Morelia,
     Michoacan, Mexico, C.P.\ 58089}
\affiliation{
Instituto de Astronomia, Universidad Nacional Autonoma de Mexico, Apartado Postal 70-264, 04510 Ciudad de Mexico, Mexico
}

\author{Karla \'Alamo-Mart\'{\i}nez}
\affiliation{Departamento de Astronomia, Instituto de F\'{\i}sica da Universidade Federal do Rio Grande do Sul, Brazil}

\author{George Heald}
\affiliation{CSIRO Astronomy and Space Science, PO Box 1130, Bentley WA 6102, Australia}

\author{Iskren Y.\ Georgiev}
\affiliation{Max-Planck Institut f\"ur Astronomie, K\"onigstuhl 17, D-69117 Heidelberg, Germany}

\author{Yasna \'Ordenes-Brice\~no}
\affiliation{Instituto de Astrof\'{\i}sica, Pontificia Universidad Cat\'olica de Chile, Av.\ Vicu\~na Mackenna 4860, 7820436 Macul, Santiago, Chile  }

\author{Ariane Lan\c con}
\affiliation{Observatoire Astronomique de Strasbourg, Universit\'e de Strasbourg, CNRS, UMR 7550, 11 rue de l'Universit\'e,  67000 Strasbourg, France}

\author{Maritza A.\ Lara-L\'opez}
\affiliation{Dark Cosmology Centre, Niels Bohr Institute, University of Copenhagen, Juliane Maries Vej 30, 2100 Copenhagen, Denmark}

\author{Luis Lomel\'{\i}-N\'u\~nez}
\affiliation{Instituto Nacional de Astrof\'{\i}sica, \'Optica y Electr\'onica, Luis Enrique Erro 1, Tonantzintla, Puebla, C.P. 72840, Mexico}

\author{Gustavo Bruzual}
\affiliation{Instituto de Radioastronomia y Astrofisica, UNAM, Campus Morelia,
     Michoacan, Mexico, C.P.\ 58089}

\author{Thomas H.\ Puzia}
\affiliation{Instituto de Astrof\'{\i}sica, Pontificia Universidad Cat\'olica de Chile, Av.\ Vicu\~na Mackenna 4860, 7820436 Macul, Santiago, Chile  }

\begin{abstract}

We present multi-object spectroscopic observations of 23 globular cluster candidates (GCCs) in the prototypical megamaser galaxy
NGC 4258, carried out with the OSIRIS instrument at the 10.4 m Gran Telescopio Canarias. The candidates have been selected based
on the ($u^*\ -\ i^\prime$) versus 
($i^\prime\ -\ K_s$) diagram, in the first application of the \uiks\ method to a spiral galaxy. 
In the spectroscopy presented here, 70\% of the candidates are confirmed as globular clusters. 
Our results validate the efficiency of the \uiks\ method in the sparser GC systems of spirals, and given the 
downward correction to the total number of GCs, the agreement of the galaxy with the correlations between 
black hole mass, and total number and mass of GCs is actually improved.
We find that the GCs, mostly metal-poor, co-rotate with the HI disk, even at large galactocentric distances.

\end{abstract}


\section{Introduction} \label{sec:intro}

It has become commonplace to state that globular clusters (GC) are powerful signposts of
the structure, dynamics, and star formation and assemblage histories of their host galaxies.   
Not only are they compact and bright, and hence traceable to large distances; 
they do correlate
with global properties and other components of their hosts, such as their luminosity
\citep[e.g.,][]{peng08,geor10}; light concentration and stellar velocity dispersion
\citep{harr13}; total stellar mass and bulge light \citep{rhod12},
and total halo \citep{spit09,harr17} and central supermassive black hole masses \citep[e.g.,][]{sado12,burk10,harr11,harr14}.
However, until now, studies of GC systems have focused on early type galaxies, 
given their significantly larger
number of GCs, and the ease of their identification in the absence of a dusty, stellar disk, 
composed of partially resolved, bright sources.

Although understandable, given the obvious difficulties, the scarcity of studies of late-type
galaxies constitutes a real obstacle to actually using GC systems as probes of galaxy formation and 
evolution, for several reasons, as expressed already a couple of decades ago \citep[e.g.,][]{kiss99}.  
Firstly, we do not really know how typical the GC system of our Galaxy is, and yet it is the standard to which all
other systems are compared. More importantly, perhaps, while there seems to be a consensus that
elliptical galaxies, and hence their GC systems, form through mergers, there is now mounting observational 
\citep[e.g.,][]{foer09,tacc13,dess15,patr18} and 
theoretical \citep[e.g.,][]{deke09,krui15} evidence that massive stellar clusters form at 
$z \geqslant 2$ by the fragmentation of
gaseous disks, rotating but turbulent, suffering intense, cold gas accretion flows.
Globular clusters would be the relics of this star formation epoch, and low mass spirals may still 
contain relatively undisturbed primeval systems, given their slightly lower major ($M_{\rm sat}/M_{\rm primary} \geqslant$ 1:4) merger rates \citep{rodr15}.

Motivated originally to explore the correlation between total number of GCs, $N_{\rm GC}$, and 
central black hole mass, $M_\bullet$ \citep[e.g.,][]{burk10,harr11} in spiral galaxies, we have embarked on a campaign to observe 9 northern galaxies 
closer than 16 Mpc, with good measurements of their central black hole masses. The first analyzed target was the 
megamaser prototype NGC~4258 (M~106). It is the largest member of the Canes II group, populated mostly by late-type spirals and irregular galaxies. Thanks to a circumnuclear maser disk, NGC~4258 has the
most precise extragalactic $M_\bullet$ measurement. It also follows several correlations established for elliptical galaxies, like those between $M_\bullet$, and bulge luminosity and mass, respectively \citep{laes16}.     

Using the \uiks\ photometric method \citep{muno14}, described below in Section~\ref{sec:sample}, \citet{gonz17} selected a sample of GC candidates (GCCs) in NGC~4258 and determined that it likewise falls on the $N_{\rm GC}$ versus $M_\bullet$ correlation. Interestingly, the projected spatial distribution of the \citeauthor{gonz17} sample appears disky, and aligned with the galaxy, although their colors rule out that they are young clusters in the stellar disk (see \citealt{gonz17} and Section~\ref{sec:sample} below). The \uiks\ is possibly the most efficient photometric method to select GCCs in elliptical galaxies. This paper analyzes spectra of the GCC sample of NGC~4258, with two main goals: quantify the efficiency of the \uiks\ method for the study of the much sparser GC systems of spirals, and investigate the kinematics of the NGC~4258 GCs.  
The paper is organized as follows: Section~\ref{sec:sample} recapitulates the sample selection process carried out in \citet{gonz17}; Section~\ref{sec:specobs} presents the spectroscopic observations, their reduction, and the determination of the line-of-sight velocities of the candidates; in Section~\ref{sec:HIetc} we derive the kinematics of the system; the metallicity of the composite spectra of the whole sample is estimated in Section~\ref{sec:Z}; the efficiency of the \uiks\ method is analyzed in Section~\ref{sec:method}; finally, the discussion and conclusions are presented in Section~\ref{sec:concl}.   

\section{GCC sample} \label{sec:sample}

\citet{gonz17} defined a 
sample of 39 GCCs in NGC~4258, from 
Canada-France-Hawaii Telescope (CFHT) 
optical (\ust, \gp, \rp, \ip)  
and near-infrared (\ks) data, %\edit1{
both archival and proprietary. 
Potential candidates had photometric errors smaller than
0.2 mag in the \ip-band, and were within $\pm$ 3.6 mag of the expected GC luminosity (LF) function turnover (LFTO) magnitude in
every filter. Also, they were neither saturated, nor blended or close to bright neighbors. These latter requirements against
crowding eliminated all sources within an elliptical region centered on NGC~4258, with semi-major axis = 3$\farcm$5, or 0.37 $R_{25}$, axis ratio = cos(67$\degr$), and the same position angle (P.A.) as the galaxy, of $= 150\degr$ \citep{rc3}.
\hyperlink{tab:photobs}{Table 1} gives a summary of the photometric observations, including exposure times, depths, and
expected location of the LFTO. Given the significantly shorter exposure time, all the analysis by \citeauthor{gonz17} was limited by the
\ks-band image; the GCLF was complete to $\sim$ 20.5 AB mag in this wavelength \citep{gonz17}. %} 

%\documentclass{aastex62}
%\usepackage{lscape}
%\usepackage{wrapfig}
%\usepackage{rotating}

%\begin{document}

%\floattable
\begin{deluxetable}{crrrcc}[h]
%\begin{table}{CRRrcc}[h]
\hypertarget{tab:photobs}{}
\tablecaption{Summary of Photometric Observations}%\label{tab:photdata}
\tablewidth{0pt}
\tablehead{
\colhead{{\rm Filter}} & \colhead{$\lambda_{\rm cen}^a$} & \colhead{ {\rm FWHM}$^b$} & \colhead{Exposure} & \colhead{Depth} &\colhead{$m_0^{\rm TO}$} \\
\colhead{           }  & \colhead{                 }  & \colhead{          } & \colhead{s} &
\colhead{AB mag} & \colhead{AB mag} 
}
\startdata
  $u^*$        &  3793\ \AA & 654\ \AA &  13360 & 26.0 & 23.1   \\
  $g^\prime$   &  4872\ \AA & 1434\ \AA &  10400 & 26.5 & 22.2  \\
  $r^\prime$   &  6276\ \AA & 1219\ \AA &   3500 & 25.4 & 21.7  \\
  $i^\prime$   &  7615\ \AA & 1571\ \AA &   8080 & 25.7 & 21.5   \\
  $K_s$  &  2.15\ $\mu{\rm m}$ & 0.33\ $\mu{\rm m}$ & 200 & 21.5 & 21.3  \\ 
\enddata
\tablecomments{\ $^a$The central wavelength between the two points defining FWMH
(\url{http://svo2.cab.inta-csic.es/svo/theory/fps3/index.php?id=CFHT/}). \\
$^b$ {\it Ibid.}
}
%\end{table}
\end{deluxetable}
%\end{document}

\begin{figure}
%\hspace{3.5cm}\includegraphics[scale=0.40,angle=-90.]{fig1.pdf}
\includegraphics[width=0.8\linewidth,angle=-90.]{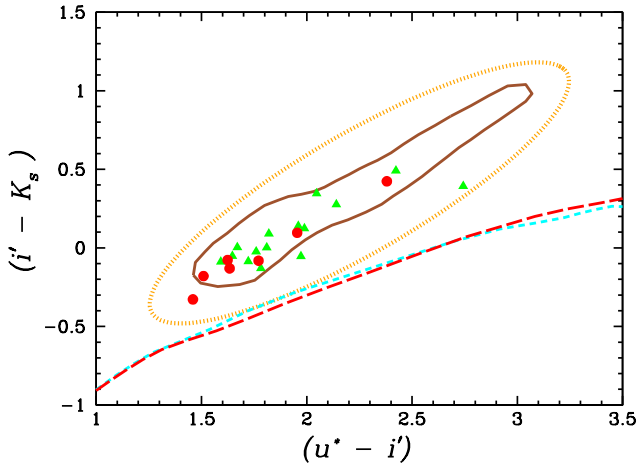}
%/Users/ragl/Documents/n4258_spec_paper/figs_w_captions
\caption{\uiks\ color--color diagram. The dotted orange line delimits the region occupied by BC03 SSPs with
ages 8--12 Gyr and metallicities $0.0004 \leqslant Z \leqslant 0.05$; the solid brown contour traces area with
the highest density of spectroscopically confirmed GCs in M87; the short-dashed cyan and long-dashed red lines are zero-age main sequences with
$Z=$ 0.008 and $Z=$ 0.02, respectively, and sketch the loci of stars in the Milky Way (MW). 
{\it Green triangles:} spectroscopically confirmed GCCs; {\it red circles:} GCCs rejected by spectroscopic observations (see text, section \ref{subsec:velocities}). 
\label{fig:regions}}
\end{figure}

Our main selection tool was the 
($u^*\ -\ i^\prime$) versus 
($i^\prime\ -\ K_s$) color-color diagram \citep{muno14}; this was the first application of the \uiks\ technique to a spiral galaxy. 
The GCC selection region in this diagram was determined from the area that  
contains the highest density (more than 2000) of spectroscopically confirmed
GCs in M~87 \citep{muno14,powa16}.
The region has approximately the shape of a drop or pennon, and
appears drawn in brown in Figure \ref{fig:regions}. 
Crucially, it is well separated from the loci occupied by
background galaxies, foreground Galactic stars, and young stellar clusters in the disk of spiral galaxies.
To further decrease the probability of contamination by background
galaxies, as well as by dwarf galaxies in NGC~4258 itself, only sources with \ip-band FWHM $\leqslant 0\farcs84$ (31 pc at a distance of 7.6 Mpc) and   
SExtractor \citep{bert96} light concentration parameter SPREAD\_MODEL $\leqslant$ 0.017 were kept.\footnote{
The SPREAD\_MODEL value for each object
results from the comparison between its best fitting point-spread-function (PSF), and the convolution of such PSF with
an exponential disk with scale length FWHM$_{\rm PSF}$/16, where FWHM$_{\rm PSF}$ is the
full width at half max of the same PSF \citep{desa12}.} Half-light radii of the objects, measured by fitting King profiles to
their radial light distributions with the program {\sc ishape} \citep{lars99}, 
were all smaller than 6 pc.

A couple of GCCs in the spectroscopic sample come from a previous, unpublished, 
photometric sample of 38 objects (the published and unpublished samples share 30 sources in common). 
In this case, the selection region in the \uiks\ diagram was
the ellipse centered at (\ust\ - \ip) $= 2.25$, (\ip\ - \ks) $= 0.35$,
with semi-major and semi-minor axes 1.25 mag and 0.345 mag, respectively,
and rotated 39$\degr$ counterclockwise from the (\ust\ - \ip) axis. 
This elliptical region contains the pennon, was delimited based on
\citet[][hereafter BC03]{bruz03} model single stellar populations (SSPs), and is drawn,
with a dotted orange line, in Figure~\ref{fig:regions}. 
The figure also shows two lines, short-dashed cyan and long-dashed red, that represent zero age main
sequences with, respectively, $Z = 0.008$ and $Z = 0.02$. They trace a band of Galactic
field stars that brushes the orange ellipse near (\ust\ - \ip) $\sim$ 1.8,
(\ip\ - \ks) $\sim$ -0.4. In view of potential contamination from these objects, 
the unpublished sample also had minimum size bounds in the \ip\ filter
(FWHM = $0\farcs65$, SPREAD\_MODEL = 0.002). 
These minimum size bounds, however, would have eliminated GCs with $r_e <$ 4.7 pc, or
about half the total in the MW.  

GCCs in either of the photometric samples are identified with the acronym GLL (``Globular Luis Lomel\'{\i}") before their
equinox J2000.0 equatorial coordinates. 
The spectroscopic sample, enumerated in \hyperlink{tab:velocities}{Table 2}, includes GLL GCCs, objects in either of the color-color
selection regions that did not fulfill the light concentration requirements, and random objects in the observed fields as additional controls.
The two latter classes of objects are listed by their equinox J2000.0 equatorial coordinates, with no preceding acronym.   

%\documentclass{emulateapj}
%\begin{document}
%\clearpage
% \begin{center}

\begin{sidewaystable*}[!t!h]
\vspace*{8cm}
%\begin{table}[t]
%\hypertarget{tab:velocities}{}
\caption{Target velocity fit parameters, {\sl HST} image information, and probable object types.}
 \begin{center}
% \begin{minipage}{280mm}
  \begin{tiny}
\begin{tabular}{@{}ccccccccccccccc@{}}
\hline
\hline
Name &   RA J2000 & DEC J2000  & SNR  & $V_{\rm proj}$& $\Delta V_{\rm proj}$ &  $r$-value &  Template &   Range & $V_{\rm HI, proj}$ & $\Delta V_{\rm HI, proj}$ & HST image & FWHM     & FWHM PSF & Object type\\
     &     deg    &    deg     &      & km s$^{-1}$   & km s$^{-1}$           &            &           &   $\AA$ &  km s$^{-1}$       &  km s$^{-1}$              &           &$\arcsec$ & $\arcsec$ &  \\ 
\hline
\multicolumn{11}{c}{Confirmed GC candidates}\\
\hline
%YES/YES
GLL J121851+471400 & 184.7166 &   47.2335    &  28  &    324.0 &   72.6  &  4.92     &   fglotemp &   4400--7100    & 378.5   & 14.9   &  hst\_11570\_97\_acs\_wfc\_f555w & 0.160 & 0.100 & GC  \\%  3_30 = 26_30new.ms_6  33  19
GLL J121852+472201 & 184.7205 &   47.3672    &  11   &    779.4 &   82.3   &    2.58   &   fglotemp &   4100--7198    & 573.4  & 36.8   & hst\_11570\_86\_acs\_wfc\_f555w & 0.130 & 0.105 & GC \\% 1_26  = 23_26.ms_4      3  10
GLL J121854+472318 & 184.7266 &   47.3885    &  31  &  496.9  &   57.4   &   5.35    &   fglotemp &   4100--7198    & 555.1  & 23.1   & hst\_11570\_91\_acs\_wfc\_f555w & 0.195 & 0.100 & GC \\% 1_31   = 27_35.ms_3     67 23
GLL J121856+471411 & 184.7345 &   47.2366    &  12  &  315.9  &   51.7   &   3.67    &   fs2temp  &   4000--7198    & 364.0  & 21.1   & hst\_11570\_97\_acs\_wfc\_f555w & 0.165  &  0.105 & GC \\%  3_26 = 26_30new.ms_2 17 11
GLL J121857+472330 & 184.7384 &   47.3918    &  23  &  338.7  &   48.3   &   8.78    &   fglotemp &   4100--7198    & 528.2  &  8.7  & hst\_11570\_91\_acs\_wfc\_f555w  & 0.155 & 0.105  & GC\\% 1_32  = 27_35.ms_5     28 16
GLL J121859+472107 & 184.7500 &   47.3520    &  52  &  662.2  &   69.6   &    6.00   &   fglotemp &   3720--7100    &  514.3  & 25.8   & hst\_11570\_86\_acs\_wfc\_f555w & 0.150 & 0.100  & GC\\% 1_21  = 19_22.ms_3     54 43
J121901+471859$^a$     & 184.7553 &   47.3166    &  9  &  577.3  &   49.4   &    4.31   &   fs2temp  &   4000--7000    &449.5   & 44.3   & hst\_11570\_88\_acs\_wfc\_f555w & 0.105 & 0.100  & GC \\% 1_4 = 4_5.ms_1 15 10, marg. resolved.     
GLL J121901+471905 & 184.7565 &   47.3183    &  15  &  260.1  &   68.7   &    3.71   &   fglotemp &   4100--7000    & 448.4  & 43.8   & hst\_11570\_88\_acs\_wfc\_f555w & 0.155 & 0.095  & GC \\% 1_5   = 4_5.ms_2   20 15
GLL J121902+472048 & 184.7585 &   47.3469    &  33  &  383.8  &   52.4   &    5.44   &   fglotemp &   4000--6950    & 488.6  & 22.9   & hst\_11570\_86\_acs\_wfc\_f555w & 0.175 & 0.095 & GC \\% 1_20  =  19_22.ms_2  42 33
GLL J121902+472125 & 184.7622 &   47.3572    &  23  &  608.3  &   23.6   &    9.82   &   fglotemp &   4000--6850    & 488.8  &  11.8  & hst\_11570\_99\_acs\_wfc\_f555w & 0.125 & 0.090  & GC\\% 1_23  = 23_26.ms_1   22 28
GLL J121902+472043 & 184.7624 &   47.3453    &  35  &  625.8  &   31.4   &    7.78   &   fglotemp &   4000--6850    & 477.0  & 9.0   & hst\_11570\_88\_acs\_wfc\_f555w & 0.155 & 0.100  & GC\\% 1_19   = 19_22.ms_1    43 34
GLL J121908+471355 & 184.7847 &   47.2322    &  27  &  257.9  &   47.5  &   5.86    &   fglotemp &   4000--7400    & 322.2  & 31.3   & hst\_11570\_96\_acs\_wfc\_f555w & 0.140 & 0.100  & GC\\%  3_18  = 15_18new.ms_4  22  25
GLL J121925+471228 & 184.8578 &   47.2079    &  36  &  282.0  &   39.0   &  7.03     &   fglotemp &   4200--7198    & 289.0   &  14.2  & hst\_11570\_94\_acs\_wfc\_f555w & 0.165 & 0.100  & GC \\%  3_7 = 1_11new.ms_5  47  36
GLL J121928+471152$^b$ & 184.8685 &   47.1978    &  17  &  505.5  &   72.8   &  4.19     &   fglotemp &   4300--7198    &290.8   & 14.2   & hst\_11570\_94\_acs\_wfc\_f555w & 0.200 & 0.100  & GC\\%  3_5 = 1_11new.ms_4  26 16, OLD ONLY
GLL J121929+471215 & 184.8711 &   47.2042    &  54  &  219.5  &   17.1   &  14.02    &   fglotemp &   4250--7198    & 298.9  &  16.3  & hst\_11570\_94\_acs\_wfc\_f555w & 0.150 & 0.100 & GC \\% 3_4 = 1_11new.ms_3     61 56
GLL J121934+471316 & 184.8944 &   47.2212    &  21  &  401.2  &   66.6   &   4.03    &   fglotemp &   3720--7198    & 344.7  &  6.0  & hst\_10399\_50\_acs\_wfc\_f606w & 0.170 & 0.080  & GC\\% 3_1  = 1_11new.ms_1    43  24
\hline
\multicolumn{11}{c}{Rejected GC candidates}\\
\hline
%YES/NO
GLL J121852+471313 & 184.7187 &   47.2204    &  12  &  -209.3  &   47.9   &  5.66     &     eltemp &   4100--7198    & 368.1  & 15.3   & hst\_11570\_97\_acs\_wfc\_f555w & 0.105 & 0.100  & star\\%  3_29 = 26_30new.ms_5 23 11 v2  11570 92 97* 555, marg. resolved
GLL J121854+472144$^c$ & 184.7261 &   47.3625    &  8  & 543.6     &   53.7   &    4.69   &  fm32temp  &   4100--7198    & 564.3  &  32.6 & hst\_11570\_90\_acs\_wfc\_f555w & 0.155 & 0.090 & dwarf gal/GC \\% 1_25  = 23_26.ms_3    8  7   
GLL J121854+472245 & 184.7278 &   47.3792    &   53 &  -199.3   &   14.6   &   18.59   &   fm32temp     &   4000--7100    & 558.8  & 25.4   & hst\_11570\_91\_acs\_wfc\_f555w & 0.095 & 0.10& star \\% 1_28 = 27_35.ms_2  79 53   v2 only, POINT SOURCE
GLL J121903+472613$^b$ & 184.7654 &   47.4372    &  20  & 470.2 & 158.1    & 1.64      & fglotemp &   4300--5500$^{d}$    & 512.6  & 7.7   &  no images & & & dwarf gal/GC  \\%    1_41    = 38_41.ms_1  23 17    v1 only 
GLL J121905+472422 & 184.7744 &   47.4063    &  86  &  750.7  &   49.7   &   5.42    &   fm32temp     &   4100--6600   & 498.4   & 6.1   & hst\_11570\_91\_acs\_wfc\_f555w & 0.285 & 0.100 & dwarf gal/GC\\% 1_35  = 27_35.ms_6  95 70 SI/NO     both
GLL J121909+471335 & 184.7909 &   47.2265    &  12  & 400.3  & 101.4   &  2.43     & eltemp   &     4300--5500$^{e}$    & 309.4  & 34.5   & hst\_11570\_96\_acs\_wfc\_f555w & 0.160 & 0.100 & dwarf gal/GC \\% 3_16 = 15_18new.ms_2 13 10 both
GLL J121910+471343 & 184.7931 &   47.2287    &  23  &  172.2  &  104.9   &   4.34    &   fglotemp &   4000--7200    & 308.9   & 37.1   & hst\_11570\_96\_acs\_wfc\_f555w & 0.250 & 0.100 & dwarf gal/GC \\%  3_15 = 15_18new.ms_1  31 19       both
\hline
\multicolumn{11}{c}{Confirmed non-candidates}\\
\hline
%NO/NO
J121853+472136     & 184.7219 &   47.3600    &   21 &  853.5  &   36.2   &    8.05   & fn7331temp &   4100--7198    & 578.3  & 45.7   & hst\_11570\_90\_acs\_wfc\_f555w & 0.235 & 0.100 & dwarf gal (?)  \\% 1_24 = 23_26.ms_2 26 22 NO/NO
J121907+472113     & 184.7795 &   47.3538    &   19 &  43004.9  &   31.1   &    7.90   &   eltemp   &   4000--6500   & 472.4  & 15.6   & hst\_11570\_99\_acs\_wfc\_f555w & 0.240 & 0.090 & gal  \\% 1_22 = 19_22.ms_4  45  40 NO/NO
J121924+471306     & 184.8518 &   47.2185    &   12 &  12576.2  &   19.8   &  5.66     &   femtemp97&   4100--7198    & 300.1  &  12.6  & hst\_11570\_94\_acs\_wfc\_f555w & 0.115 & 0.100 & gal\\%  3_9 = 1_11new.ms_6  9 12
%NOT IN ANY SAMPLE
    J121849+471357 & 184.7042 &   47.2325    &   2  &  38418.3 &   46.6   &  3.26     &   femtemp97&   3720--7198    & 386.2  & 14.6   & hst\_11570\_97\_acs\_wfc\_f555w & 0.100  & 0.100 & gal (?)\\%  3_32 = 26_30new.ms_7 3 2.3 Diva  
    J121853+471330 & 184.7232 &   47.2250    &   19 &  329.3  &   16.4   &  10.32    &   femtemp97&   3720--7198    & 366.4   & 16.2   & hst\_11570\_97\_acs\_wfc\_f555w & 0.117 & 0.100 & HII reg\\%  3_28 = 26_30new.ms_4 34 19 Diva HII saturated?  
    J121854+472235 & 184.725  &   47.3764    &   24 &  618.8  &   13.9   &   12.74   &   femtemp97&   3720--7198    & 564.4   & 25.3  & hst\_11570\_91\_acs\_wfc\_f555w & 0.100 & 0.100 & HII reg\\% 1_27 = 27_35.ms_1 52 19 HII reg Diva
    J121854+471324 & 184.7284 &   47.2235    &   18 & 10015.5  &   99.5   &   2.67    &   eltemp   &   4100--7198    & 363.2   &  17.0  & hst\_11570\_97\_acs\_wfc\_f555w & 0.115  & 0.100 & gal\\%  3_27  = 26_30new.ms_3 18 16 Diva 
    J121856+471342 & 184.7374 &   47.2286    &   12 &   380.4 &   16.0   &   10.70   &   femtemp97&   3720--7198    & 356.8  & 20.8   & hst\_11570\_90\_acs\_wfc\_f555w & 0.120 & 0.090 & HII reg\\%  3_25 26_30new.ms_1  11 15 Diva 
J121920+471332     & 184.8365 &   47.2256    &   12 & 104927$^{e}$  &     &      &     &        & 302.8  & 12.2   & hst\_11570\_94\_acs\_wfc\_f555w & 0.310 & 0.100 & gal\\%  3_11 = 1_11new.ms_7 11 13 Diva z=0.35 
J121930+471254     & 184.8786 &   47.2152    &    3 & 43234.8  &   51.8   &   2.17    &   femtemp97&   4000--7198    & 321.3  & 11.1   & hst\_11570\_94\_acs\_wfc\_f555w & 0.267 & 0.100 & gal (?)\\% 3_3 = 1_11new.ms_2 5 3 Diva, looks like sp. galaxy  11570 94* 555
\hline
\vspace*{-0.5cm}
\end{tabular}
\end{tiny}
%\end{minipage}
\end{center}
\begin{scriptsize}
$^a$ Rejected from unpublished sample due to small size, and from published \citep{gonz17} sample because of its colors (see text).

%\vspace{-3pt}

$^b$ In unpublished sample, then rejected from published sample because of its colors.

%\vspace{-3pt}

$^c$ Should not have been in sample, owing to axis ratio $<$ 0.7.

%\vspace{-3pt}

$^d$ With full spectral range (4000--6800 $\AA$), best correlation ($r$-value 2.04) found for template sptemp, with $V_{\rm proj} = 18461\pm85$ km s$^{-1}$. However, spectral lines are consistent with 

%\vspace{-3pt}

an old stellar population at the systemic velocity of NGC~4258 (Figure 4).

%\vspace{-3pt}

$^e$ With full spectral range (4000--7400 $\AA$), best correlation ($r$-value 2.86) found for template sptemp, with $V_{\rm proj} = 12141\pm79$ km s$^{-1}$. However, spectral lines are consistent with  

%\vspace{-3pt}

a stellar population at the systemic velocity of NGC~4258 (Figure 4).

%\vspace{-3pt}

$^f$ No available {\it rvsao} template fits the very obvious emission lines. The highest $r$-value is 3.29 for sptemp, with $V_{\rm proj} = 1071\pm54$ km s$^{-1}$. fn7331temp comes closest to the real 

%\vspace{-3pt}

measurement; it yields $z =$ 0.31, with $r =$ 0. 

%\vspace{-5pt}

\end{scriptsize}
%\label{tab:velocities}
%\end{table}
\end{sidewaystable*}
% \end{center}

%\end{document}

\section{Spectroscopic observations} \label{sec:specobs}

The dataset used in this work 
was obtained with the Optical System for Imaging and low-Intermediate-Resolution
Integrated Spectroscopy (OSIRIS), in multi-object spectroscopy (MOS) mode, at the 
Nasmyth-B focus of the Gran Telescopio Canarias (GTC).\footnote{
Gran Telescopio Canarias is a Spanish initiative with the participation
of Mexico and the University of Florida, USA, and is installed
at the Roque de los Muchachos Observatory in the island of La Palma.
This work is based on the proposal GTC17-9AMEX, using Mexican share of the GTC time.} 
OSIRIS has two mosaicked Marconi CCD44-82 detectors, with a $9\farcs4$ gap
between them. Each detector has 2048$\times$4096 pixels,
whose physical size is 15$\mu$m on the side, for a scale of $0\farcs127$ on the sky. 
In standard observing modes, however, pixels are binned 2$\times$2, for an
effective spatial scale of $0\farcs254$.
We observed with the R1000B grism, whose wavelength range goes from $\sim$3700~\AA\ to 7500~\AA.
The slit width of 1\arcsec\ yielded a resolution of 5.51~\AA.
OSIRIS offers a maximum field-of-view (FOV) of $7.5^\prime \times 6.0^\prime$ for slitlet placement.
However, we used only a FOV of $7.5^\prime \times 3.0^\prime$ in each of our two 
pointings in order to ensure the coverage of the complete spectral range of 
3700~\AA--7500~\AA\ for all observed targets.

The dataset consists of spectra in two fields, located, respectively, north (NS1) and south (EW4) of the minor axis of NGC~4258.
The NS1 field was oriented N-S, with the dispersion axis running E-W; conversely, the EW4 field was
oriented E-W, with the dispersion axis along the N-S direction. Their centers were, respectively,
 RA = 12h18m52.10s, Dec=+47$\degr19\arcmin51\farcs8$, (J2000.0) for field NS1, and 
RA=12h19m27.17s, Dec=+47$\degr13\arcmin16\farcs4$ (J2000.0) for field EW4.
 
\begin{figure*}
\begin{tabular}{ll}
%\hspace*{0.3cm}\includegraphics[scale=0.80]{fig2_a1.pdf}
\includegraphics[width=0.4\linewidth]{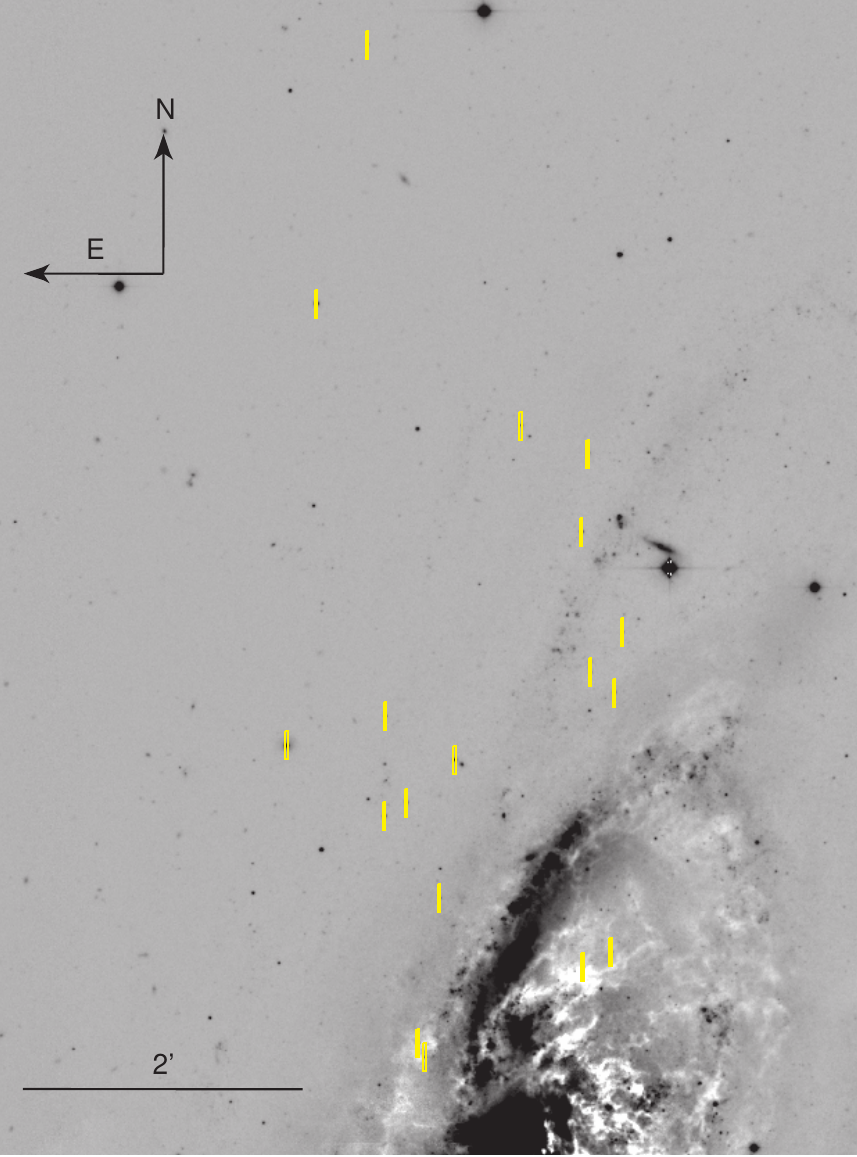}
%/Users/ragl/Documents/cat_gtc/maskDesigner_v325_2017/2017A/NS1
&
\includegraphics[scale=0.62]{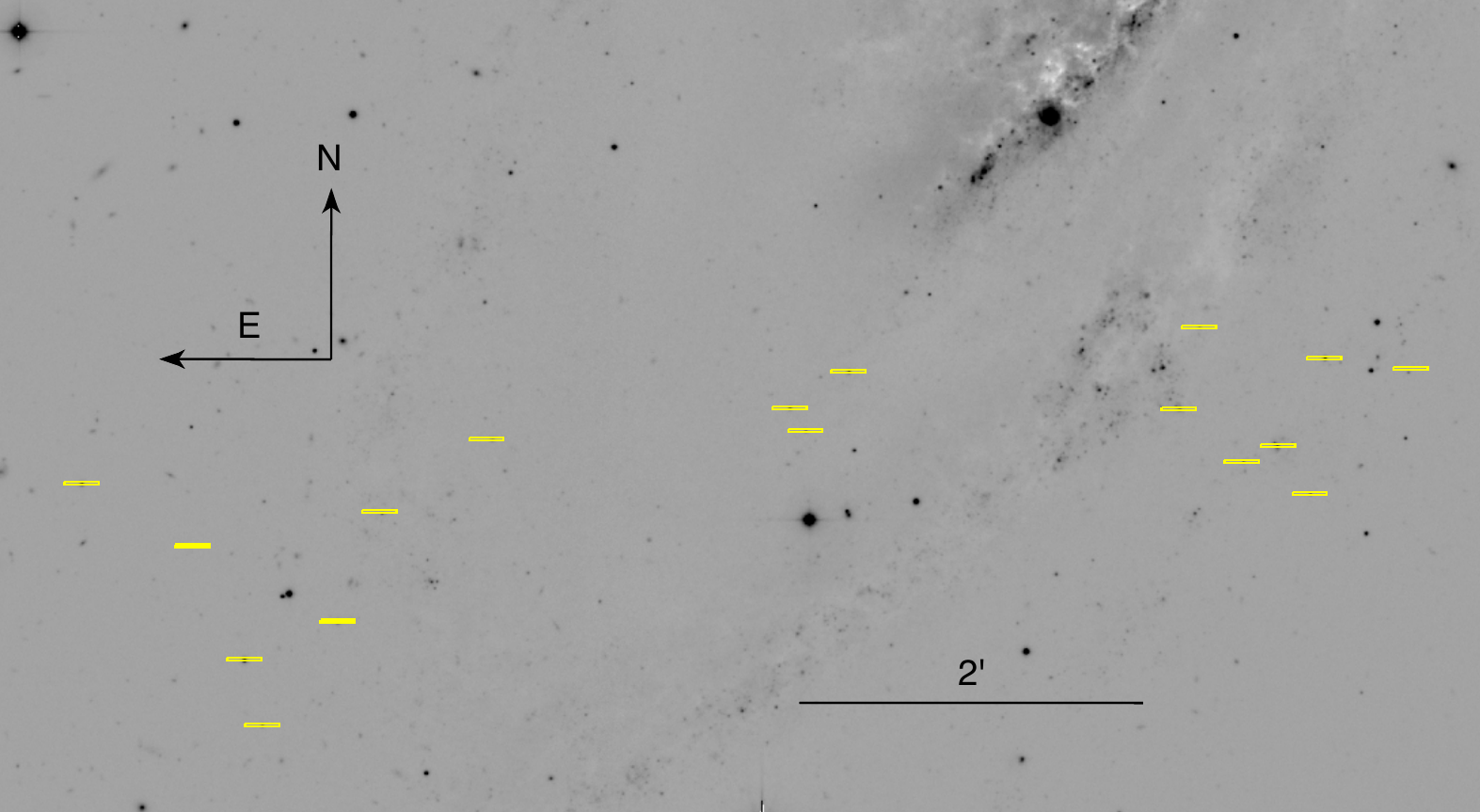}
%/Users/ragl/Documents/cat_gtc/maskDesigner_v325_2017/2017A/EW4
%cropped_ed only in /Users/ragl/Documents/n4258_spec_paper
\end{tabular}
\caption{Sections of archival CFHT \rp-band image of NGC 4258 showing, in yellow, the location of slitlets
observed with GTC/OSIRIS. {\it Left:} field NS1, centered at RA = 12h18m52.10s, Dec=+47$\degr19\arcmin51\farcs8$ (J2000.0); 
{\it right:} field EW4, centered at RA=12h19m27.17s, Dec=+47$\degr13\arcmin16\farcs4$ (J2000.0).
\label{fig:fields}}
\end{figure*}

Spectra were obtained for slightly over 50\% of the GCC sample defined by \citet{gonz17}.
The positions of the science slits in each field are shown 
as yellow boxes, over grayscales 
of the CFHT \rp\ image of the galaxy, in
Figure~\ref{fig:fields}. For illustration purposes, all slits have been drawn
with the same size; in actuality,  
slitlets had varying lengths, so as to not overlap, and included object-free regions to allow for
the subtraction of sky spectra. 
Observations of reference stars for astrometry were also carried out (positions not shown).

Each field was observed for a total of 3975 s, divided in 3 exposures of 1325 s to
facilitate cosmic ray cleaning. Standard stars, arc lamps, bias, and flat-field frames were also
obtained for each pointing. 
Field NS1 was secured on 2017 May 24, i.e., a clear, dark night; field EW4 was observed during 2017 May 30 and 31,
which were gray nights with some cloud cover. Consequently,
while the average signal-to-noise ratio (SNR) of the sources in field NS1 is $\sim$ 29,
for those in field EW4 it is only $\sim$ 19, as gauged by the routine DER\_SNR.\footnote{Stoehr, F., 
ST-ECF Newsletter, Issue \# 42, \url{www.stecf.org/software/ASTROsoft/DER\_SNR/}.} 
A summary of the spectroscopic observations is given in \hyperlink{tab:obs}{Table 3}.
%A summary of the \edit1{spectroscopic} observations is given in \hyperlink{tab:obs}{Table 3}.

%\documentclass{emulateapj}
%\begin{document}
\begin{table*}[t]
%\hypertarget{tab:obs}{}
\caption{Observation Log}
 \begin{center}
 \begin{minipage}{280mm}
  \begin{tiny}
\begin{tabular}{@{}ccccccccccc@{}}
\hline
\hline
Run/Field &  Principal Investigator    & Date      & Position angle  & Slit Width & Grism & Exp.\ time    & Airmass  & Seeing     & Night           & Standard star \\
         &                            &            & $\degr$         & $\arcsec$  &       &     s         &          &  $\arcsec$ &                 &     \\ 
\hline
2017A-NS1& R.\ Gonz\'alez-L\'opezlira & 2017-05-24 & 270             &  1.0       & R1000B& 3$\times$1325 &  1.42    &  $0\farcs7$            &  Dark, clear    & Ross\ 640    \\
2017A-EW4& R.\ Gonz\'alez-L\'opezlira & 2017-05-30 & 0               &  1.0       & R1000B& 2$\times$1325 &  1.24    &     $0\farcs8$         &  Gray, clouds   & GD153    \\
2017A-EW4& R.\ Gonz\'alez-L\'opezlira & 2017-05-31 & 0               &  1.0       & R1000B& 1325 &  1.39    &     $0\farcs8$       &  Gray, clouds   & GD153    \\
\hline
\vspace*{-0.5cm}
\end{tabular}
\end{tiny}
\end{minipage}
\end{center}
%\label{tab:observations}
\end{table*}
%\end{document}

\subsection{Data reduction} \label{subsec:specred}

The reduction of the spectroscopic data was performed with {\it GTCMOS},\footnote{
{\url http://www.inaoep.mx/$\sim$ydm/gtcmos/gtcmos.html}
} an ad hoc pipeline based on IRAF\footnote{
IRAF is distributed by the National Optical Astronomy Observatories, 
    which are operated by the Association of Universities for Research 
    in Astronomy, Inc., under cooperative agreement with the National 
    Science Foundation.
} \citep{tody86,tody93} tasks. The reduction procedure has been described at length by 
\citet{gome16}, but we give a brief summary here.

The pipeline tiles the two OSIRIS CCDs and corrects for geometrical distortions. All tiled bias frames
in an observing block are combined into a master bias image with a median algorithm; the master bias is
then subtracted from all the images in the block. All images of each target 
are then combined, also with a median algorithm, a procedure that cleans cosmic rays from the
final frames. Arc spectra are also tiled and combined into a single arc image, which is used
for wavelength calibration. The pipeline also corrects the curvature of the slit images in the
spatial direction and obtains independent dispersion solutions for each slitlet, using a
spline3 function of order 2. Final rms errors are better than 0.5 ${\rm \AA}$. The best solution for
each slitlet is used to create a wavelength calibrated 2-D image, with the spectral axis linearly
resampled to a dispersion of 2.1 ${\rm \AA}$ pixel$^{-1}$. The dispersion-corrected spectra are also
shifted linearly, in order to force the centroid of the [OI]$\lambda$5577 sky line in every spectrum
to lie at its rest wavelength.  

Standard star spectra are reduced individually in a similar fashion. Once spectra are extracted, 
sensitivity tables between the flux and count rate at all available spectral
bands are obtained with the IRAF task {\it standard}. Atmospheric extinction corrections are
performed, for both standard stars and targets, by combining the extinction curve for the 
observatory with the registered airmasses of the observations. 

The IRAF task {\it apall} was used for the extraction of the spectra of individual targets from
the wavelength- and flux-calibrated 2-D images.
The background (sky plus underlying galaxy) spectrum was also subtracted during the extraction procedure. 
For all the targets the continuum is well registered so as to enable in-situ tracing. 
The width of the extraction window is kept between 4--6~pixels (1--1.5\arcsec), which has been found 
to maximize the SNR of the extracted spectrum. Most of the slitlets were 
long enough to include object-free pixels for the extraction of the background spectrum. 
In a few cases, where this was not possible, background zones were selected from slitlets 
that were spatially closest to the object.

The sky plus underlying galaxy background is as bright or, very often, brighter than the
object within the extraction window. Furthermore, sky noise increases monotonically towards
the blue, i.e., almost by a factor of two between 5000~\AA\ to 4000~\AA.
Thus, the SNR of the extracted spectrum, especially blueward of 5000~\AA, is mainly 
governed by the photon noise of the background spectrum. We considered only those 
parts of each spectrum where the SNR pixel$^{-1}$ (of 2.1~\AA\ width) is above at least 10.
In \hyperlink{tab:velocities}{Table 2} we give the SNR, again as estimated by the routine DER\_SNR,
as well as the useful wavelength range for each spectrum.

\subsection{Line of sight velocities} \label{subsec:velocities}

Projected heliocentric velocities of the GCCs and other targets were determined with the 
Fourier cross-correlation method implemented in the {\it xcsao} task of the 
{\it rvsao} package \citep{kurt98}, within IRAF. For comparison with our
spectra, we used some of the templates provided by the Smithsonian Astrophysical
Observatory (SAO; \url{http://tdc-www.harvard.edu/software/rvsao/Templates/}).
We employed the two templates derived from M~31
GCs (called {\it fglotemp} and {\it fs2temp}); a synthetic galaxy emission line spectrum 
({\it femtemp97}) that
provides a good match to HII regions; four galaxy templates, i.e., a composite 
spiral galaxy template, a composite elliptical galaxy template, one derived from M~32 and one produced from NGC~7331 
(respectively, {\it sptemp, eltemp, fm32temp, fn7331temp}). All templates were resampled to
the resolution of the spectra obtained with OSIRIS/GTC and reduced with {\it GTCMOS}. 
The goodness of fit is gauged through the $r$-value, also provided by the {\it xcsao} task for each comparison; the
higher the $r$-value, the better the fit. 
The sky region, between 5500 ${\rm \AA}$ and 5650 ${\rm \AA}$, was always excluded from the fits, 
as were the edges of the spectra in the cases where they looked noisy.

For an object to be considered a GC, the highest $r$-value should occur for the M~31 GC templates 
%\edit1{in the whole observed spectral window},  
in the whole observed spectral window,  
and the derived heliocentric velocity had to lie in the expected range of objects in NGC 4258, i.e., 
between $\sim$ 300 km s$^{-1}$ and $\sim$ 750 km s$^{-1}$. Objects that had a good match
in the same velocity range with the template femtemp97 are likely HII regions.
All other objects were considered either
as foreground stars (velocity less than the minimum allowed, and FWHM similar
to the PSF FWHM --see \hyperlink{tab:velocities}{Table 2}, columns 13 and 14), dwarf satellite galaxies (velocity
in the range), or background galaxies (velocity larger than the maximum allowed).

\begin{figure*}
\begin{tabular}{ll}
%\hspace*{0.3cm}\includegraphics[scale=0.477,angle=0.]{footprints_hla_ed_cropped.pdf}
%\hspace*{0.3cm}\includegraphics[scale=0.477,angle=0.]{fig3_a1.pdf}
\includegraphics[scale=0.477,angle=0.]{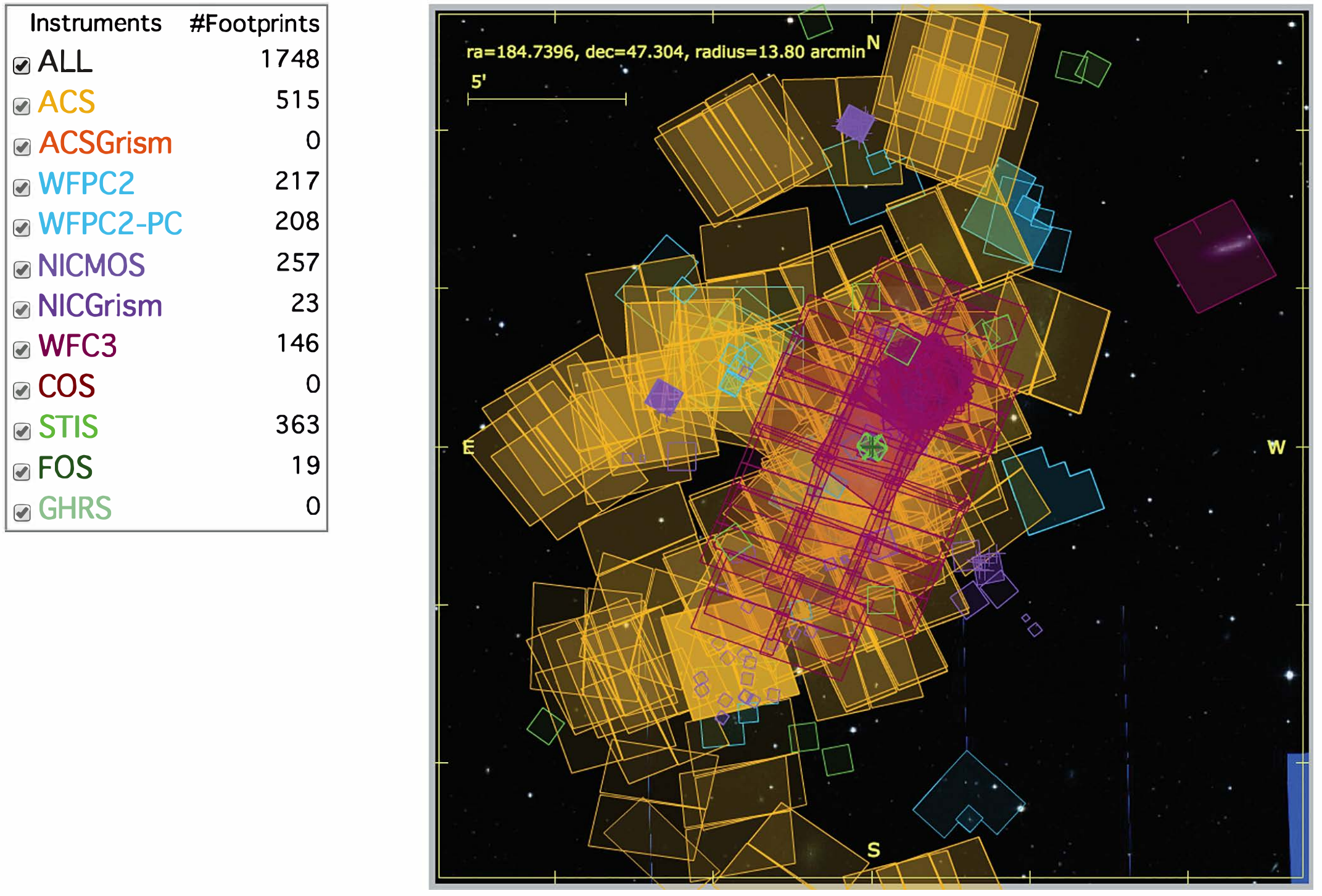}
&
%\hspace*{0.3cm}\raisebox{0.1cm}{\includegraphics[scale=0.470]{footprints_cropped.pdf}}
%\hspace*{0.3cm}\raisebox{0.1cm}{\includegraphics[scale=0.470]{fig3_a2.pdf}}
\includegraphics[scale=0.470]{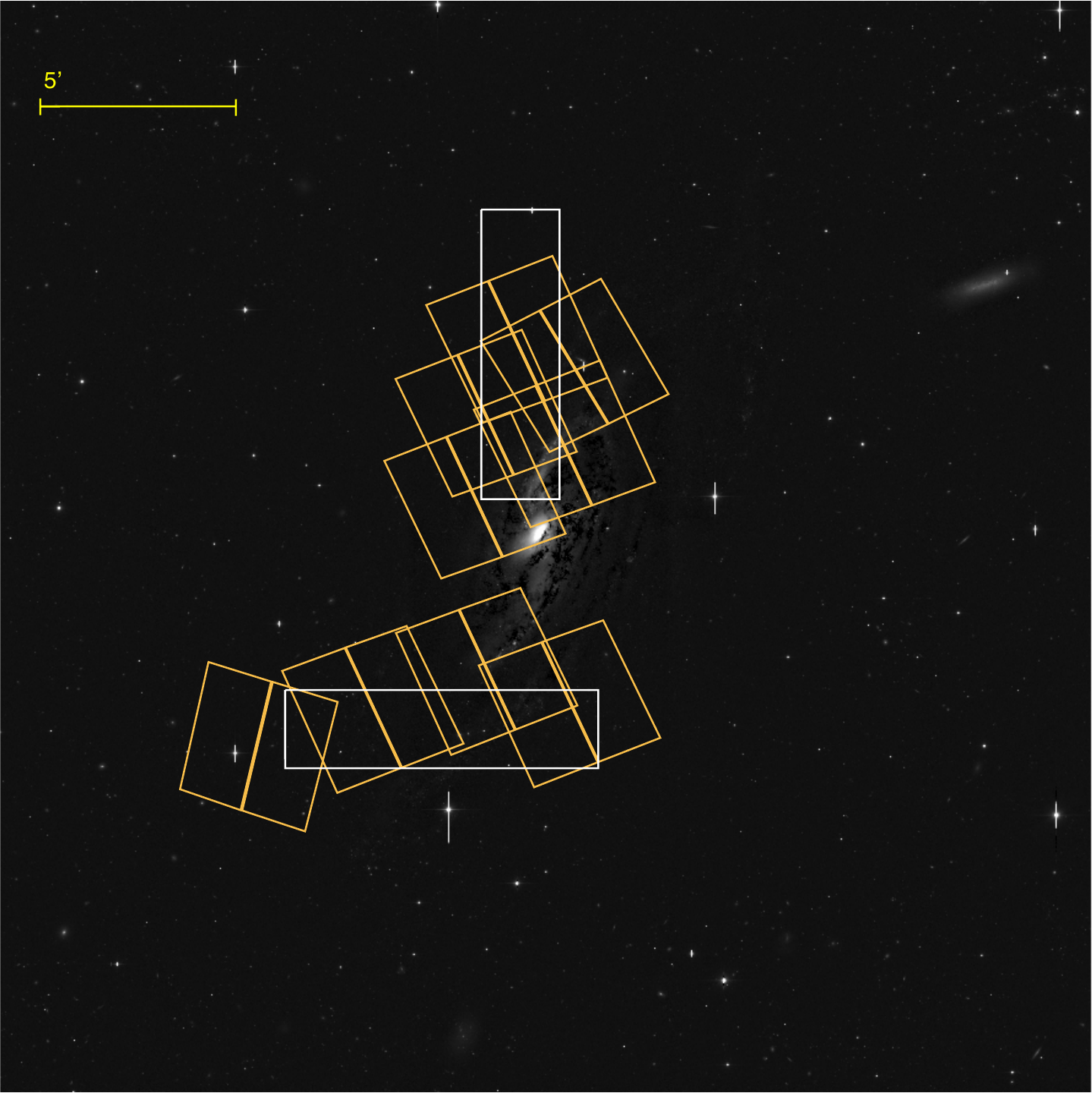}
\end{tabular}
\caption{{{\sl HST} footprints of NGC~4258 data. {\it Left:} all available Hubble Legacy Archive data, superimposed on a DSS image. The footprints for each instrument are coded by color; the legend indicates the number of images obtained by each camera or spectrograph. The data are so plentiful that
the galaxy is not visible under the footprints. {\it Right:} footprints of ACS frames used in this work ({\it golden yellow}), superimposed on the CFHT \ip-band data of the galaxy; the outlines of the fields observed with the GTC are shown in white. The physical and display scales in both panels have been roughly matched.} 
\label{fig:footprints}}
\end{figure*}

The classification of all objects was further visually confirmed through the inspection of 
Hubble Space Telescope ({\sl HST}) archival images, 
%\edit1{
as NGC~4258 has one of the largest Hubble Legacy Archive (HLA)\footnote{\url{https://hla.stsci.edu}.} 
data collections available (Fig.~\ref{fig:footprints}, 
left panel\footnote{The graphic was produced using the HLA site, with the footprints superimposed on a
Digitized Sky Survey (DSS; \url{http://archive.stsci.edu/dss/}) image.}).
All the frames used in this work were acquired with the Advanced Camera for Surveys (ACS); their footprints are
illustrated in golden yellow, in the right panel of Fig.~\ref{fig:footprints}, superimposed on the CFHT \ip-band data of NGC~4258.  Except in one case, the images were obtained through the
%} 
F555W filter and proceed from proposal GO 11570 (P.I.\ A.\ Riess).  
For GLL J121934+471316, we used an F606W image from proposal GO 10399 (P.I.\ L.J.\ Greenhill); no {\sl HST} image 
was found for GLL J121903+472613. 
%\edit1{
Via this comparison with ACS data, for example, we assessed that GLL J121854+472245 is likely a star; we
also rejected GLL J121910+471343, given its very elongated aspect, in spite of having a best match with a GC spectral template.  %}
Confirmed and rejected GCC candidates are shown as, respectively, green triangles and red circles in Figure~\ref{fig:regions}.

For each observed source, together with its name and J2000 coordinates, \hyperlink{tab:velocities}{Table 2} lists its projected velocity $V_{\rm proj}$
and error $\Delta V_{\rm proj}$; the $r$-value of the cross-correlation; the velocity template with the best fit; the spectral
range over which the cross-correlation was performed; the projected velocity of the HI gas at the position of the 
source $V_{\rm HI, proj}$ and its error $\Delta V_{\rm HI, proj}$ (see below, Section~\ref{sec:HIetc}); the archival {\sl HST} image inspected; the FWHM of the 
object; the FWHM of the point-spread-function (PSF) of the image; the probable object type.   The table has three sections: confirmed candidates; 
rejected candidates; confirmed non-candidates. Among the rejected candidates, given their spectra 
and projected velocities, 
GLL J121854+472144, GLL J121903+472613, GLL J121905+472422, GLL J121909+471335, and
GLL J121910+471343 are consistent with being satellite dwarf galaxies of NGC~4258. 
The first three objects in the latter section (confirmed non-candidates) were
within the elliptical selection region in the \uiks\ diagram described in Section~\ref{sec:sample}, i.e., their colors resemble those of GCs. J121907+472113 had been discarded mainly on account of its angular size, and indeed it is a spiral galaxy at $z= 0.14$. J121924+471306 was blended in the CFHT data and has a small axis ratio (0.5); it appears to be a galaxy at $z= 0.04$, 
with emission lines in its spectrum. J121853+472136, however, although blended in the CFHT data and with a small axis ratio (0.6), could be associated with NGC~4258, given its spectrum and projected velocity. 
The remaining non-candidates were observed only as a consistency check because they fell on the two fields targeted with the GTC.   

Figure~\ref{fig:specim_conf} shows, for each confirmed candidate, its spectrum in observed wavelength and {\sl HST} image.
For most objects, the spectra are presented in two windows, one shortward of 5450 $\AA$, and one showing the H$\alpha$ region. 
Spectra have been smoothed with a boxcar of size 5 pixels.
The most characteristic spectral lines of old stellar populations or HII regions are indicated.\footnote{
We note here that lines are not conspicuous in either GLL J121856+471411 or GLL J121901+471859; however, 
the SNRs of their spectra are not high enough to confidently eliminate them from the list 
of confirmed candidates on this basis. 
}

Spectra and images of rejected candidates 
and confirmed non-candidates are presented, respectively, in Figure~\ref{fig:specim_rej} and Figure~\ref{fig:specim_non}.

The spectrum and image of one object within each category are given in the main body of the text; 
all three figures are continued in the Appendix.

\begin{figure*}[t]
\begin{tabular}{ll}
%\hspace*{0.3cm}\raisebox{-0.57cm}{\includegraphics[scale=0.5,angle=0.]{fig4_a1.pdf}}
\raisebox{-0.57cm}{\includegraphics[scale=0.5,angle=0.]{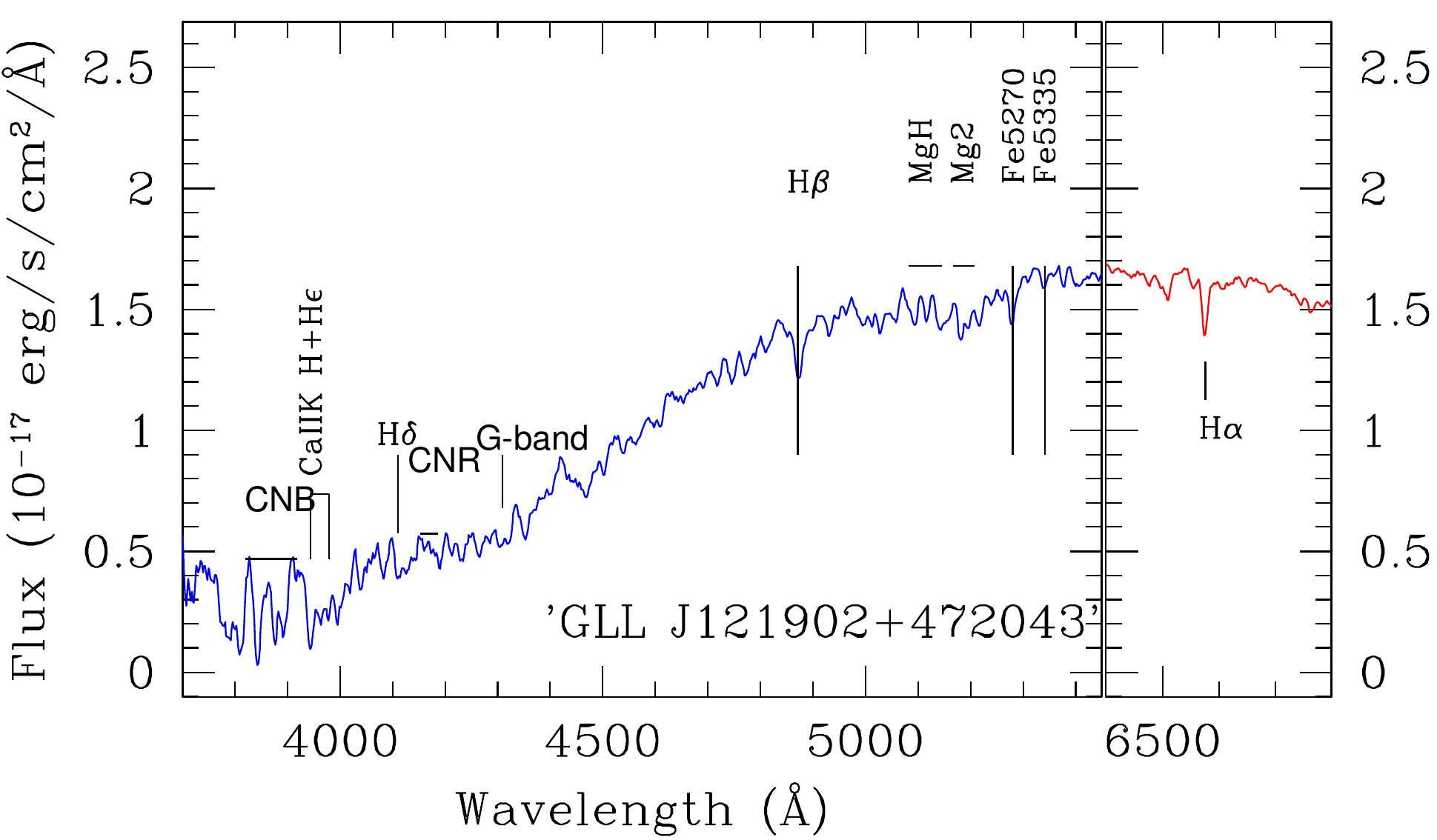}}
&
%\hspace*{0.3cm}\includegraphics[scale=0.413]{fig4_a2.pdf}
\includegraphics[scale=0.413]{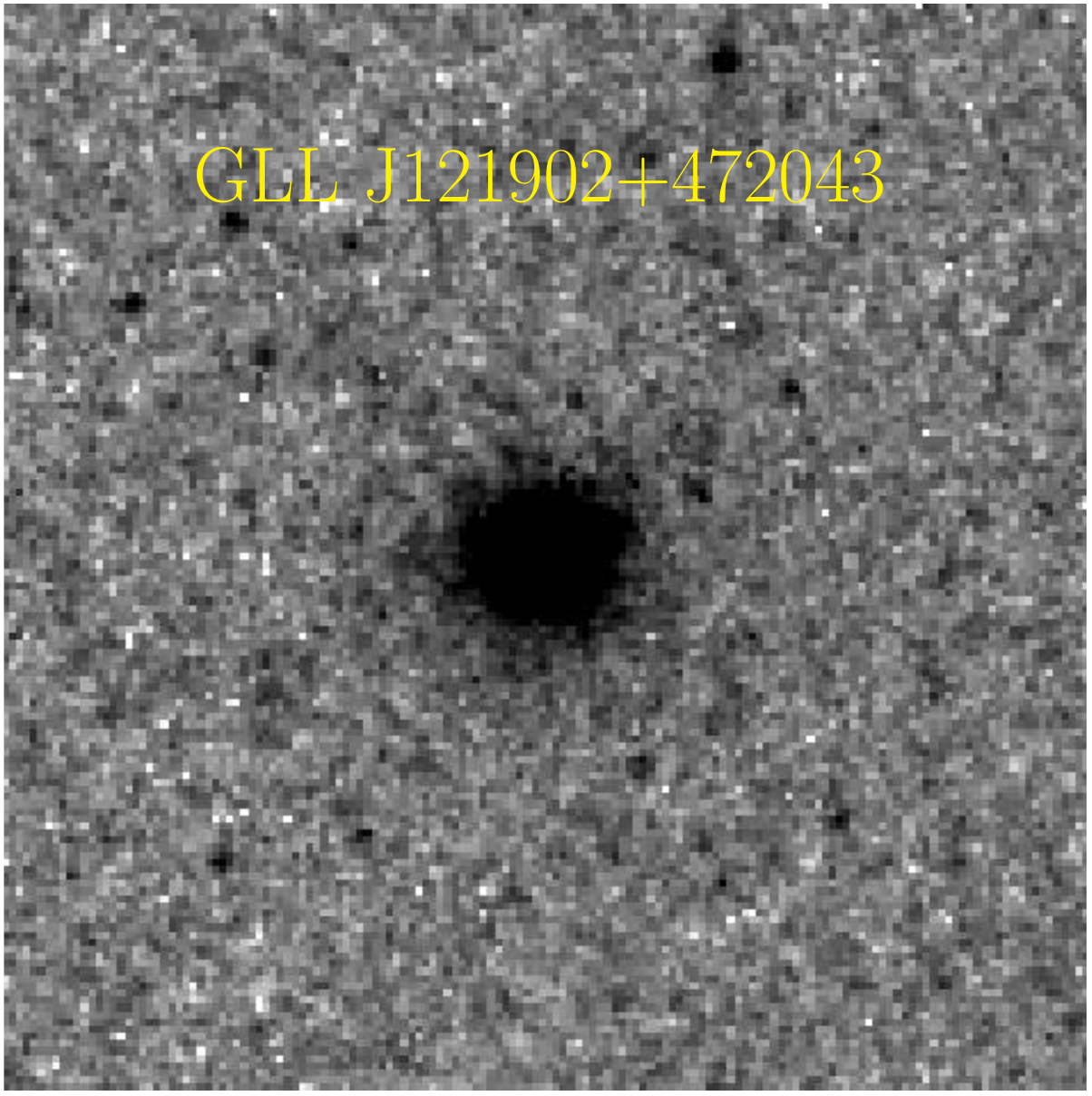}
\end{tabular}
\caption{Spectra ({\it left}) and grayscales ({\it right}) of confirmed GC candidates.
\label{fig:specim_conf}}
\end{figure*}

\begin{figure*}
\begin{tabular}{ll}
\raisebox{-0.57cm}{\includegraphics[scale=0.5,angle=0.]{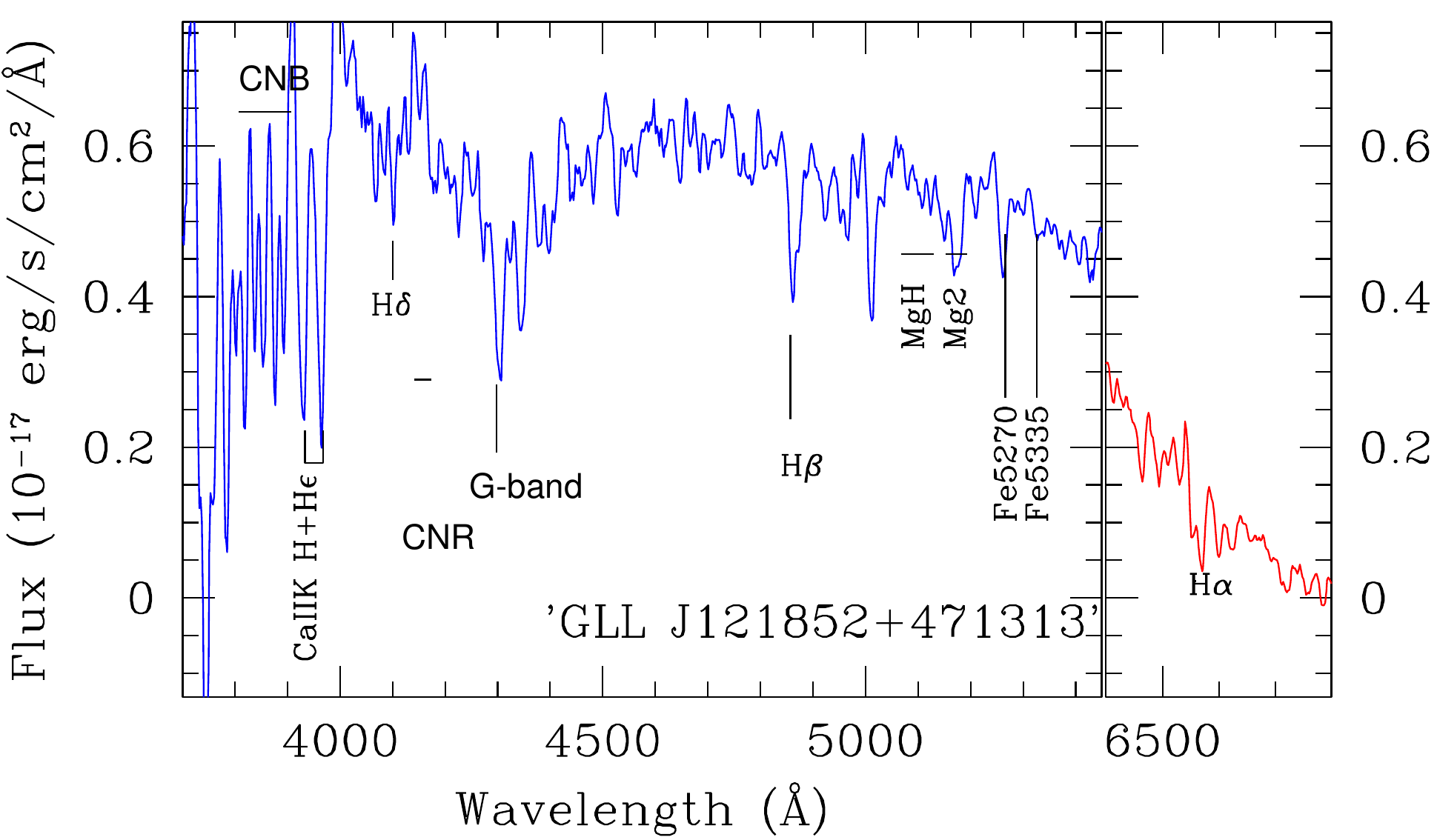}}
&
\includegraphics[scale=0.413]{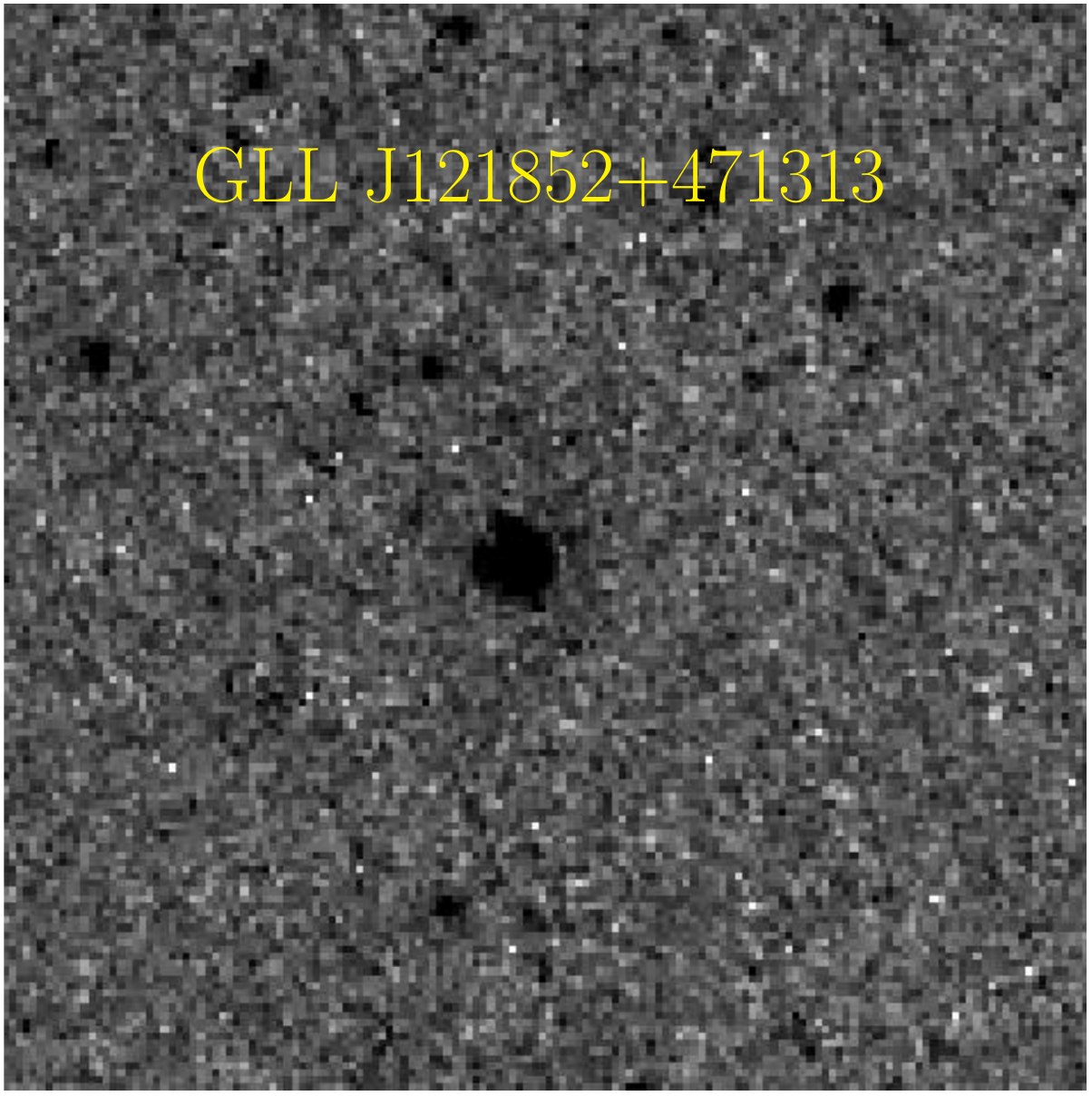}
\end{tabular}
\caption{Spectra ({\it left}) and grayscales ({\it right}) of rejected GC candidates.
{The spectrum of GLL J121852+471313 correlates best with {\it eltemp}, a non-GC 
template, and shows a negative radial velocity.}
\label{fig:specim_rej}}
\end{figure*}

\begin{figure*}
\begin{tabular}{ll}
\raisebox{-0.57cm}{\includegraphics[scale=0.5,angle=0.]{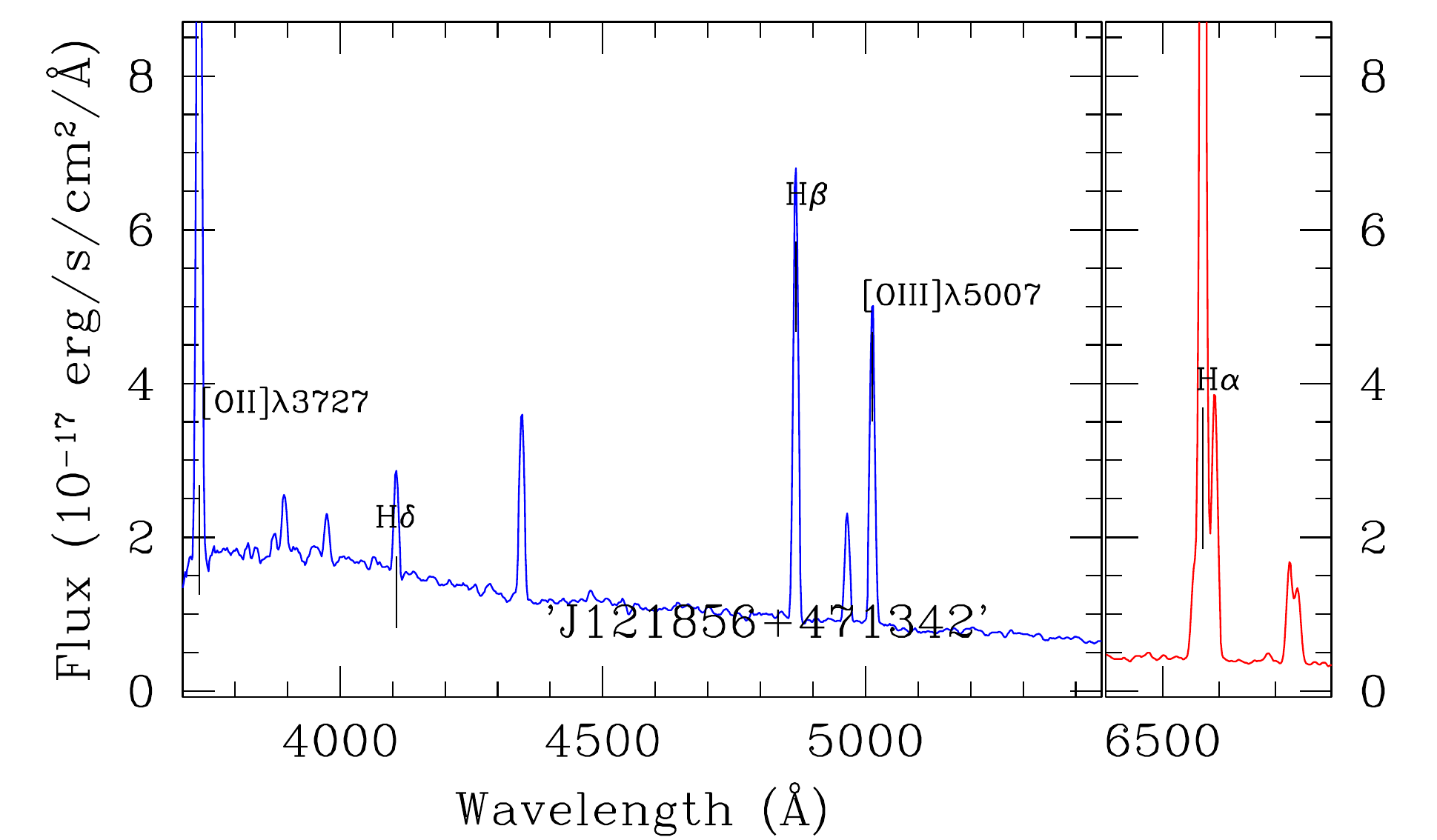}}
&
\includegraphics[scale=0.413]{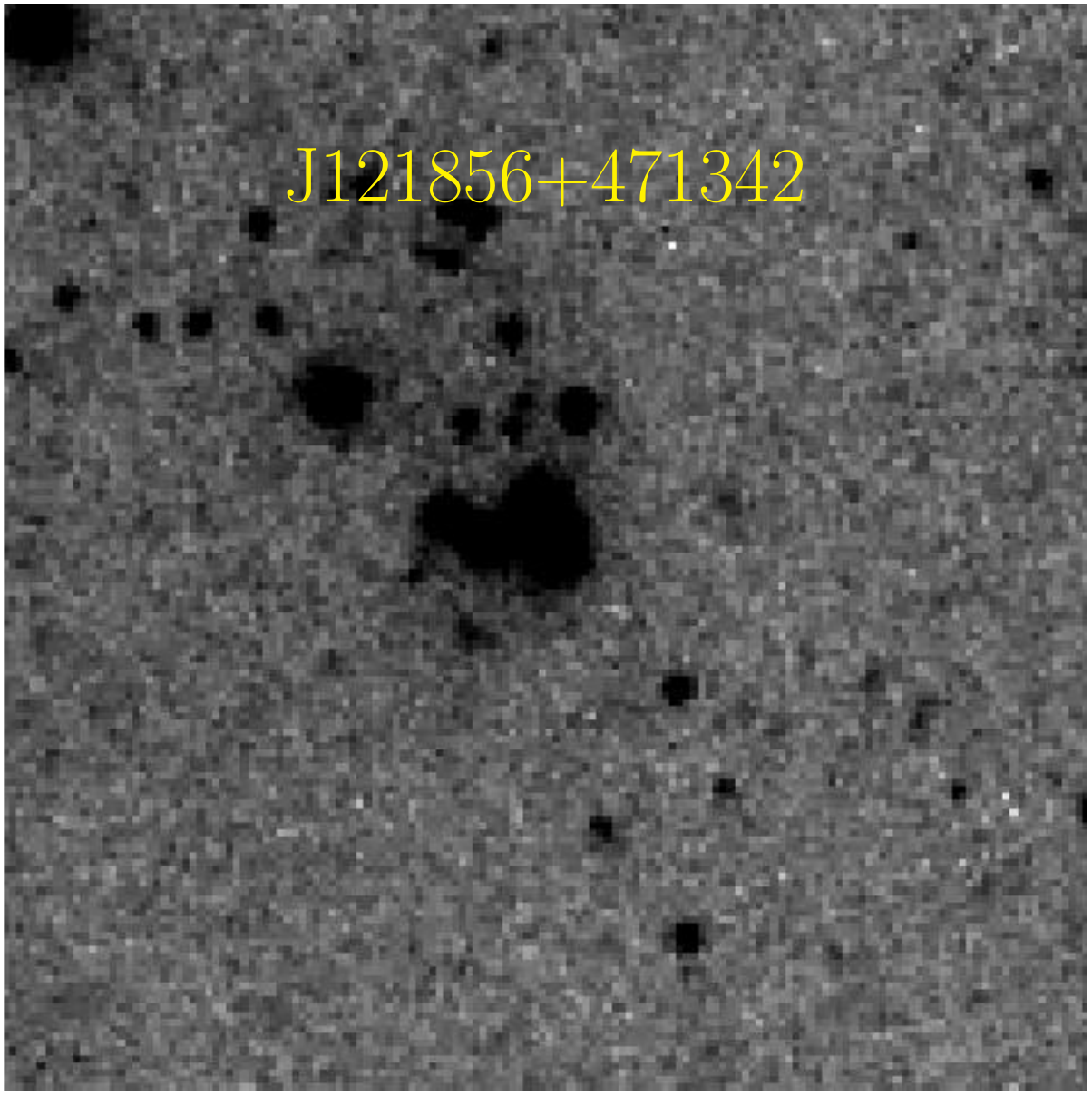}
\end{tabular}
\caption{Spectra ({\it left}) and grayscales ({\it right}) of non-candidates.
\label{fig:specim_non}}
\end{figure*}

\section{Rotation and velocity dispersion} \label{sec:HIetc}

The projected spatial distribution of the GCC sample determined in \citet[][{their Figure 10}]{gonz17} suggests a disk.
We can now investigate 
the kinematics of the system, and 
whether there is evidence of rotation.

\begin{figure}
%/Users/ragl/Documents/GTC17_velocities/vel_vs_projdist_best5.mac
\includegraphics[width=0.90\columnwidth,angle=0.]{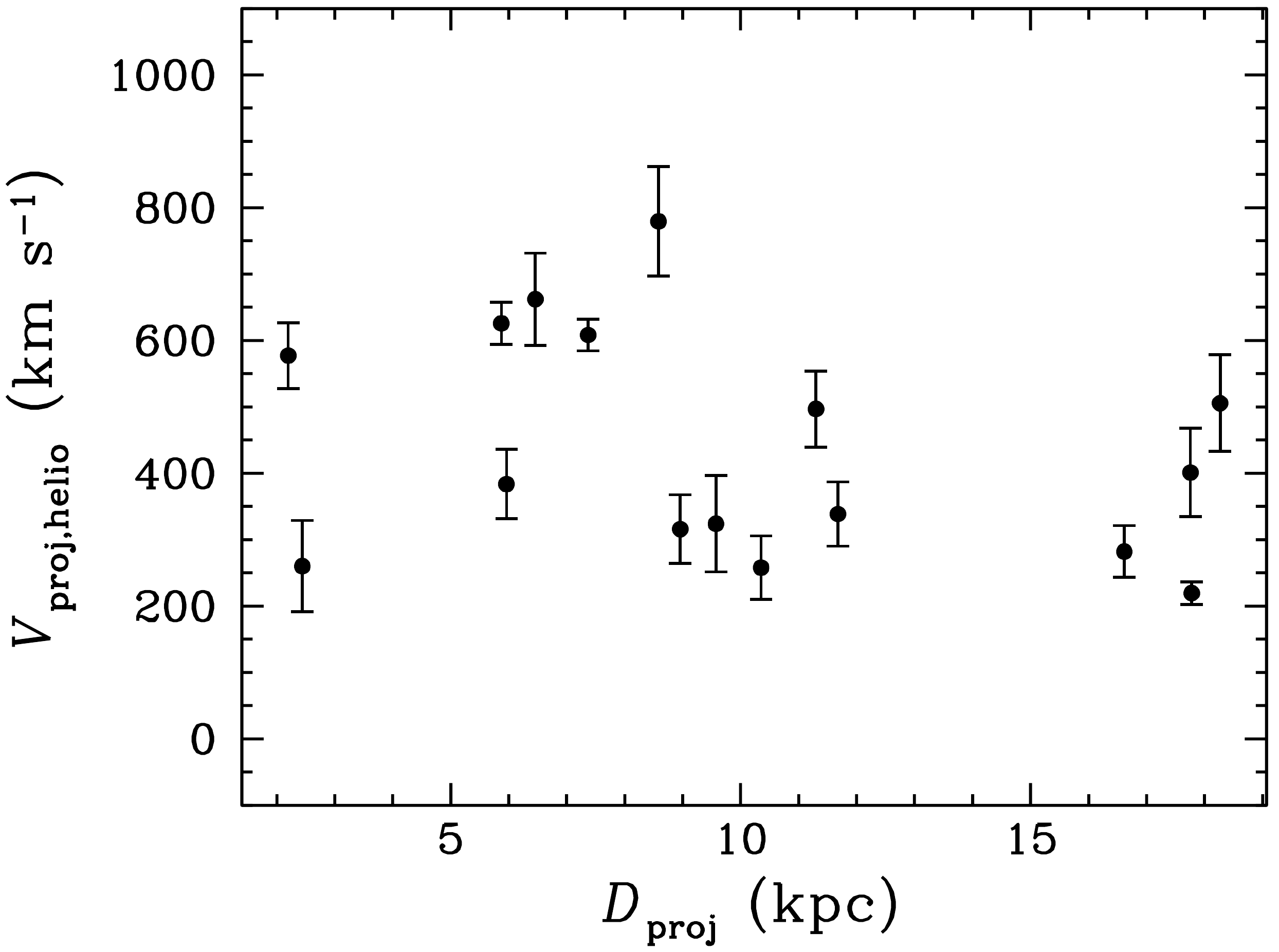}
\caption{Projected LOS heliocentric velocities of confirmed GCCs versus projected galactocentric
distance. 
\label{fig:projdist}}
\end{figure}

\begin{figure}
\includegraphics[scale=0.35,angle=0.]{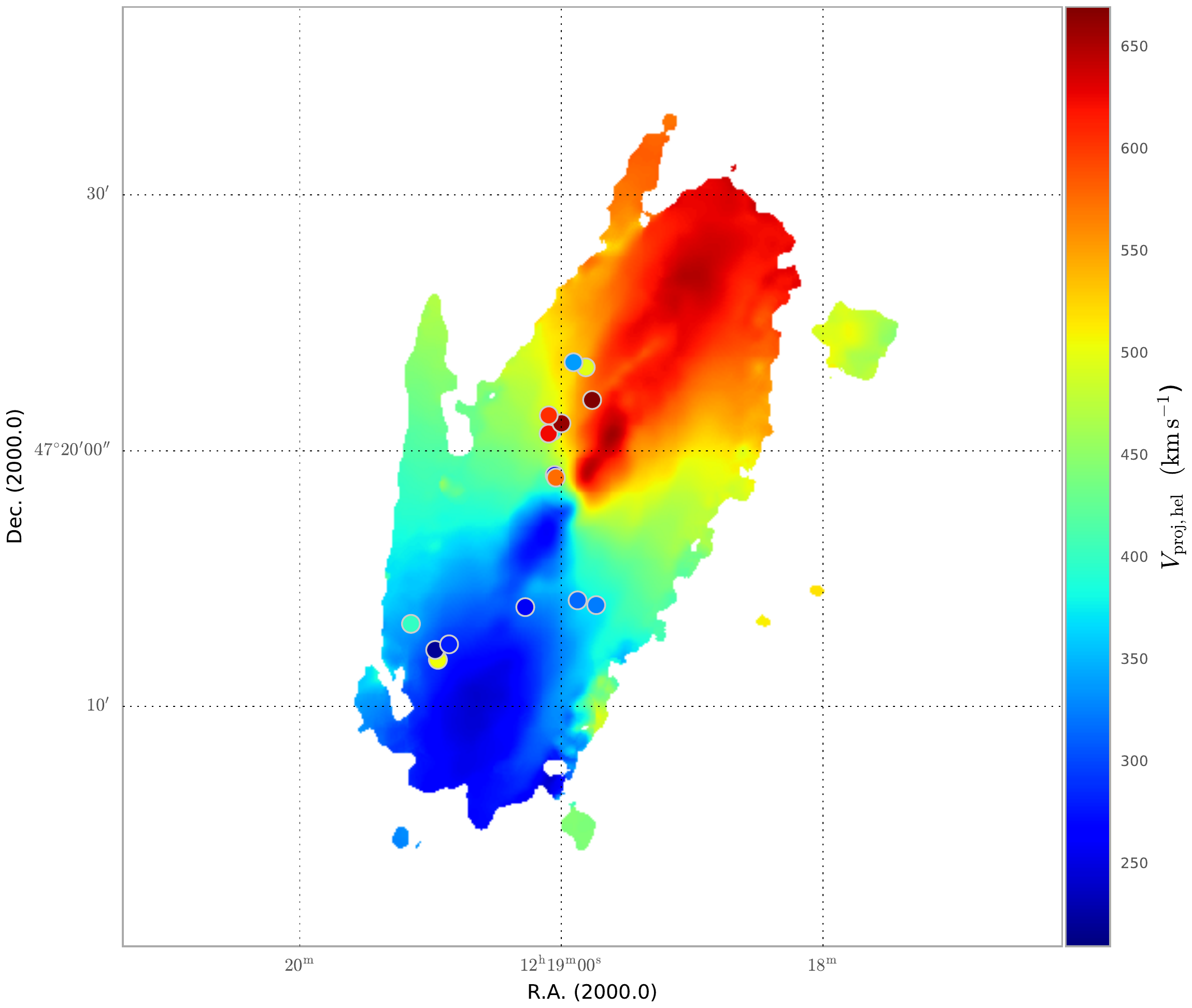}
\caption{Projected LOS heliocentric velocities of confirmed GCCs ({\it solid circles}) and NGC~4258 HI disk. 
Velocity scale is indicated in the side-bar. {\em This figure was made with the Kapteyn software package.} 
\label{fig:onvelfield}}
\end{figure}

Figure~\ref{fig:projdist} shows the projected line-of-sight (LOS) velocities of the confirmed GCCs versus their
projected distances from the center of the galaxy; there is no correlation between these two parameters
(correlation coefficient of -0.33). 
Figure~\ref{fig:onvelfield}, on the other hand, overplots the highly probable GCCs 
(filled circles, confirmed GCCs in \hyperlink{tab:velocities}{Table 2}) on the NGC~4258 HI velocity field map from the 
Westerbork Synthesis Radio Telescope (WSRT) Hydrogen Accretion in LOcal GAlaxieS (HALOGAS) survey
\citep{heal11}. LOS velocity is coded by color, as indicated in the side-bar, and the same scale is used for both
the HI and the GCCs. This graphic shows that the GCCs south of the rotation axis are approaching us, while most of the
ones in the observed northern field are receding.

\begin{figure}
\includegraphics[scale=0.8,angle=0.]{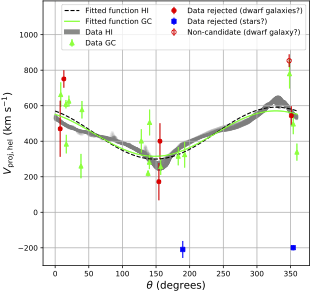}
\caption{Projected LOS heliocentric velocities versus azimuthal angle. {\it Green triangles:} confirmed GCCs; 
{\it red circles:} rejected candidates ({\it solid}) and non-candidate ({\it empty}), possibly dwarf satellite galaxies of NGC~4258; {\it blue squares:} rejected candidates, likely MW foreground stars;  
{\it thick gray line:}
HI; {\it solid green line:} sinusoidal fit to GCCs; {\it dashed black line:} sinusoidal fit to HI.
\label{fig:rotcurves}}
\end{figure}

Finally, Figure~\ref{fig:rotcurves} displays the measured LOS projected velocities versus azimuthal angle, measured
counterclockwise from North, for the confirmed 
GCCs (solid green triangles), rejected candidates that could be satellite galaxies of NGC~4258 
(solid red circles), rejected candidates that appear to be foreground stars (solid blue squares), and 
one non-candidate that could also be an object associated with NGC~4258 (empty red circle).  
Our GTC objects are overplotted on the NGC~4258 projected HI gas heliocentric velocities (gray curve) in the elliptical 
annulus with center at RA=12h18m57.505s, Dec=+47$\degr18\arcmin4\farcs3$; 
position angle PA = 150$\degr$; axis ratio $b/a$ = 0.389 (i.e., cos 67$\fdg$1),\footnote{
{The inclination $i$ is, actually, 68$\fdg$3, from cos$^2 i = [(b/a)^2 - q_0^2]/[1 - q_0^2]$, with $q_0 =$ 0.13 
\citep{giov94}.}  
}
 and inner and
outer semi-major axes, respectively, 6$\arcmin$ and 12$\arcmin$ 
(13.3 and 26.5 kpc at the distance of NGC~4258). The projection parameters of the galaxy have
been taken from \citet{rc3}. 
The HALOGAS moment-1 pixels are 4$\arcsec$ on the side, but the actual resolution is 30$\arcsec$. Given the data quality
and depth, the map is reliable to a column density level of $N_{\rm HI} = 2\times\ 10^{19}$ atom cm$^{-2}$, at which it has
been clipped. Sinusoidal unweighted fits to the projected GC and HI velocities are shown with, respectively, a solid green and a dashed black lines.

\hyperlink{tab:fits}{Table 4} lists the amplitude, phase, and systemic velocity of the fits. 
While the amplitude is slightly smaller for the
GCs,\footnote{We note that the sinusoidal fit to the HI data does not reach the maximum projected 
velocity of $\sim$ 203 km s$^{-1}$. 
} the phases and systemic velocities are virtually identical for the two systems. Moreover, 
the uncertainty-corrected velocity dispersion of the GC system, $\sigma_{\rm GC}$, goes down from 162.1 km s$^{-1}$ to 
119.8 km s$^{-1}$ when the HI velocity at its projected position is subtracted from each one of the clusters, 
and to 99.5 km s$^{-1}$ when the value of the fit to the GC projected velocities is
used instead.\footnote{The average uncertainty of the
fit, 88.5 km s$^{-1}$, was estimated by means of a Monte Carlo simulation, i.e., we performed 50,000 
realizations of the fit around the values of the amplitude, phase, and systemic velocity in 
\hyperlink{tab:fits}{Table 4}, to probe their Gaussian uncertainties.}

{
In order to estimate the probability of detecting actual rotation from a sample as small as 16 objects, 
we simulated various cluster systems with different structure and kinematics, and examined which one better matched the observations. 
The radial distribution of the clusters was chosen to reproduce that of the Milky Way GC system. Specifically, we used the Galactocentric radii in \citet{harr96}, and fitted them with a Lorentzian profile.\footnote{{${\mathcal L}(r) = h/\pi[1 + (r/\gamma)^2]$, where $\gamma =$ 5.72 kpc, and $h$ is a normalization factor that depends on the number of clusters.}
} This function provides a very good fit to the data, far superior to a Gaussian or a power-law distribution. For the angular distribution, we considered first a system with spherical symmetry, mimicking a spherical halo, and, as a second possibility, a thin disk. Finally, for the kinematics, we worked with two options. Firstly, a rotating system with a flat rotation curve with $V_{\rm max} = 220$ km s$^{-1}$ beyond a
radius of 1.5 kpc, and a solid body rotation within that radius; superposed on this organized rotation, we added a random component with a dispersion of 10 km s$^{-1}$ in each of the three directions. The second velocity choice was a random distribution, also with $V_{\rm max} = 220$ km s$^{-1}$. In all cases, we included errors on the simulated radial velocities of 20 to 80 km s$^{-1}$, consistent with the observational errors of the actual data. Summarizing, then, we had three different types of simulations: a random halo, a rotating halo, and a rotating disk.} 

\begin{figure}[h!]
%\begin{tabular}{ll}
%\hspace*{0.3cm}\includegraphics[scale=0.72,angle=0.]{fig_rot_disk_sim_cropped.pdf}
\includegraphics[scale=0.72,angle=0.]{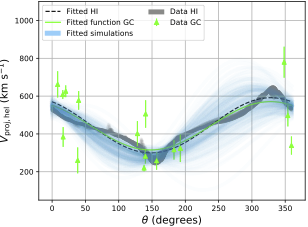}
%\hspace*{0.3cm}\includegraphics[scale=0.72]{fig_rot_halo_sim_cropped.pdf}
\includegraphics[scale=0.72]{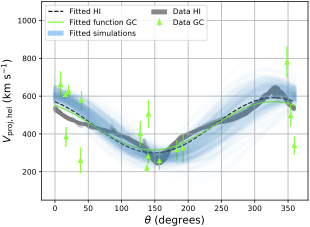}
%\end{tabular}
%\begin{tabular}{ll}
%\hspace*{0.3cm}\includegraphics[scale=0.72,angle=0.]{fig_random_halo_sim_cropped.pdf}
\includegraphics[scale=0.72,angle=0.]{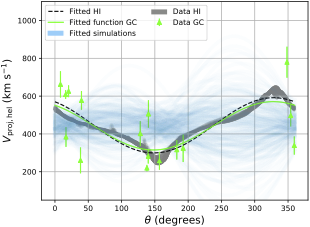}
%\end{tabular}
\caption{{Simulations. Projected LOS velocities versus azimuthal angle for a rotating disk ({\it top}), 
a rotating halo ({\it middle}), and a random halo ({\it bottom}). {\it Thick light-blue lines:} sinusoidal 
fits to 16 ``clusters" drawn from same areas as spectroscopic data (see text).
Other symbols and lines as in Figure~\ref{fig:rotcurves}.} 
\label{fig:simulations}}
\end{figure}

{
We ran each kind of simulation 200 times.
For every run, we extracted positions on the plane of the sky and radial velocities (i.e., projected along the line of sight), and selected the same number of clusters as in the observations (i.e., 16), restricted to the two areas where the actual spectroscopic data were obtained 
(see right panel of Figure~\ref{fig:footprints}). We then fitted the simulated velocities as a function of position angle with a sinusoid, as we did 
with the real observations.
The fits to the observed HI and GC data presented in Figure~\ref{fig:rotcurves}, above, agree with each other to about an rms of 13 km s$^{-1}$.\footnote{This is the rms of the difference between the two fits, calculated in 1$\degr$ bins.}  We consider that a fit to one realization of the simulations is consistent with the data if it agrees to better than 3$\sigma$ (defined in the same way as the quoted rms) with the HI fit. We find that this occurs for 51\% of the simulations of a rotating disk, for 24\% of the realizations for a rotating halo, and only 1\% of the time for a halo with randomly distributed velocities (see Figure~\ref{fig:simulations}). This exercise indicates that the measured velocities are incompatible with a non-rotating system, and significantly more compatible with a rotating disk than with a rotating spherical system.
}

Hence, we conclude that there is strong evidence of rotation among the confirmed GC candidates
of NGC~4258, and adopt a value of $\sigma_{\rm GC} = 110\pm$ 10 km s$^{-1}$ (average of the two previous
estimates) for the galaxy. We also observe that 
probable satellite galaxies seem to share the rotation of 
the GC system. If we include them in the fit to the projected GC velocities the amplitude actually
{\em increases}, while remaining consistent with the measurement for GCs only, as shown in \hyperlink{tab:fits}{Table 4}.  
{Finally, if these objects are included in the calculation of the velocity dispersion, 
we get $\sigma=120\pm9$ km s$^{-1}$, consistent with the measurement without them.} 

\begin{table}[ht]
\hypertarget{tab:fits}{}
\caption{Fits to GC and HI projected velocities}
 \begin{center}
 \begin{minipage}{280mm}
  \begin{scriptsize}
\begin{tabular}{@{}cccc@{}}
\hline
\hline
System & Amplitude &  Phase  &    Systemic velocity \\
       & km s$^{-1}$ & $\degr$ &  km s$^{-1}$      \\  
\hline
HI     &  146.1 $\pm$  0.2   & 121.6 $\pm$ 0.1      &   445.6  $\pm$  0.2      \\
GCs   &  127.9 $\pm$ 60.9  & 122.1 $\pm$    45.9    &   443.1  $\pm$  50.9   \\
GCs+dwarfs & 168.1 $\pm$ 66.0  &   132.3 $\pm$ 31.2 & 467.7 $\pm$ 44.5 \\

\hline
\vspace*{-0.5cm}
\end{tabular}
\end{scriptsize}
\end{minipage}
\end{center}
%\label{tab:fits}
\end{table}

We can now compare the measured rotation-corrected velocity dispersion of the GC system to the expected $\sigma_{\rm GC}$ from 
the correlation with the central supermassive black hole mass ($M_\bullet$) derived by \citet{sado12}. These authors find 
log ($M_\bullet$) $= \alpha + \beta$ log ($\sigma _{\rm GC}/200$ km s$^{-1}$), with $\alpha = 8.63 \pm 0.09$ and 
$\beta = 3.76 \pm 0.52$. Taking $M_\bullet = (4.00\pm0.09)\times 10^7  M_\odot$ \citep{hump13} and propagating
the errors in both the correlation coefficients and the $M_\bullet$ measurement,     
the expected velocity dispersion is $\sigma_{\rm GC} = 107\pm13$ km s$^{-1}$, which is in excellent agreement with the value of 110$\pm$10 km s$^{-1}$ derived by us.

\section{Metallicity} \label{sec:Z}

As a result of the {generally} low SNR, we did not attempt to
estimate the ages and metallicities of the GC candidates from their individual spectra. Instead,
we stacked the spectra of all the confirmed GC candidates into a single spectrum, and hence derived
a mean age and metallicity for the GC population.
To this end, we used the Penalized Pixel-Fitting algorithm \citep[pPXF;][]{capp04}, which extracts
stellar population properties by fitting a linear combination of different templates to an observed spectrum in pixel space.
The available templates are 182 
solar-scaled isochrones with a Salpeter IMF; initial masses from 0.15 $M_\odot$ to 7$M_\odot$; 
a range of metallicity -2.3 $<$ [M/H] $<$ 0.2; a helium fraction of $Y = 0.23 + 2.25 Z$; and ages between 1 and 17 Gyr.
The isochrones also incorporate the thermally pulsing AGB regime to the point of complete envelope ejection 
\citep{gira00}. 

\begin{figure}
\includegraphics[scale=0.37,angle=0.]{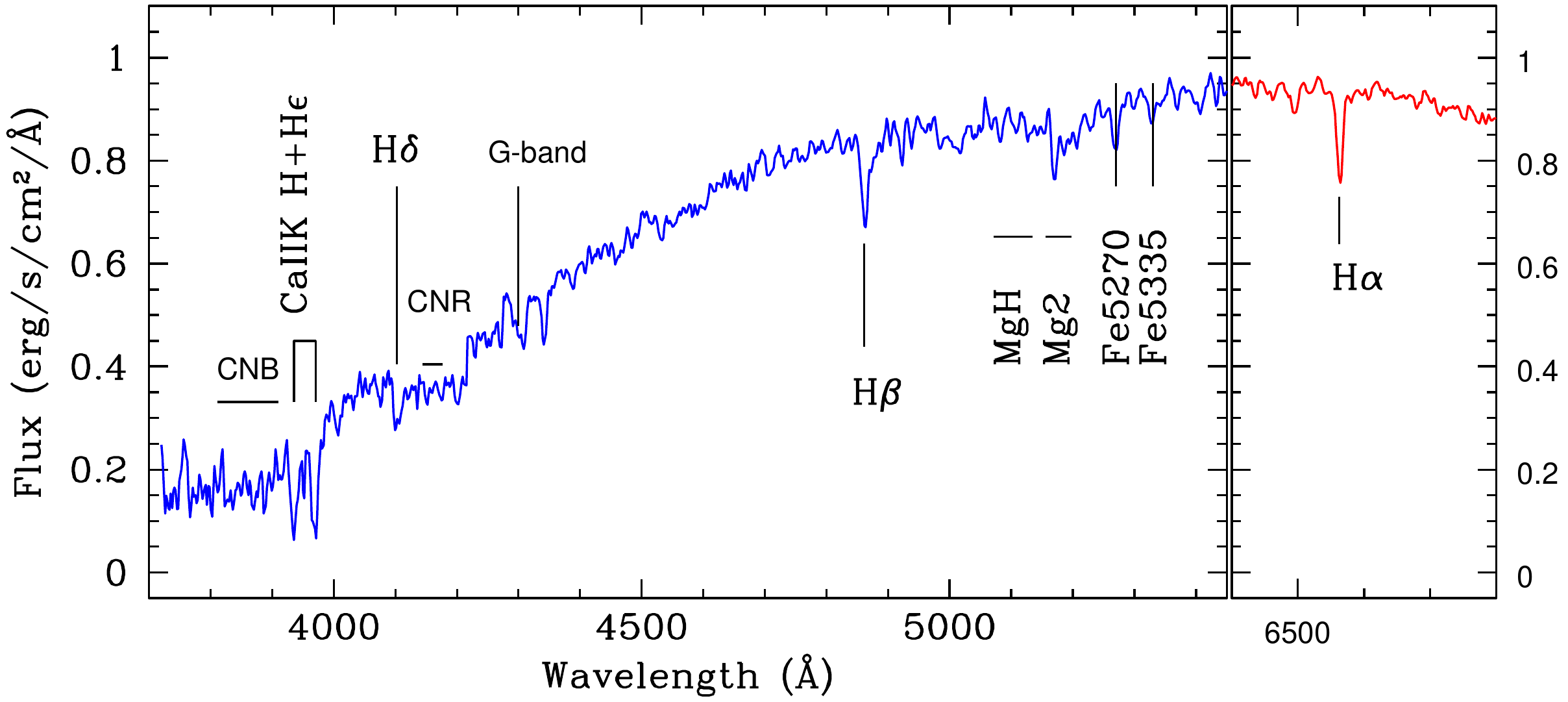}
%/Users/ragl/Documents/GTC17_metallicity/best4/scomp_best5_norm_diva_ed_cropped.pdf
\caption{Stacked, normalized spectrum of confirmed GC candidates.
\label{fig:stacknorm}}
\end{figure}

Before stacking, each individual GCC spectrum was multiplied by a mask with
a value of zero in noisy regions and one otherwise;\footnote{Noisy regions are those also excluded from the
Fourier cross-correlation for the radial velocity estimate.}
the coadded spectrum was divided by the sum of the masks, then normalized to
a maximum value of 1. The resulting spectrum, in rest wavelength, is shown in Figure~\ref{fig:stacknorm};
in view of the improved SNR ($\sim$ 80) with respect to the individual GCC spectra, we did not apply additional smoothing.

Given the lack of a perfect flux calibration along the full spectral range, we carried out separate fits in windows 200$\AA$ wide around the Mg\,$b$ triplet and H${\alpha}$,
respectively, a metallicity-sensitive indicator and a widely used age-sensitive feature.
For each region, a first pPXF fit was performed in order to estimate the noise factor, i.e., such that multiplied by
the noise one would obtain a reduced $\chi^2$ ($\chi^2$ per degree of freedom) = 1.
No polynomial (either multiplicative nor additive) was included in the fit,
since the continuum shape contains important information about the population
\citep{wolf07,capp04}.
Subsequently, a second iteration of the fit was done with the previously obtained noise factor, and with linear regularization (smoothing) of the weights
in the age and metallicity axes; both the continuum and the absorption features were fit, assuming constant noise per pixel. Milky Way foreground extinction was not considered, given its very low value of
$A_V = 0.04$ \citep{schl11} in the line of sight towards NGC~4258.

\begin{figure}[t]
\includegraphics[scale=0.4]{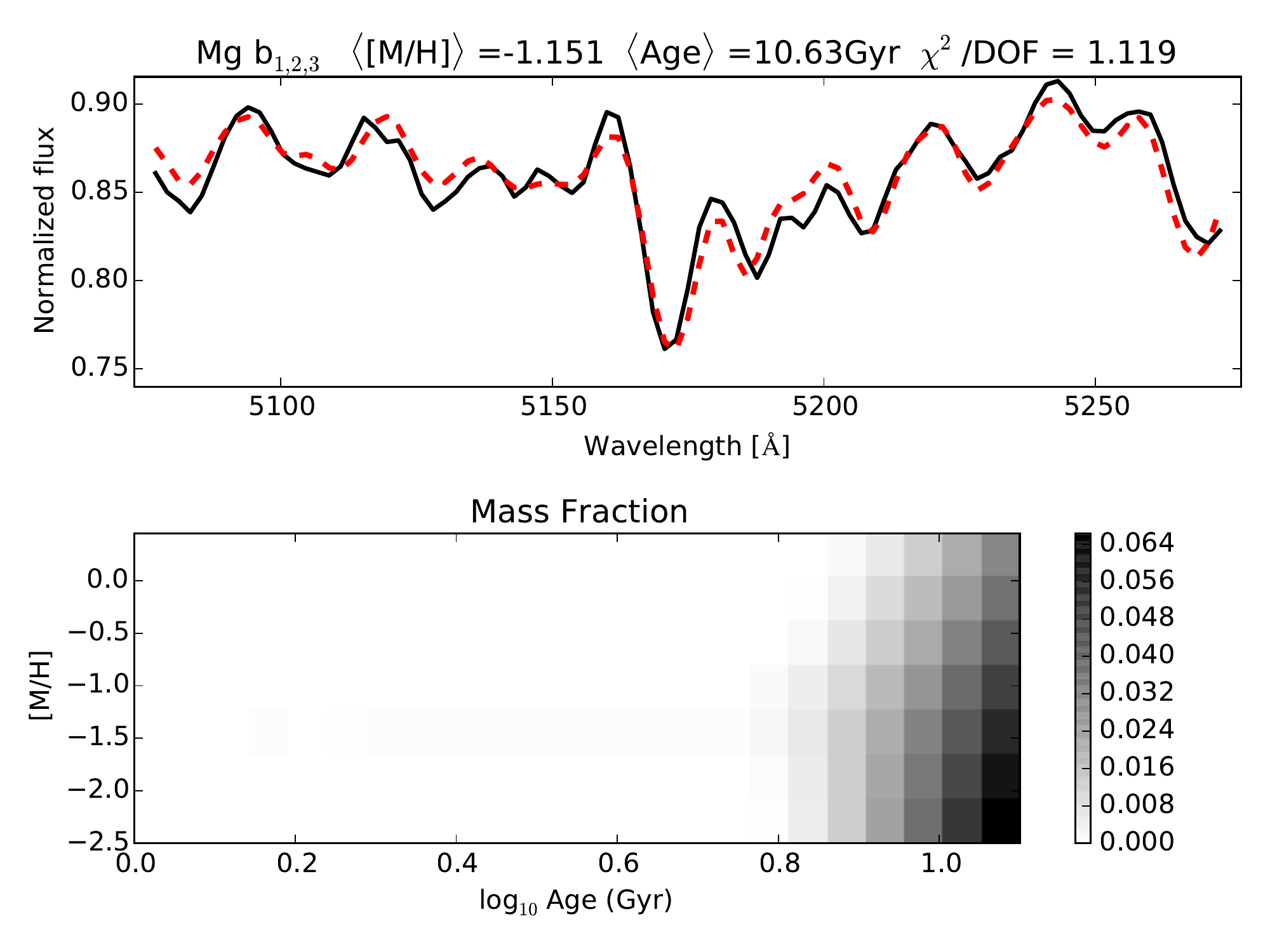}
%/Users/ragl/Documents/n4258_spec_paper/collabs/contrib/karla
\caption{Mass-weighted metallicity and age of GC system from fit to the Mg\,$b$ triplet spectral window. 
{\it Top:} observed ({\it black solid line}) and model ({\it red dashed line}) spectra; {\it bottom:} mass fraction 
distribution as a function of age and metallicity 
({\it grayscale}).  
The derived [M/H] and age, as well as the reduced $\chi^2$, are written above the top panel.
\label{fig:avgZMgb}}
\end{figure}

\begin{figure}[t]
\includegraphics[scale=0.4]{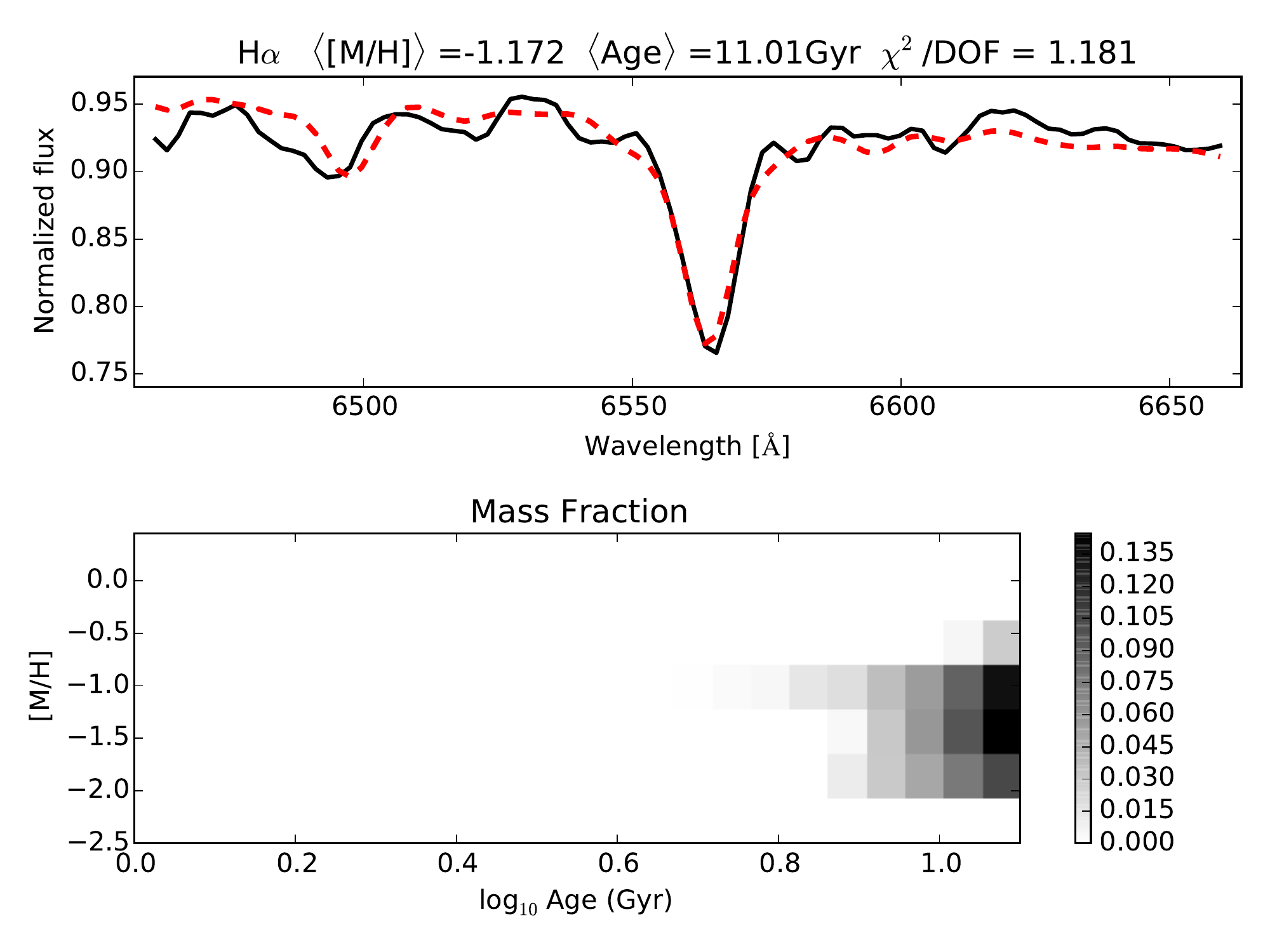}
%/Users/ragl/Documents/n4258_spec_paper/collabs/contrib/karla
\caption{Mass-weighted metallicity and age of GC system from fit to the H$\alpha$ spectral window. 
Panels as in Figure~\ref{fig:avgZMgb}.
\label{fig:avgZHa}}
\end{figure}

In Figures~\ref{fig:avgZMgb} and \ref{fig:avgZHa}, we show the best fit for the two spectral regions, and their corresponding 
{mass fraction distributions as a function of age and metallicity.}
The values above the top panels are mass-weighted metallicity [M/H] and age, respectively.
{We note that the mass fraction distribution with 
metallicity is quite different for the two spectral windows.
In the case of the Mg$b$ index, it is very extended, and goes from 1/20 solar to supersolar;  
almost half (44\%) of the mass has [M/H] $>$ -1.2.
For H$\alpha$, the mass distribution is significantly narrower in [M/H]. However,
the mass-weighted mean metallicities are very similar for both ranges.}
While there are limitations of the data and the models (the resolution can modify absortion features, calibration of wavelength and continuum, fraction of blue stragglers, among others) and we cannot set strong contraints on the age or the metallicity, we can certainly conclude that the recovered populations are old 
($\sim$ 11 Gyr) and metal-poor ($<[M/H]> \sim$ -1.2).

%\clearpage

\section{Photometric method validation and the $N_{\rm GC}$ versus $M_\bullet$ correlation} \label{sec:method}

Our spectroscopic observations of about half the total sample of NGC~4258 GCCs show 16 confirmed candidates, 10 confirmed 
non-candidates, and 7 rejected candidates, i.e., no false negatives and a sample contamination of 30\%. 
The 7 contaminants are fully consistent with the results of \citet{powa16}, who found, with the \uiks\ technique, between 50 and 100 contaminants among the GCCs of 
M87, in a FOV 9 times larger. Both M~87 and NGC~4258 lie at high Galactic latitudes (respectively, $b = 74\fdg5$ and $b= 68\fdg8$),    
and indeed the Besan\c con model of Galactic structure \citep{robi86} predicts an almost equal number of foreground stars in their lines of
sight (355 versus 329 brighter than apparent $u$ = 20 mag within one square degree). The number of background galaxies, on the other hand,
should be about the same, within a few times the Poisson error \citep[e.g.,][]{elli07}.

\begin{figure*}
%\begin{tabular}{ll}
\hspace*{2.1cm}\includegraphics[scale=0.35]{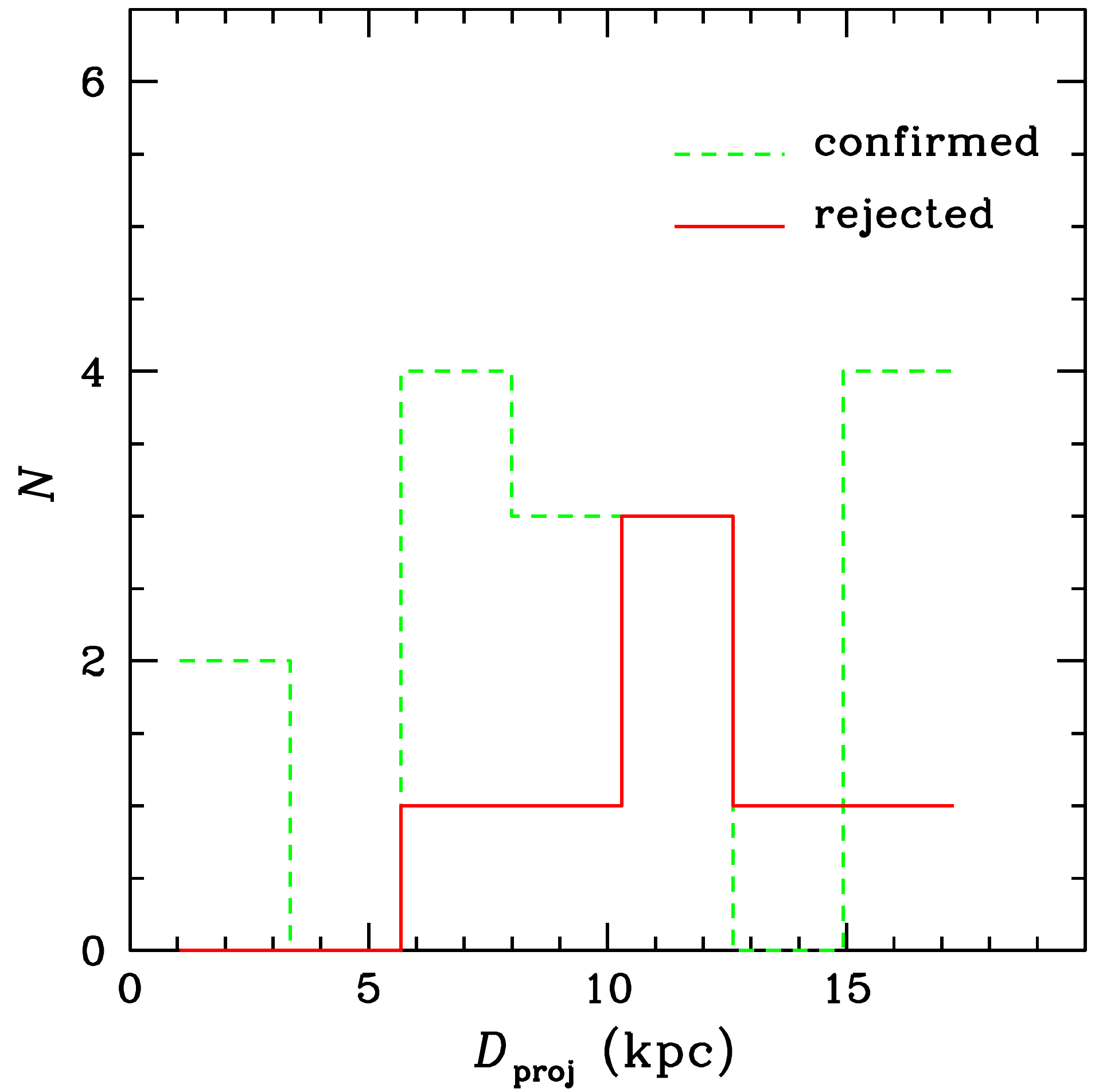}
%&
\includegraphics[scale=0.35]{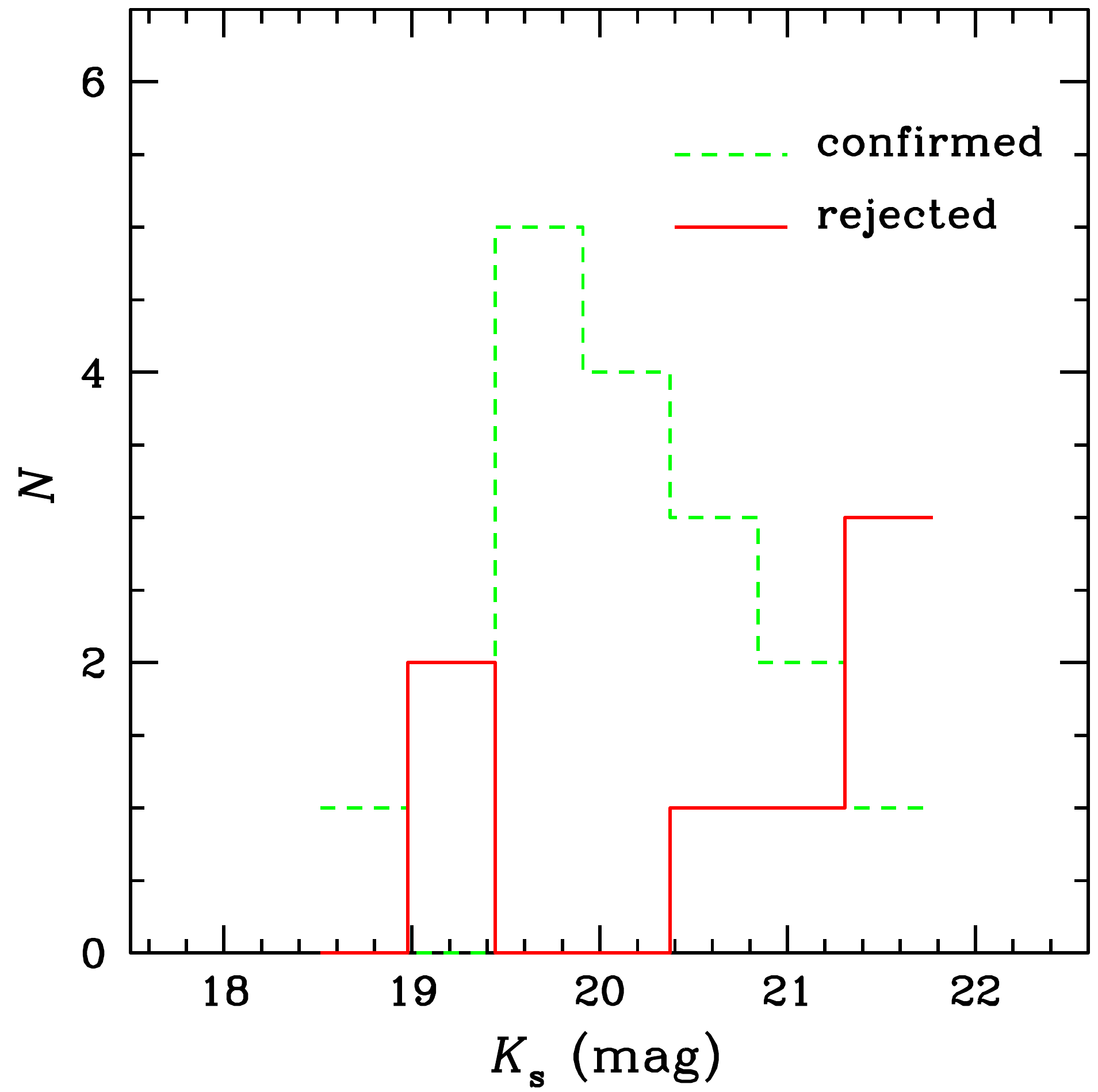}
%\end{tabular}
\caption{Distributions of confirmed ({\it green dashed lines}) and rejected ({\it red solid lines}) candidates.
{\it Left:} galactocentric projected distances; 
{\it right:} \ks-band magnitudes. 
\label{fig:histmeth}}
\end{figure*}

The left panel in Figure~\ref{fig:histmeth} displays the distributions of galactocentric projected distances of both
confirmed (dashed green line) and rejected (solid red line) candidates; the latter lie on average slightly further away than most confirmed GCCs, so 
crowding and background light should not be important contributing factors to contamination. 
Figure~\ref{fig:method} shows the FWHM (top left), SPREAD\_MODEL($\times$ 100; top right),
axis ratios (middle left), \ks\ magnitude values (middle right), (\ust\ - \ip) (bottom left), and (\ip\ - \ks) (bottom right) colors of the confirmed (green bars) and 
rejected (red bars) candidates. GLL J121905+472422 
has both the largest FWHM and SPREAD\_MODEL, and the bluest (\ust\ - \ip) and (\ip\ - \ks) colors; together with GLL J121854+472245, it is also among the 3 brightest objects in the \ks-band. 
GLL J121910+471343 has the second largest SPREAD\_MODEL value, whereas GLL J121903+472613
holds the second bluest (\ust\ - \ip) and (\ip\ - \ks) colors. 
GLL J121854+472144 has an axis ratio of 0.63, which is clearly smaller than the minimum observed axis ratio of 0.7 
for the GCs in both 
the MW \citep{whit87} and M~31 \citep{lupt89}. The inclusion of GLL J121854+472144 in our sample was definitely an oversight, since we had
actually measured the axis ratios of our candidates with {\sc galfit} \citep{peng10}. 
Given the shallowness of our near-infrared data of NGC~4258, we hesitate at this point to propose moving the faint cut in \ks\ to a 
brighter limit. However, a histogram of the \ks\ magnitudes of the confirmed and rejected candidates 
(Figure~\ref{fig:histmeth}, right panel) indicates they do have different distributions,
and the middle right panel of Figure~\ref{fig:method} shows that 3 rejected candidates cluster around \ks\ = 21.4; 
only one confirmed GCC is fainter. 
Of the 3 very faint rejected candidates, GLL J121909+471335 alone does not stand out as a near-outlier in any other gauge (GLL J121903+472613 is 
quite blue and GLL J121854+472144 has a small axis ratio). Finally, GLL J121852+471313 is the singular unremarkable source in every one 
of the inspected parameters. 

We could easily depurate the sample of 5 (out of 7) spurious candidates, while losing solely one confirmed GC, by
slightly changing the limits of our selection parameters in future works. As indicated by the dashed lines,
we would suggest using SPREAD\_MODEL $\leqslant$ 0.015, axis ratio $\geqslant$ 0.7,  LFTO - 2 (or LFTO - 1.7 $\sigma_{\rm LFTO}$)
$\leqslant$ MAG\_AUTO\footnote{Kron-like \citep{kron80} elliptical aperture magnitude measured by SExtractor.} $\leqslant$ LFTO + 3 $\sigma_{\rm LFTO}$, 
FWHM $\leqslant$ 29 pc, (\ust\ - \ip) $\la$ 1.54, (\ip\ - \ks) $\la$ -0.11.  

\begin{figure*}
%all in /Users/ragl/Documents/n4258_spec_paper/figs_w_captions
\includegraphics[scale=0.30]{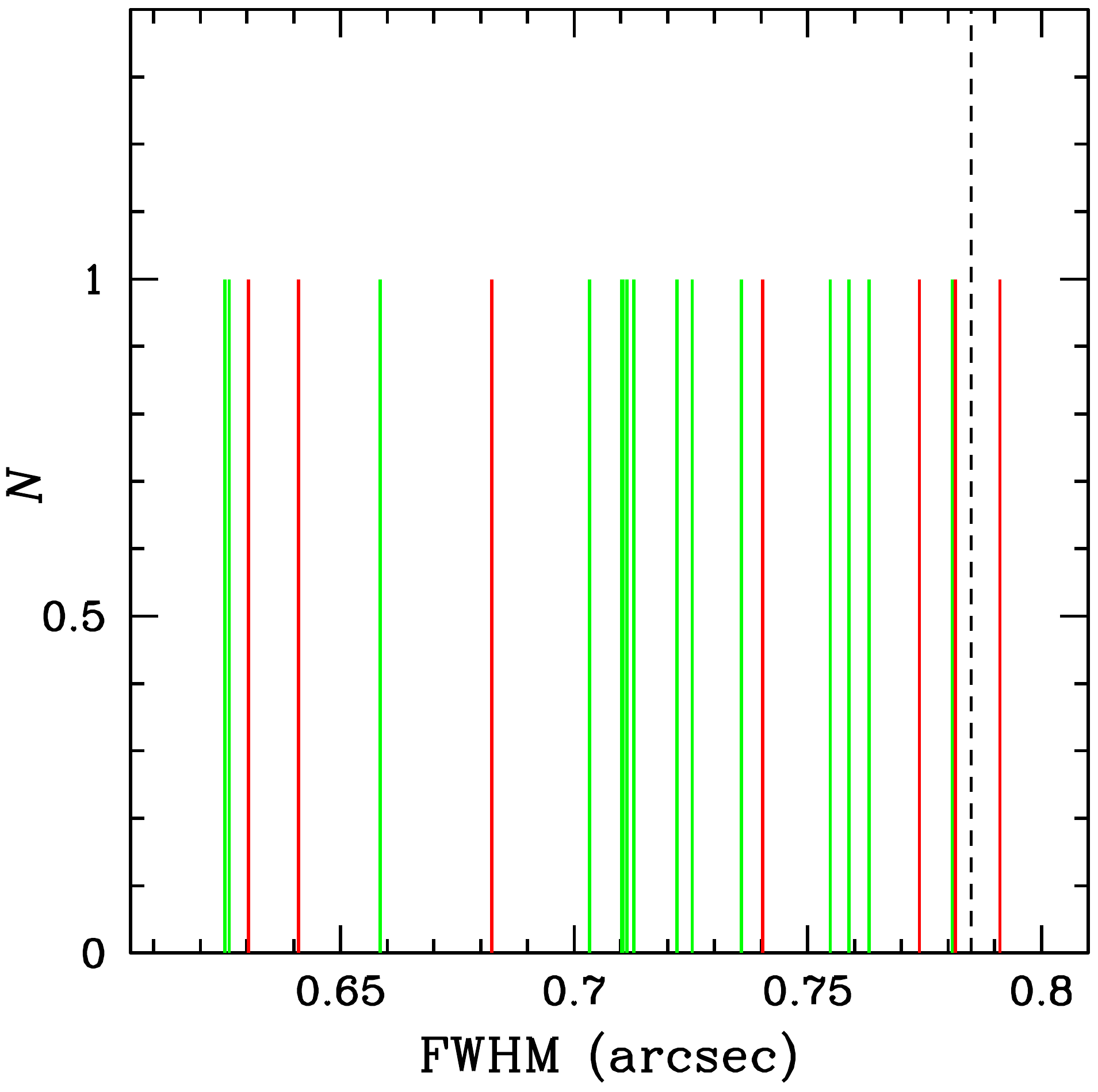}
\includegraphics[scale=0.30]{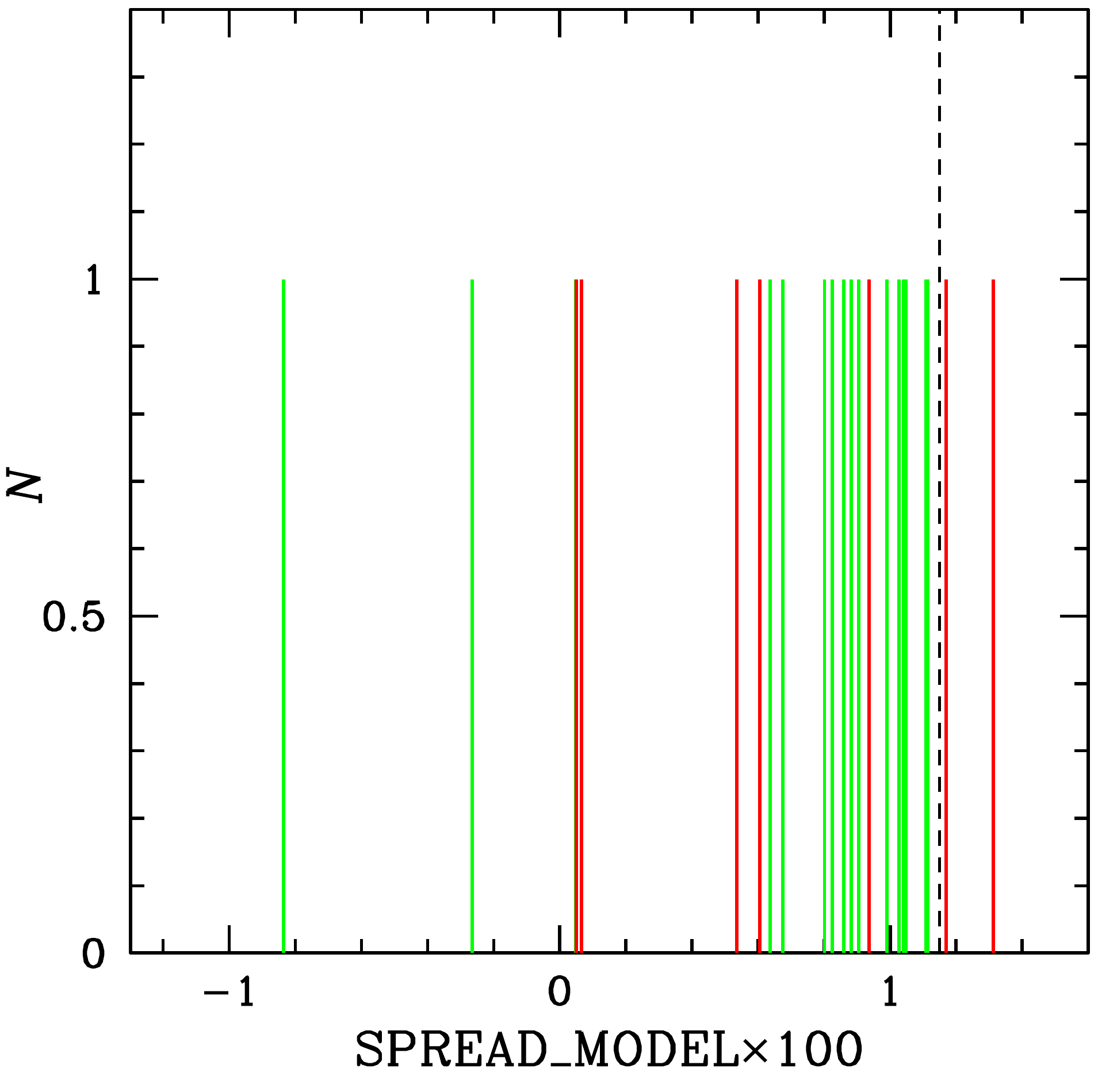}
\includegraphics[scale=0.30]{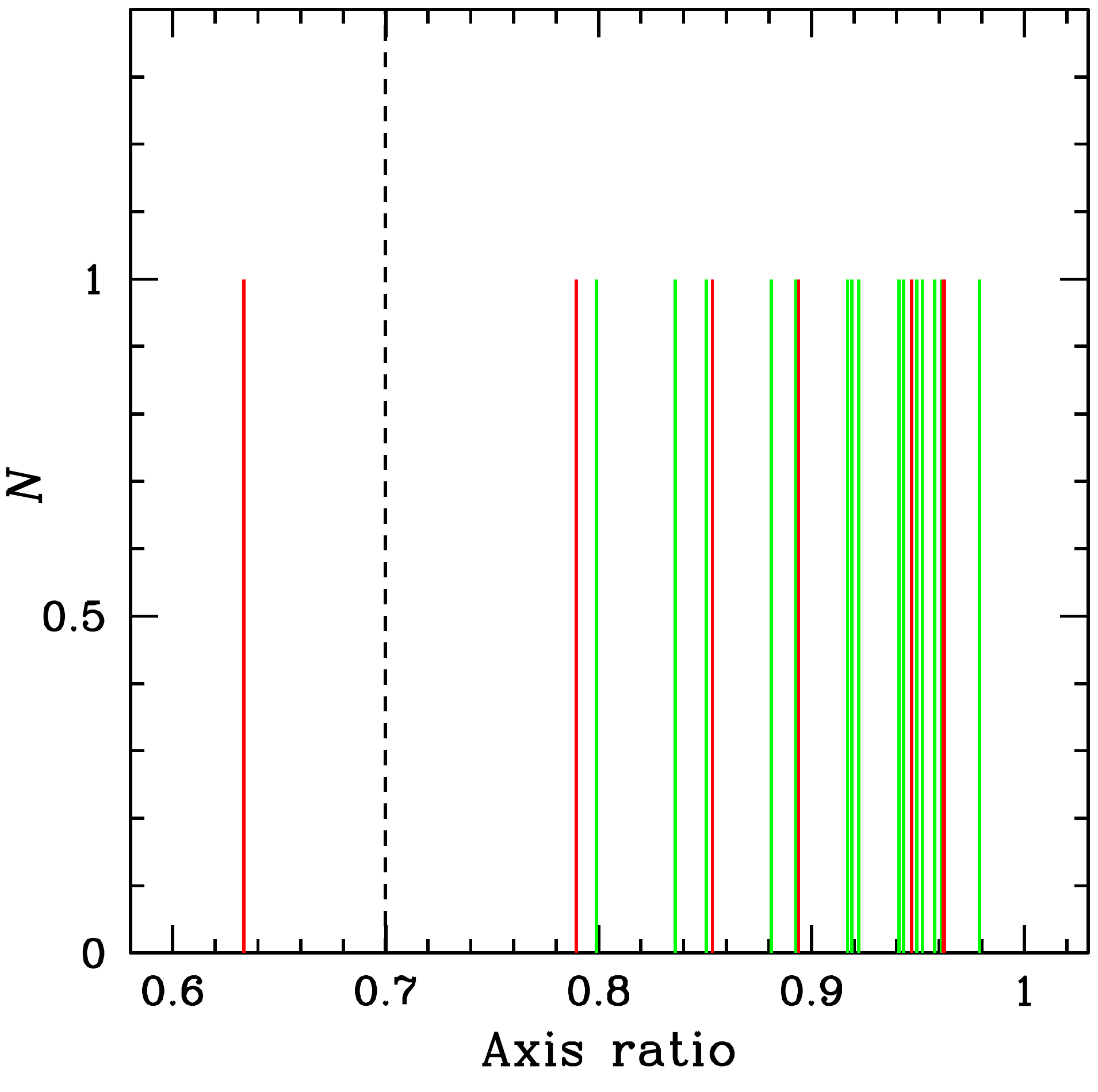}
\includegraphics[scale=0.30]{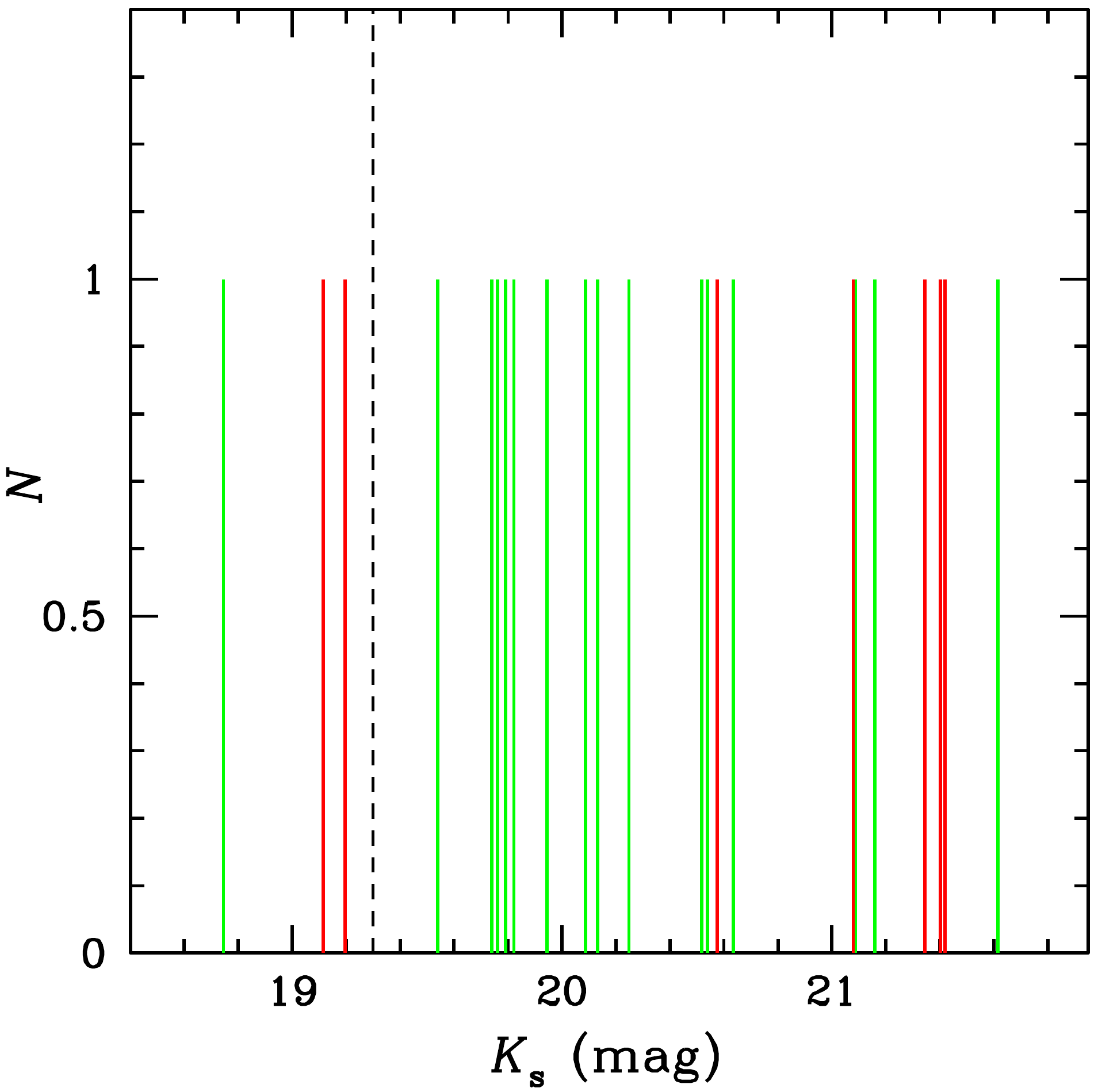}
\hspace*{0.205cm}\includegraphics[scale=0.30]{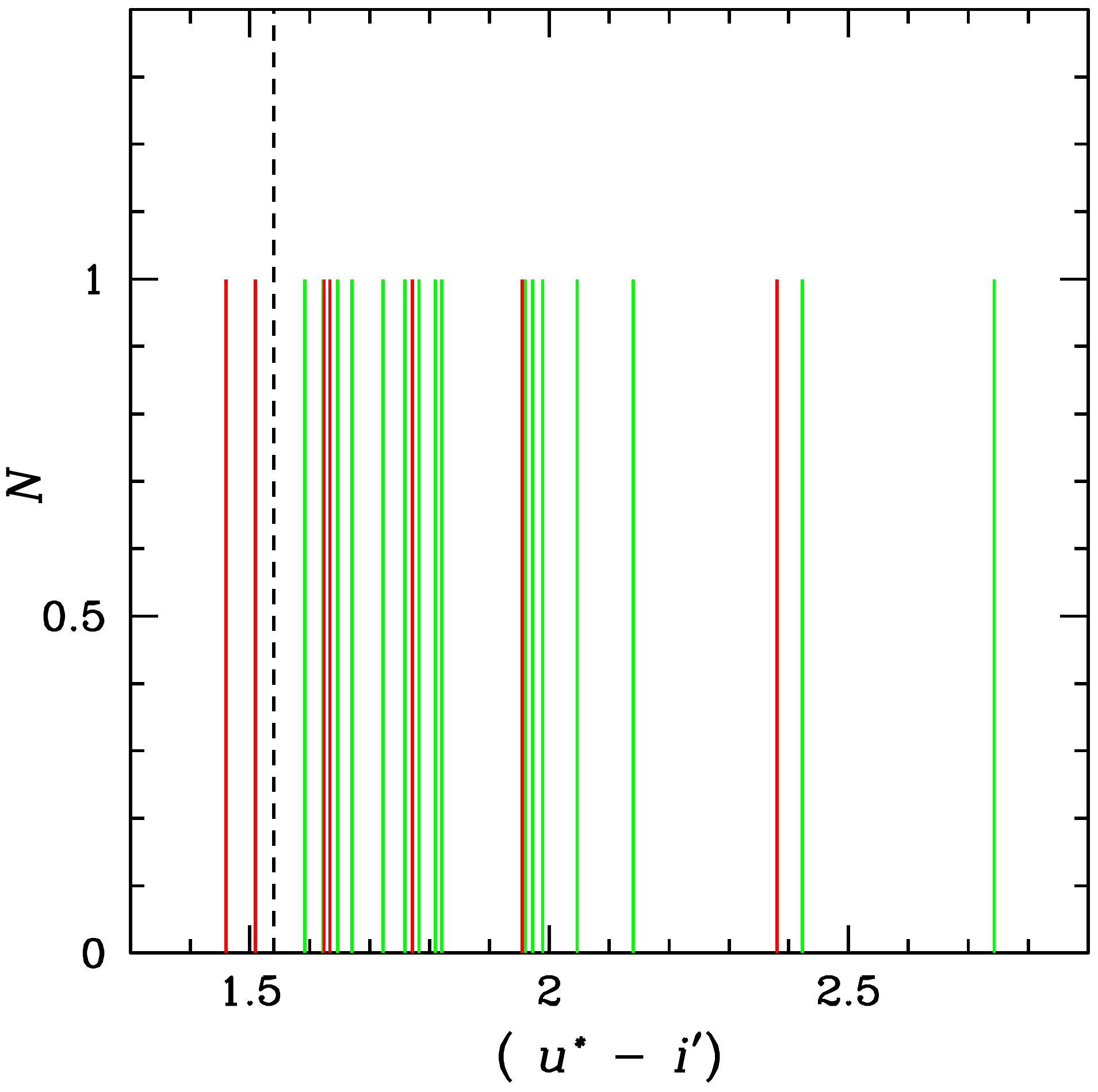}
\hspace*{0.205cm}\includegraphics[scale=0.30]{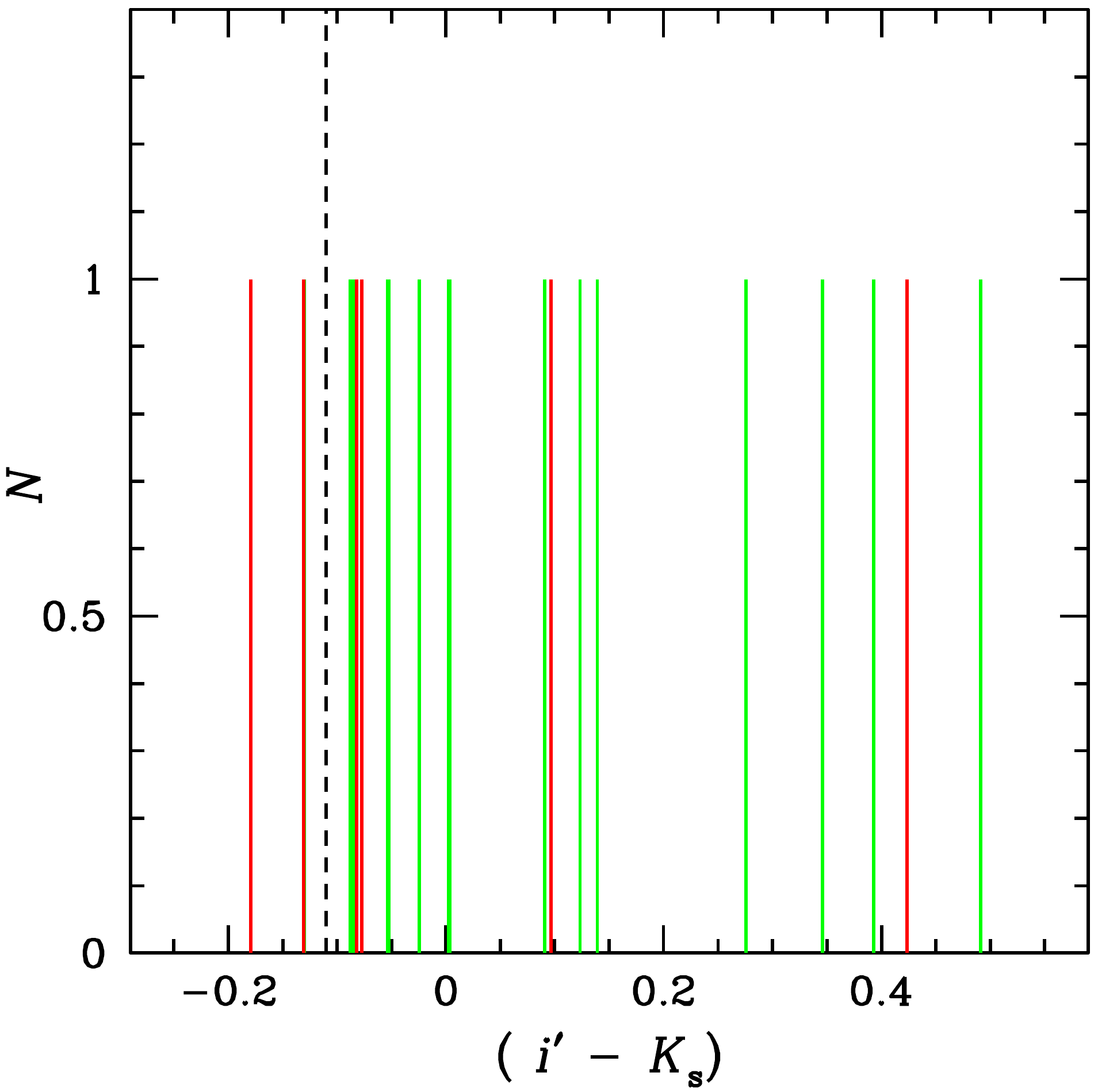}
%\caption{Sample selection parameters. {\it Top left:} FWHM; {\it top right:} SPREAD\_MODEL $\times$ 100; 
%{\it middle left:} axis ratio; {\it middle right:} \ks\ magnitude; {\it bottom left:} ($u^* - i^\prime$); {\it bottom right:} ($i^\prime - K_{\rm s}$). {\it Green bars:} confirmed GCCs;
%{\it red bars:} rejected GCCs. {\it Dashed lines:} suggested new value limits for improved selection.
\caption{Sample selection parameters. {\it Top left:} FWHM; {\it top center:} SPREAD\_MODEL $\times$ 100; 
{\it top right:} axis ratio; {\it bottom left:} \ks\ magnitude; {\it bottom center:} ($u^* - i^\prime$); {\it bottom right:} ($i^\prime - K_{\rm s}$). {\it Green bars:} confirmed GCCs;
{\it red bars:} rejected GCCs. {\it Dashed lines:} suggested new value limits for improved selection.
\label{fig:method}}
\end{figure*}

{
From their final sample of 39 GC candidates, and 
considering two (5\%) contaminants,
\citet{gonz17} extrapolated to a total of 
$N_{\rm GC}$ = 144$\pm$31$^{+38}_{-36}$ for NGC~4258, with the first error statistical
and the second, systematic. These numbers were obtained following the procedure introduced
by \citet{kiss99}. The method, based on a comparison with the
GC system of the MW, simultaneously accounts for photometric incompleteness, extrapolation over the 
luminosity function, and incomplete spatial coverage. The total number of clusters in a galaxy, $N_{\rm GC}$,
is equal to the number of MW clusters, multiplied by the ratio of the observed GCCs to the numbers that
would have been detected in the MW if our Galaxy were observed at the distance and with the orientation of the galaxy of interest, 
with the same instrument and to the same depth.   
}

{Here}, applying the 30\% contamination correction to the {total number, total mass, and 
specific frequency derived by \citet{gonz17} for the GC system of NGC 4258, we arrive at 
$N_{\rm GC} = 105\pm26\pm31$, 
log $M_{\rm GC} = 7.5\pm0.1\pm0.1$, 
$S_N = 0.3 \pm 0.1$ (random uncertainty only). After the contamination correction, NGC~4258 is} in even closer agreement
than estimated by \citet{gonz17} with the scaling relations 
between {GC system parameters} and $M_\bullet$ derived for elliptical galaxies.\footnote{
The relations can be written as log $N_{\rm GC} = (-5.78 \pm 0.85) + (1.02 \pm 0.10)$ log $M_\bullet/M_\odot$, and 
log $M_{\rm GC} = (-1.40 \pm 0.79) + (1.15 \pm 0.09)$ log $M_\bullet/M_\odot$. For the mass of the 
central black hole of NGC~4258, these correspond to 
$N_{\rm GC} =$ 100 and log $M_{\rm GC} = $ 7.3.
} 

\section{Discussion and conclusions} \label{sec:concl}

Using multi-object spectroscopy, we have investigated both the validity of the \uiks\ method plus light concentration parameters to select
GCC samples in spiral galaxies, and the kinematics of the GC system of the megamaser prototype NGC~4258.

Regarding the \uiks\ method, we have confirmed that it works very efficiently for spirals, with the only, expected, limitation that the 
fractional contamination will be slightly higher than for ellipticals, given the sparser GC systems of disk galaxies. In our working case there
were no false negatives, and the absolute number of contaminants in the sample was consistent with the Galactic
latitude of the observed field and the cosmic variance of background galaxies.
We suggest here a few distance-independent refinements to the limits on the light-concentration parameters and magnitude values used for the sample selection that should help reduce contamination by 30\% -- 50\%.

Once corrected for contamination, NGC~4258 falls right on the $N_{\rm GC}$ and $M_{\rm GC}$ (total mass of the GC system) versus $M_\bullet$ relations followed by elliptical galaxies, within significantly less than one sigma. 

We have also determined that the GC system seems to actually have a disky configuration, and to be rotating with a projected maximum velocity of 128 km s$^{-1}$, in phase with the HI disk. This merits further investigation, for example, through a full dynamical modeling. In a few GC systems for which rotation has been measured, it is most common for the metal-rich population (MRP) with small real or projected galactocentric distances. For example, in the MW,\footnote{
\url{https://ned.ipac.caltech.edu/level5/Harris2/Harris1.html}.} 
the MRP ([Fe/H] = -0.6) within 9 kpc from the Galaxy center has an average rotation velocity 
$V_{\rm rot, MRP}$ = 116$\pm$24 km s$^{-1}$ ($\sim$ 50\% of disk rotation; Reid et al.\ 2014), and a ratio $V_{\rm rot,MRP}/\sigma_{\rm LOS,MRP}$ = 1.4$\pm$0.3,
where $\sigma_{\rm LOS}$ is the line-of-sight velocity dispersion.
However, the metal-poor population (MPP, [Fe/H] = -1.6) is pressure-supported, with an average $V_{\rm rot, MPP}$ = 31$\pm$26 km s$^{-1}$, $V_{\rm rot,MPP}/\sigma_{\rm LOS,MPP}$ = 0.3$\pm$0.3 between 0 and 12 kpc.     

On the other hand, in M~81 \citep{nant10} the MRP ([Fe/H] $>$ -1.06) lies mostly within 4 kpc from the center
and has $V_{\rm rot, MRP}$ = 122$\pm$18 km s$^{-1}$ ($\sim$ 60\% of disk rotation, \hyperlink{fabe79}{Faber \& Gallagher 1979}), $V_{\rm rot,MRP}/\sigma_{\rm LOS,MRP}$ = 1., 
whereas the MPP ([Fe/H] $<$ -1.06) reaches beyond 8 kpc and is, once again, pressure supported, 
with $V_{\rm rot, MPP}$ = 67$\pm$ 38 km s$^{-1}$, $V_{\rm rot,MPP}/\sigma_{\rm LOS,MPP}$ = 0.5.
In M~31, \citet{lee08} find that the MRP ([Fe/H] = -0.6) rotates within 3 kpc with an average 
$V_{\rm rot,MRP}$ = 221$\pm$35 km s$^{-1}$ ($\sim$ 95\% of disk rotation, \hyperlink{fabe79}{Faber \& Gallagher 1979}), $V_{\rm rot,MRP}/\sigma_{\rm LOS,MRP}$ = 1.9 $\pm$ 0.3.
In addition, \citet{perr02} notice that the MRP of M~31 is not flattened, hence bulge-like; that its rotation
axis is tilted 5$\degr$--10$\degr$ with respect to the minor axis of M~31; and that its velocity dispersion is 
quite similar to the bulge's. Beyond 3 kpc, there is   
no evidence of rotation for the metal-rich clusters. 
The MPP in M~31 rotates with basically the same speed as the MRP, $V_{\rm rot,MPP}$ = 
217$\pm$ 33 km s$^{-1}$, to a projected distance of 5 kpc, albeit with a slightly
larger velocity dispersion, i.e.,  $V_{\rm rot,MPP}/\sigma_{\rm LOS,MPP}$ = 1.6$\pm$0.5. Further away than 5 kpc, 
the MPP in M~31 is pressure supported, with $V_{\rm rot,MPP}/\sigma_{\rm LOS,MPP}$ = 0.7 $\pm$ 0.7. 

The GC system of NGC~4258 is different. 
It does not have strong signs of bimodality \citep{gonz17}; its members seem to have low metallicity, around $Z =$ 0.001 ([Fe/H] $\sim$ -1.2); its spatial distribution seems flattened; 
and its projected rotation velocity of 128 km s$^{-1}$ is $\sim$ 65\% of the projected HI disk rotation velocity, {\em up to a projected distance of more than 17 kpc}.
Assuming that the GC system has an intrinsic anisotropy similar to the HI disk, and that the viewing angle is the same,
the deprojected velocity would be $V_{\rm rot} \sim$ 140 km s$^{-1}$, and $V_{\rm rot}/\sigma_{\rm GC}$ = 1.3; 
without deprojection, $V/\sigma$ = 1.2.

These characteristics evoke the clumpy, gaseous, high pressure, disks at $z \geqslant$ 2 that
constitute nowadays the more favored environment for GC formation \citep[e.g.][]{ager09,krui15}. 
These disks rotate, but are highly turbulent, with $V/\sigma \sim$ 1, and their existence is
supported by strong observational evidence \citep[e.g.,][]{foer09,wisn15,charr17,turn17,dess18}. 

Beyond validating the \uiks\ plus light-concentration parameters method for the identification of GCCs in spirals, the
present work highlights the importance of studying GC systems and scaling relations in spirals. While these
relations in ellipticals seem to be showing the end result of star formation and galaxy assembly, in
lower mass galaxies we may be witnessing the fingerprints of the individual steps of the process, 
and hence fundamental clues to
the understanding of galaxy formation.

\acknowledgements
R.A.G.L. \& L.L. acknowledge the financial support of DGAPA, UNAM (respectively, projects IN108518 and IN112417),
and of CONACyT, Mexico. {We sincerely thank the anonymous referee for the very constructive and insightful comments.}

{Based on observations made with the NASA/ESA Hubble Space Telescope, and obtained from the Hubble Legacy Archive, which is a collaboration between the Space Telescope Science Institute (STScI/NASA), the Space Telescope European Coordinating Facility (ST-ECF/ESA) and the Canadian Astronomy Data Centre (CADC/NRC/CSA).}

\vspace{5mm}
%\facilities{GTC:OSIRIS,HST,CFHT,Westerbork Synthesis Radio Telescope}

%\software{
%          SExtractor \citep{bert96},
%          IRAF,
%          rvsao \citep{kurt98},
%          sm \citep{lupt91},
%          der\_snr, 
%          Kapteyn \citep{terl15}
%          }

\clearpage

\clearpage

\appendix

\setcounter{figure}{3}
\begin{figure}[h!]
\begin{tabular}{ll}
\hspace*{0.3cm}\raisebox{-0.57cm}{\includegraphics[scale=0.5,angle=0.]{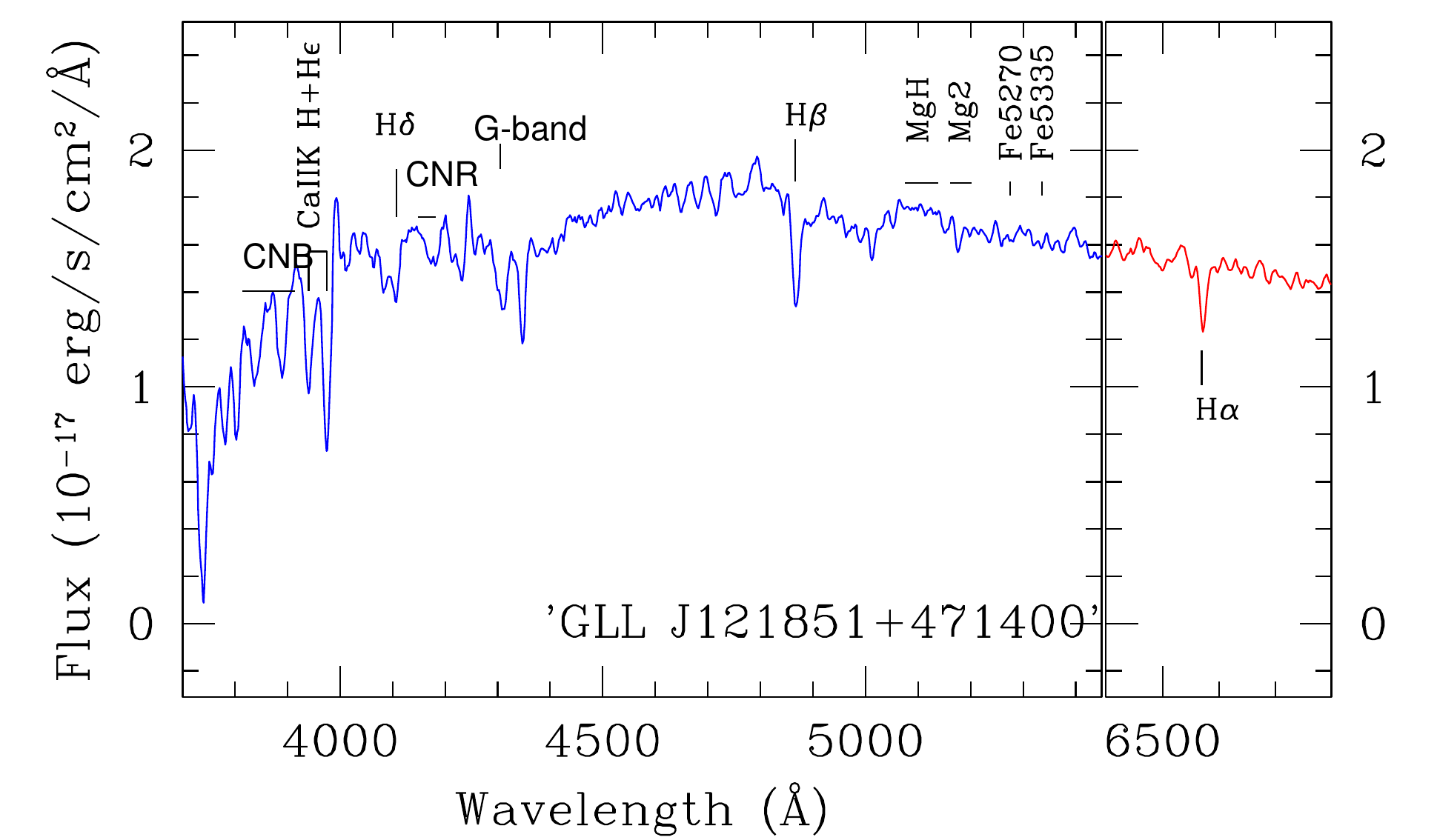}}
&
\hspace*{0.3cm}\includegraphics[scale=0.413]{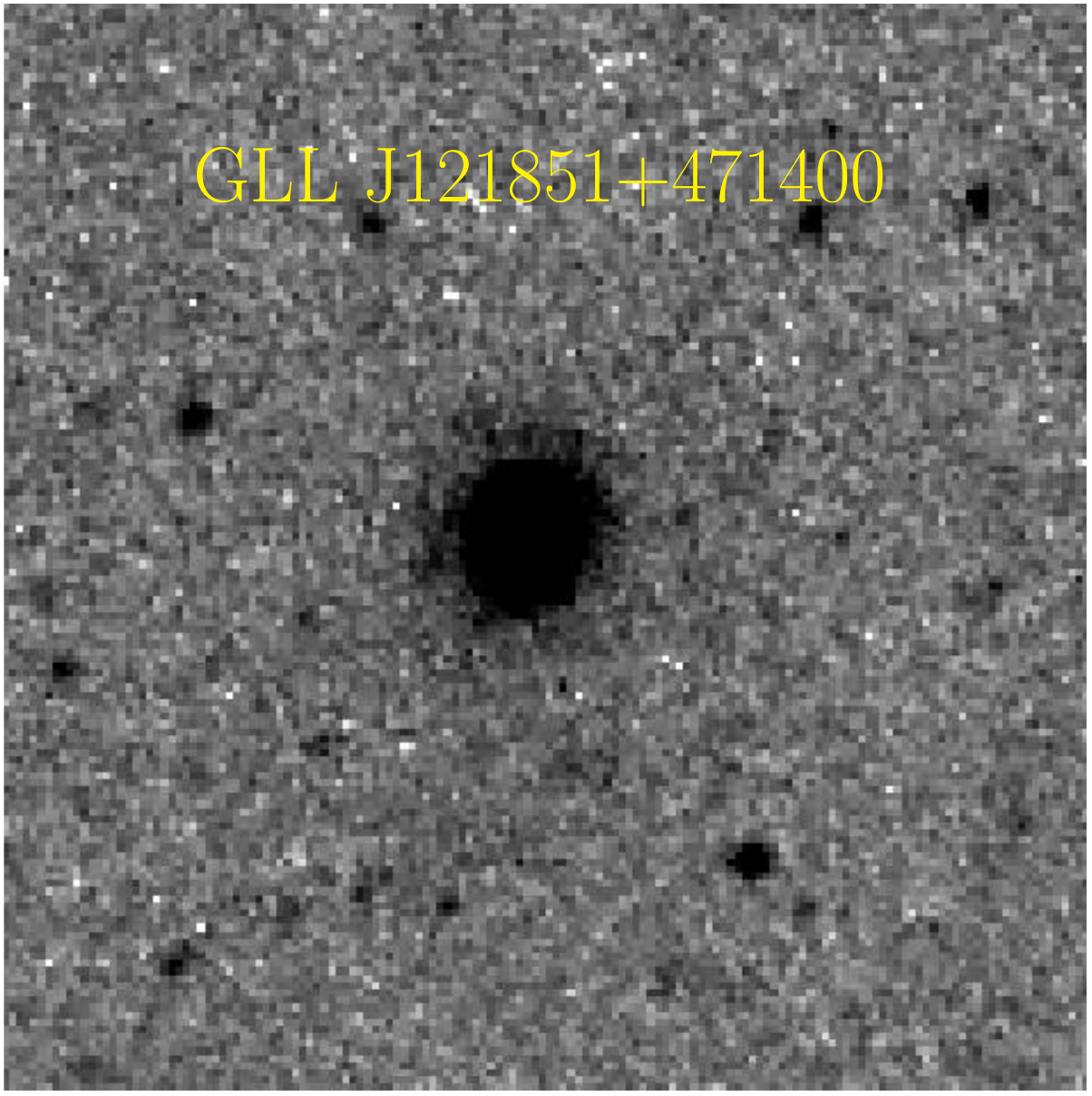}
\end{tabular}
\begin{tabular}{ll}
\hspace*{0.3cm}\raisebox{-0.57cm}{\includegraphics[scale=0.5,angle=0.]{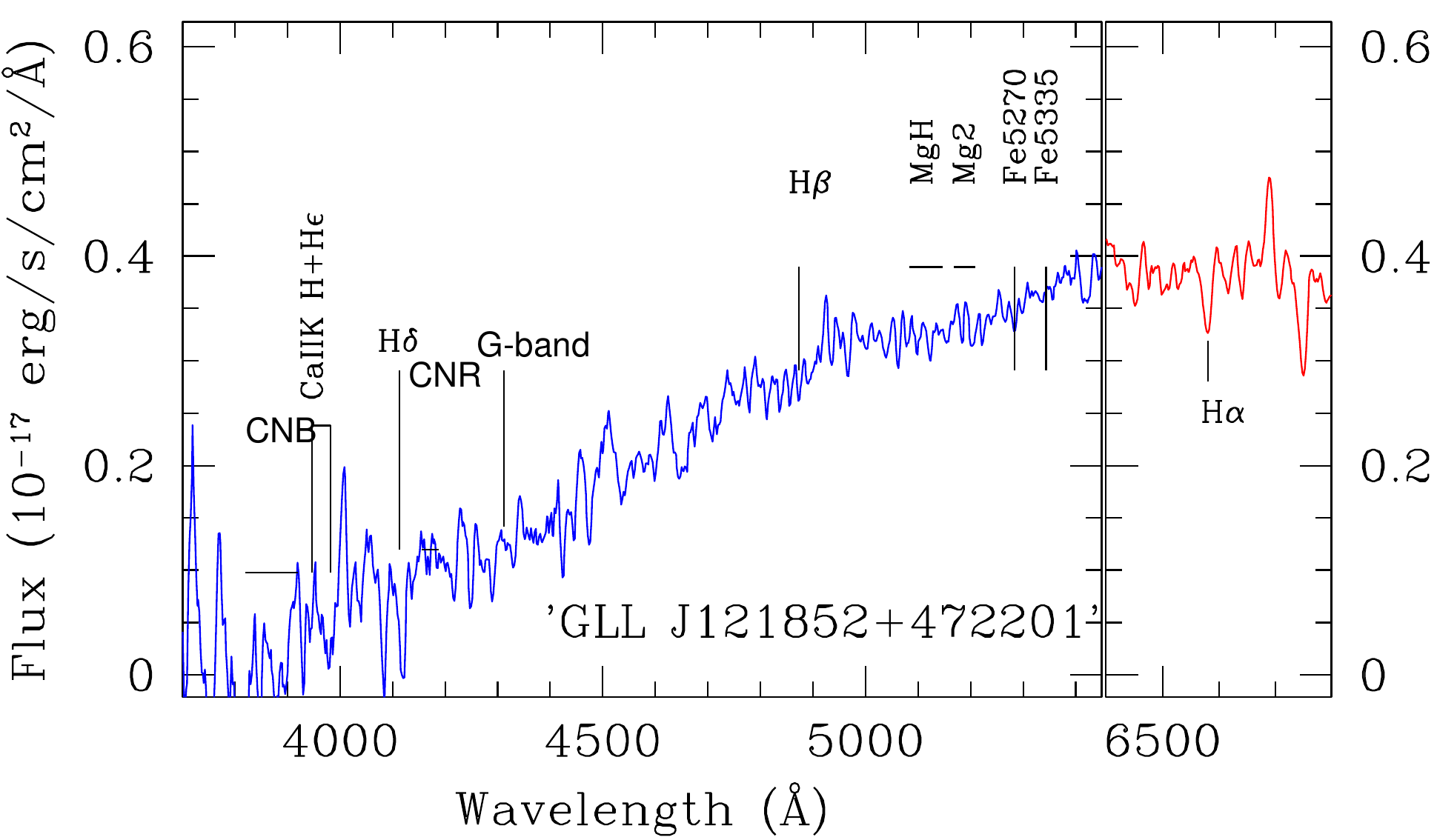}}
&
\hspace*{0.3cm}\includegraphics[scale=0.413]{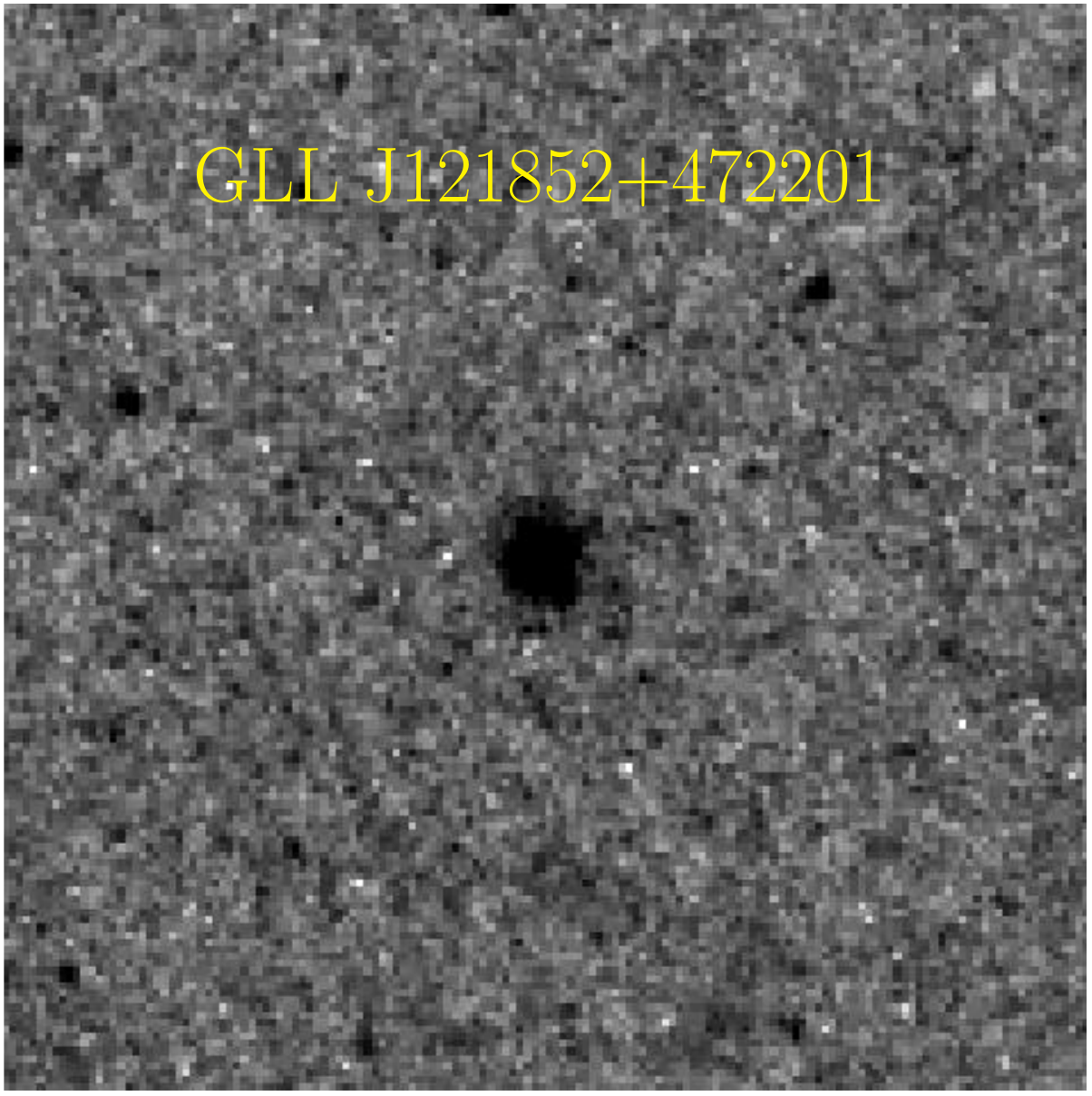}
\end{tabular}
\begin{tabular}{ll}
\hspace*{0.3cm}\raisebox{-0.57cm}{\includegraphics[scale=0.5,angle=0.]{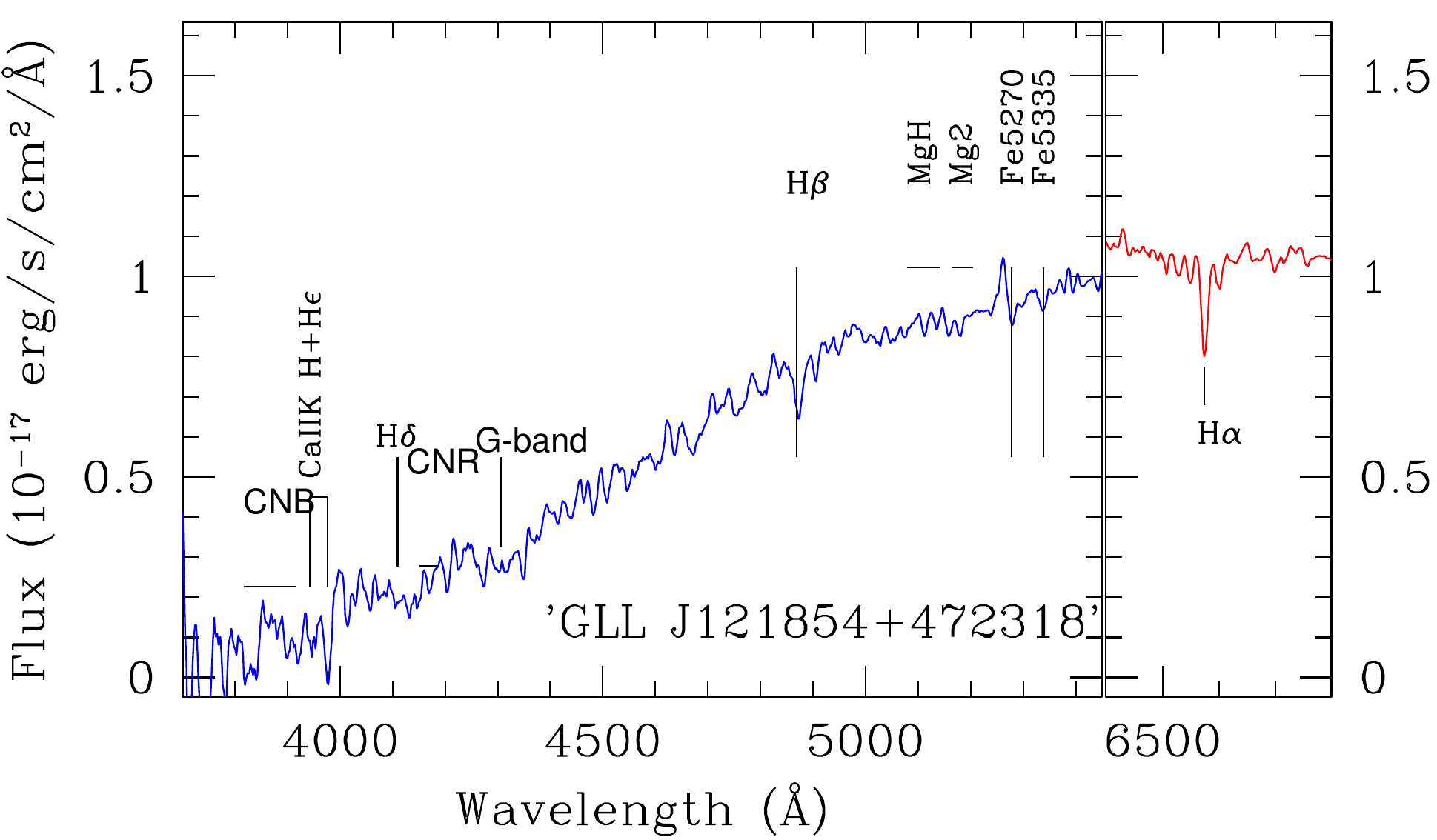}}
&
\hspace*{0.3cm}\includegraphics[scale=0.413]{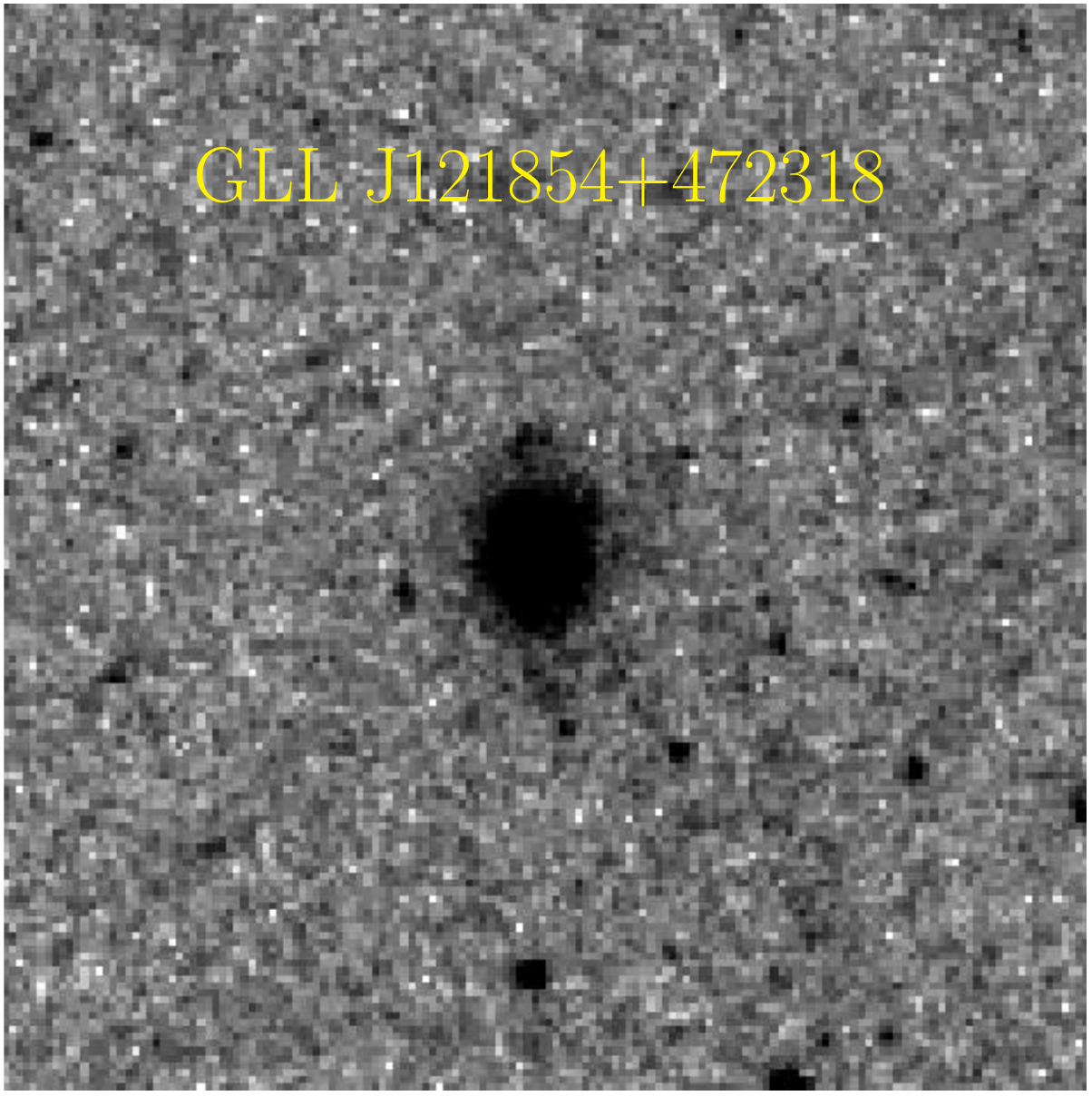}
\end{tabular}
\caption{{\it (Continued--)} Spectra ({\it left}) and grayscales ({\it right}) of confirmed GC candidates.
\label{fig:specim_conf}}
\end{figure}

\clearpage

\setcounter{figure}{3}
\begin{figure}
%\ContinuedFloat
\begin{tabular}{ll}
\hspace*{0.3cm}\raisebox{-0.57cm}{\includegraphics[scale=0.5,angle=0.]{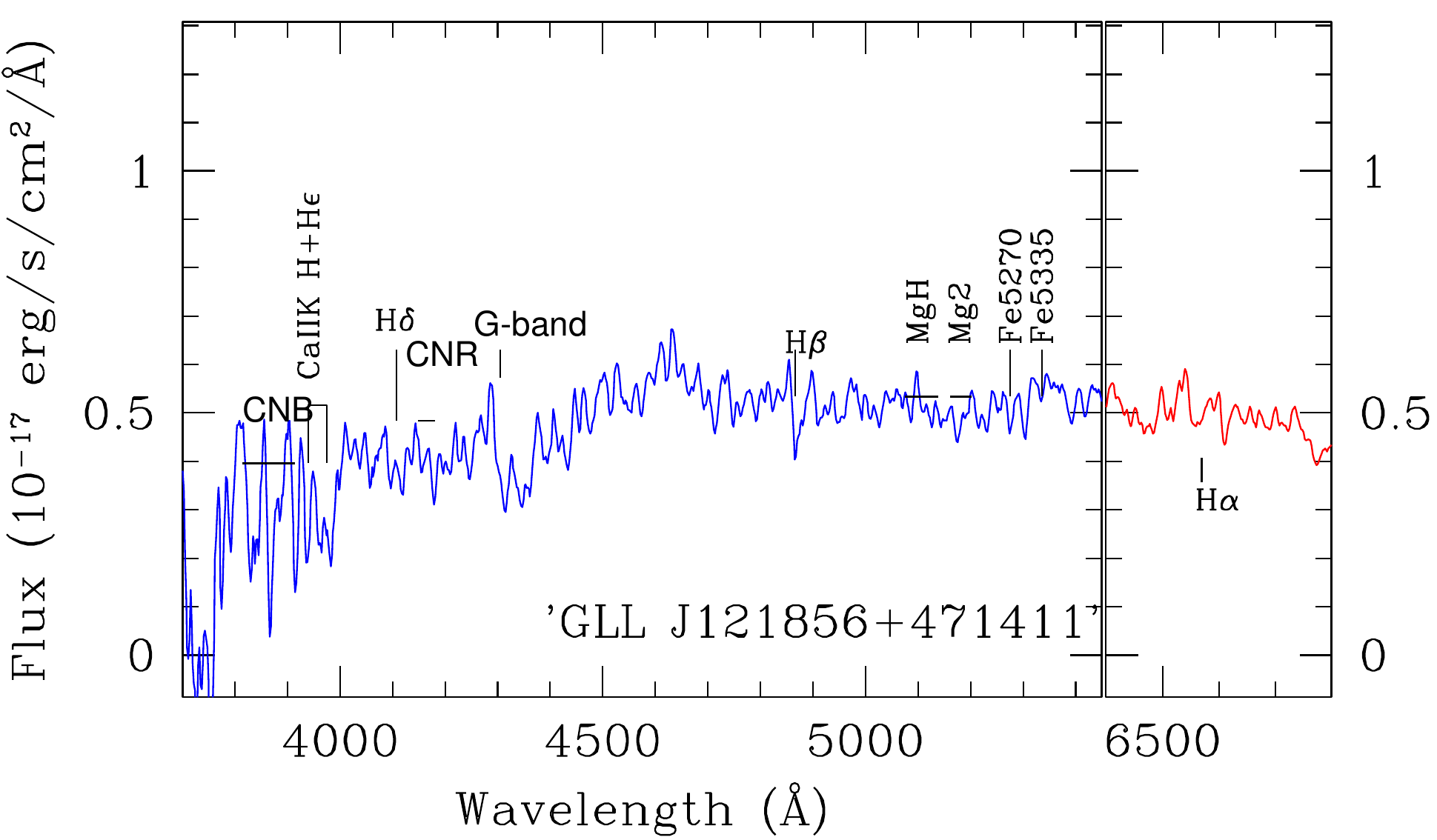}}
&
\hspace*{0.3cm}\includegraphics[scale=0.413]{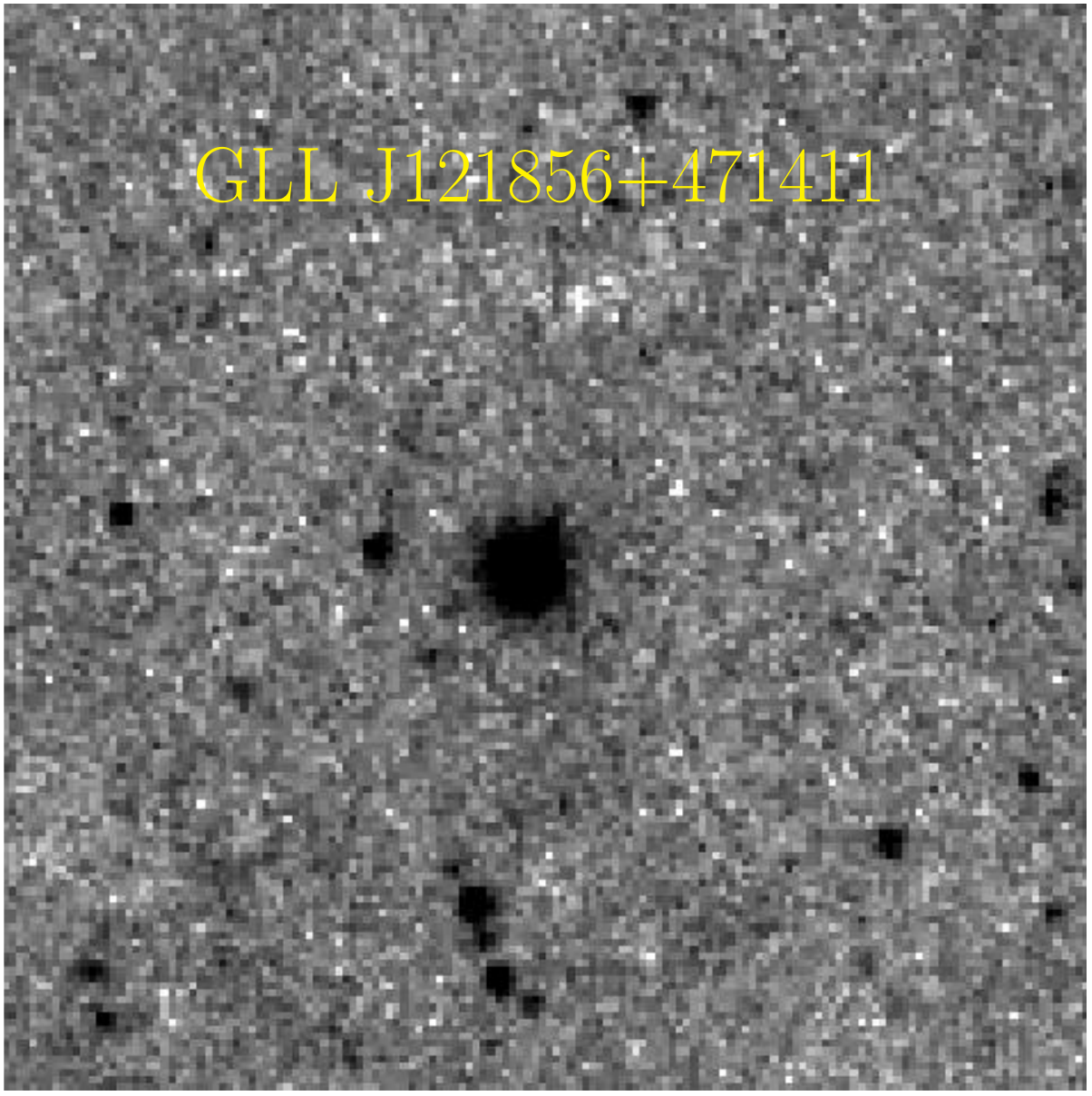}
\end{tabular}
\begin{tabular}{ll}
\hspace*{0.3cm}\raisebox{-0.57cm}{\includegraphics[scale=0.5,angle=0.]{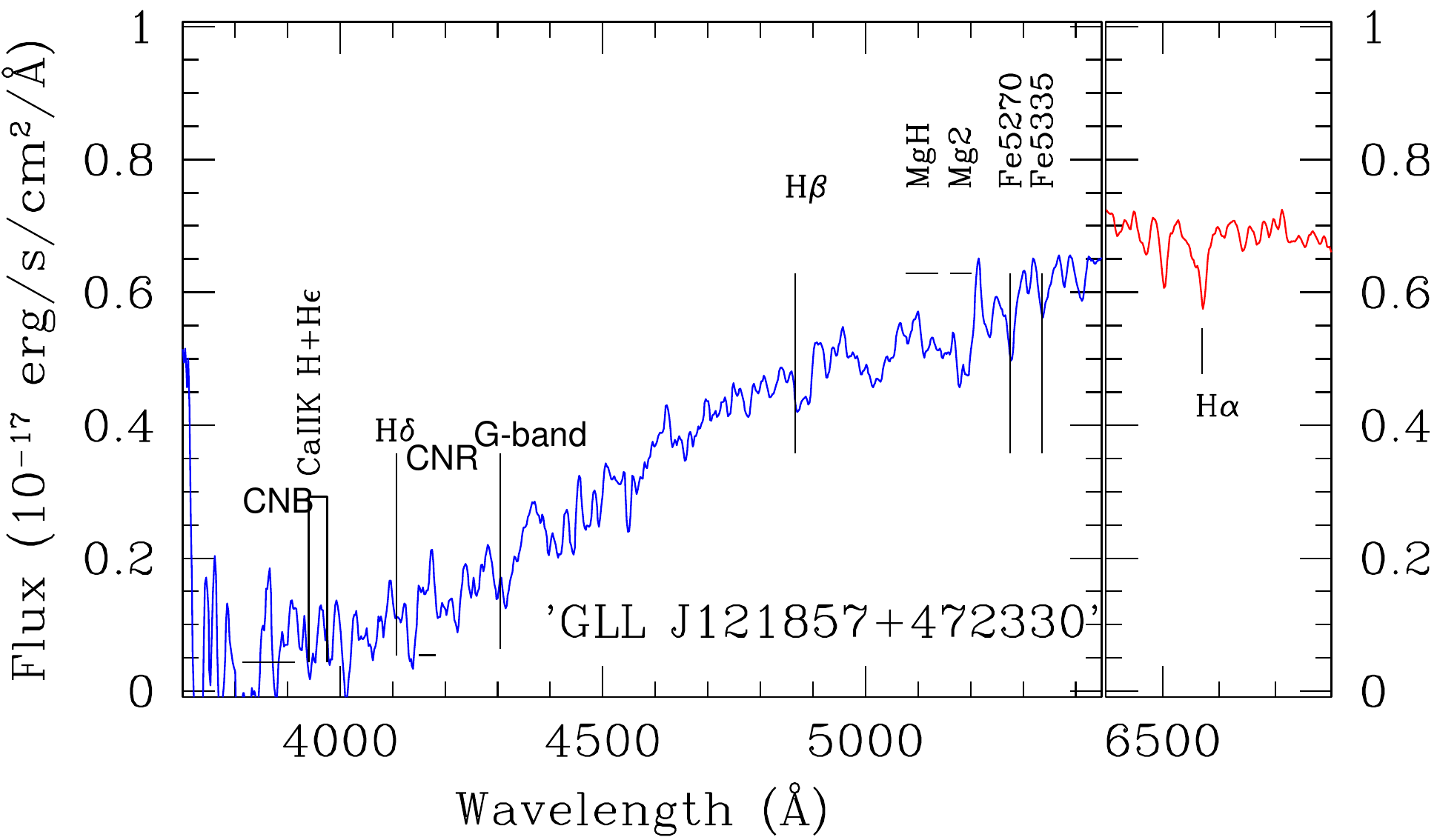}}
&
\hspace*{0.3cm}\includegraphics[scale=0.413]{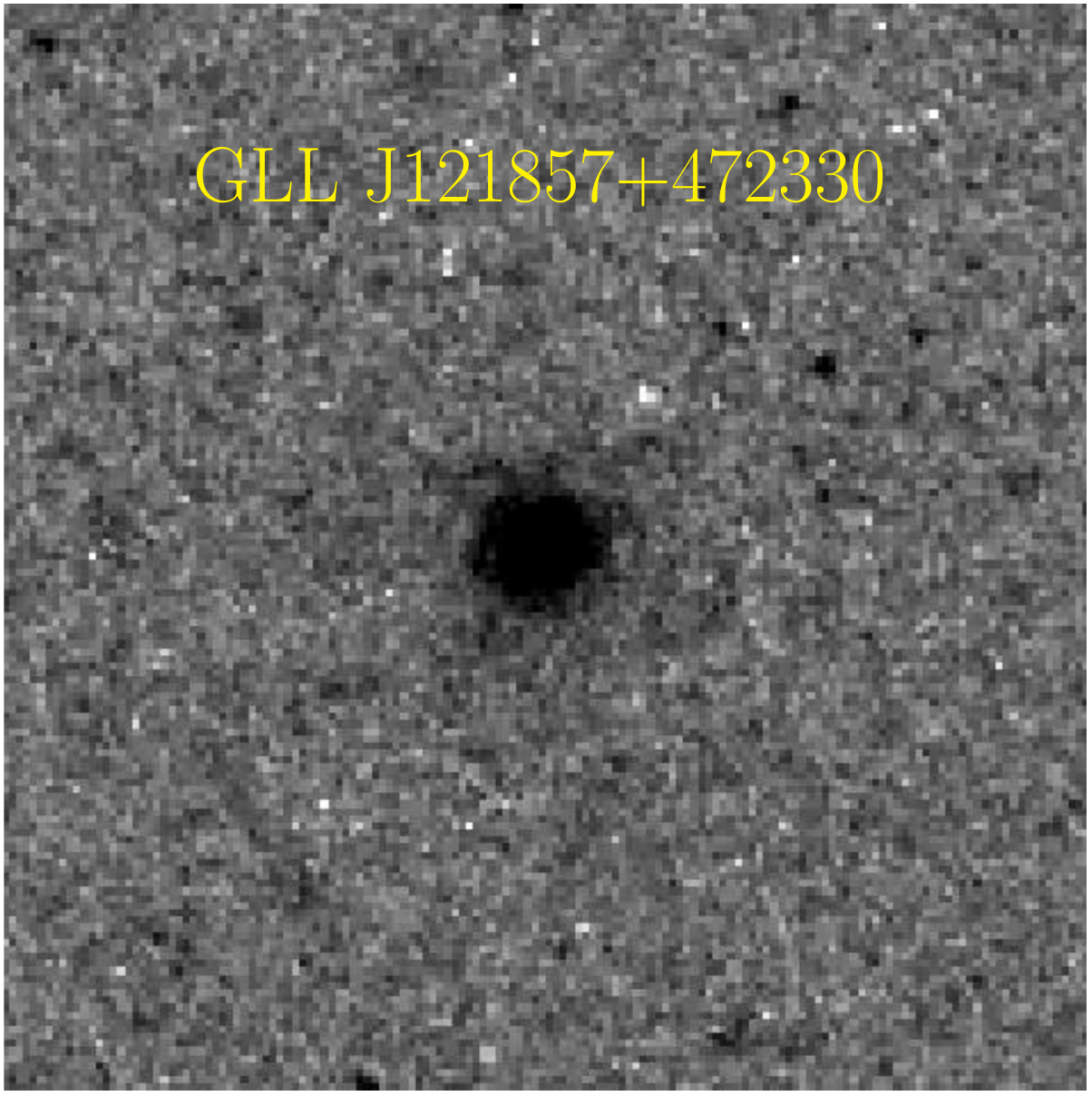}
\end{tabular}
\begin{tabular}{ll}
\hspace*{0.3cm}\raisebox{-0.57cm}{\includegraphics[scale=0.5,angle=0.]{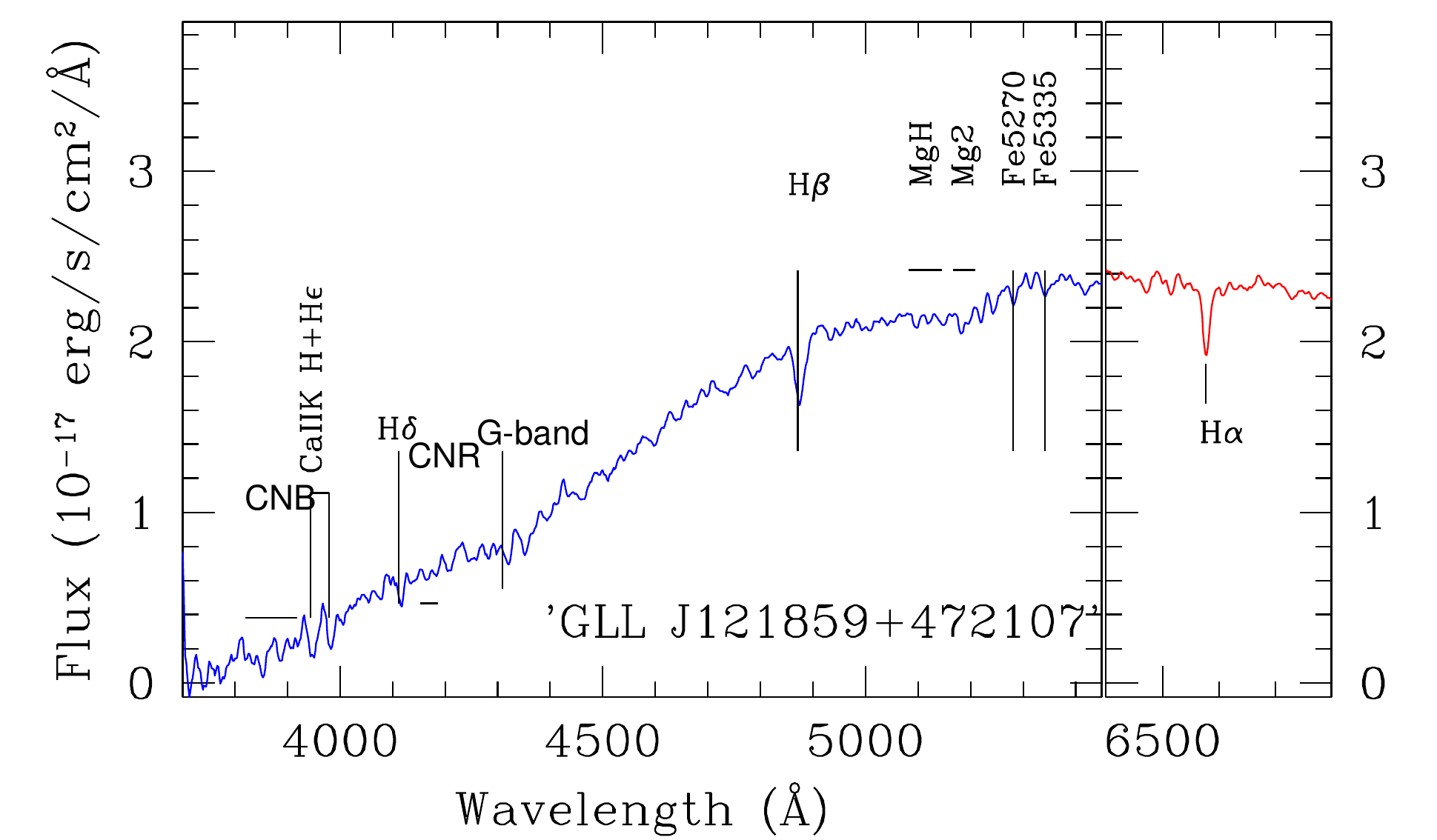}}
&
\hspace*{0.3cm}\includegraphics[scale=0.413]{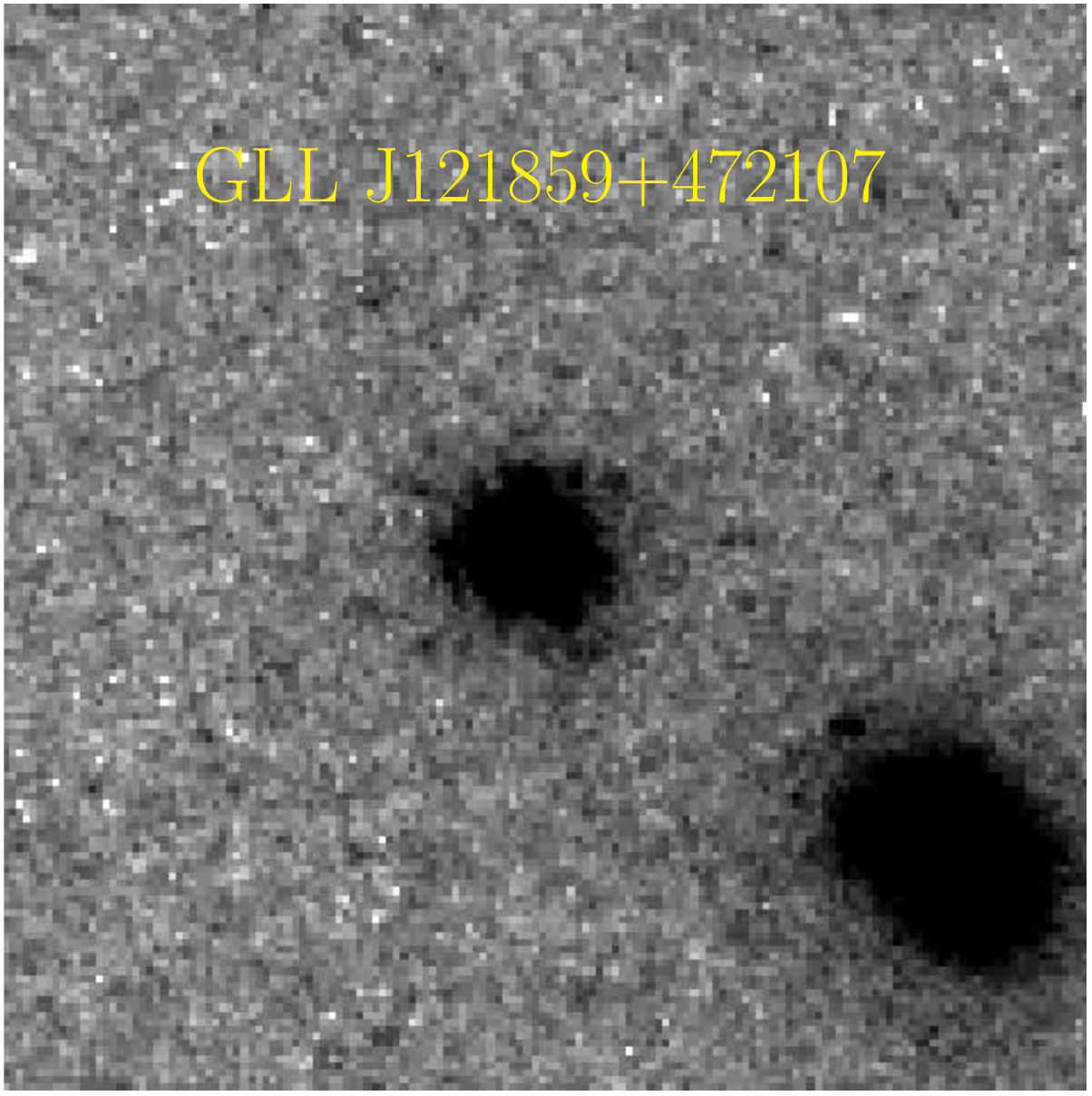}
\end{tabular}
\caption{{\it (Continued--)} Spectra ({\it left}) and grayscales ({\it right}) of confirmed GC candidates.
\label{fig:specim_conf}}
\end{figure}

\clearpage

\setcounter{figure}{3}
\begin{figure}
%\ContinuedFloat
\begin{tabular}{ll}
\hspace*{0.3cm}\raisebox{-0.57cm}{\includegraphics[scale=0.5,angle=0.]{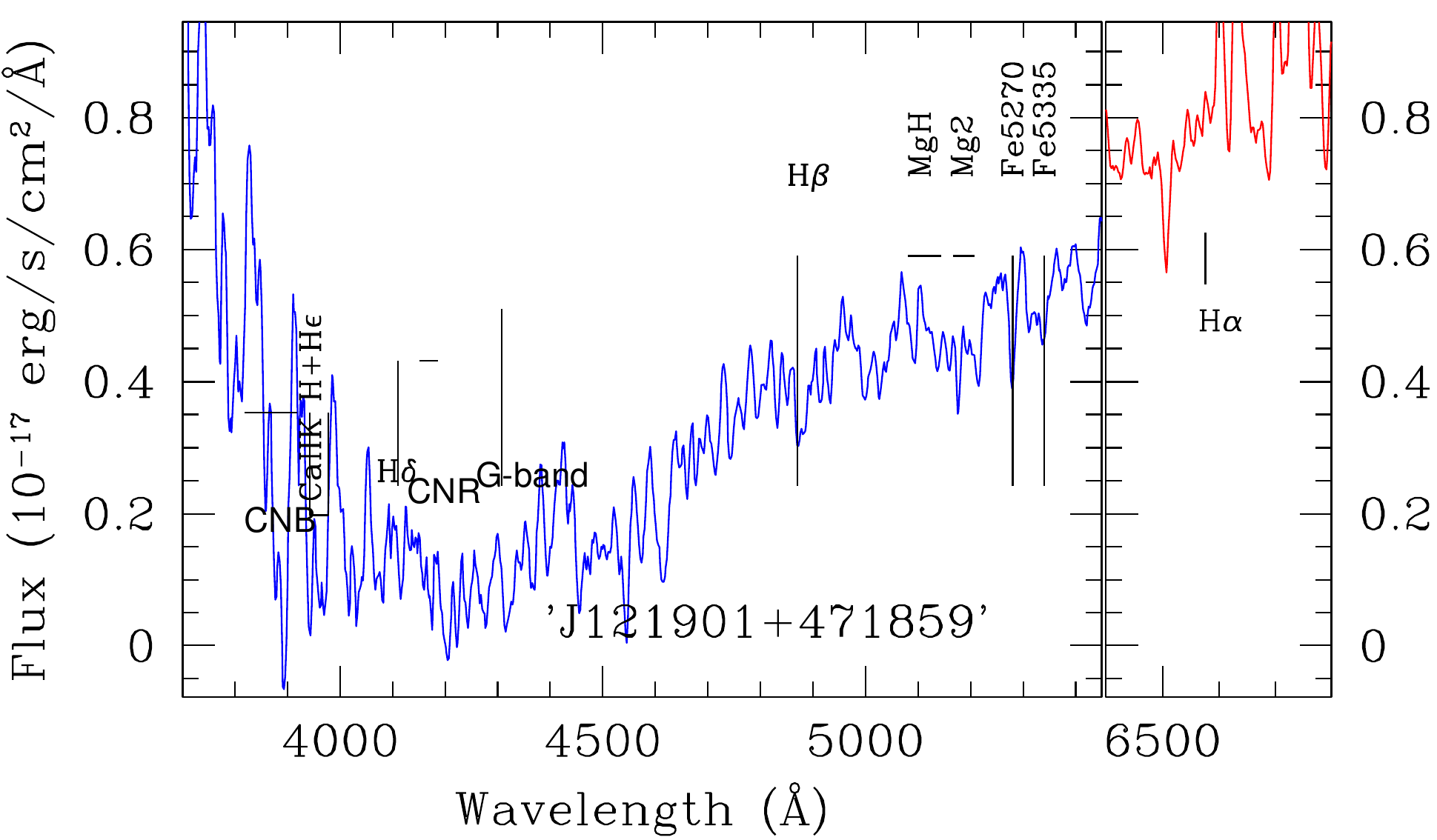}}
&
\hspace*{0.3cm}\includegraphics[scale=0.413]{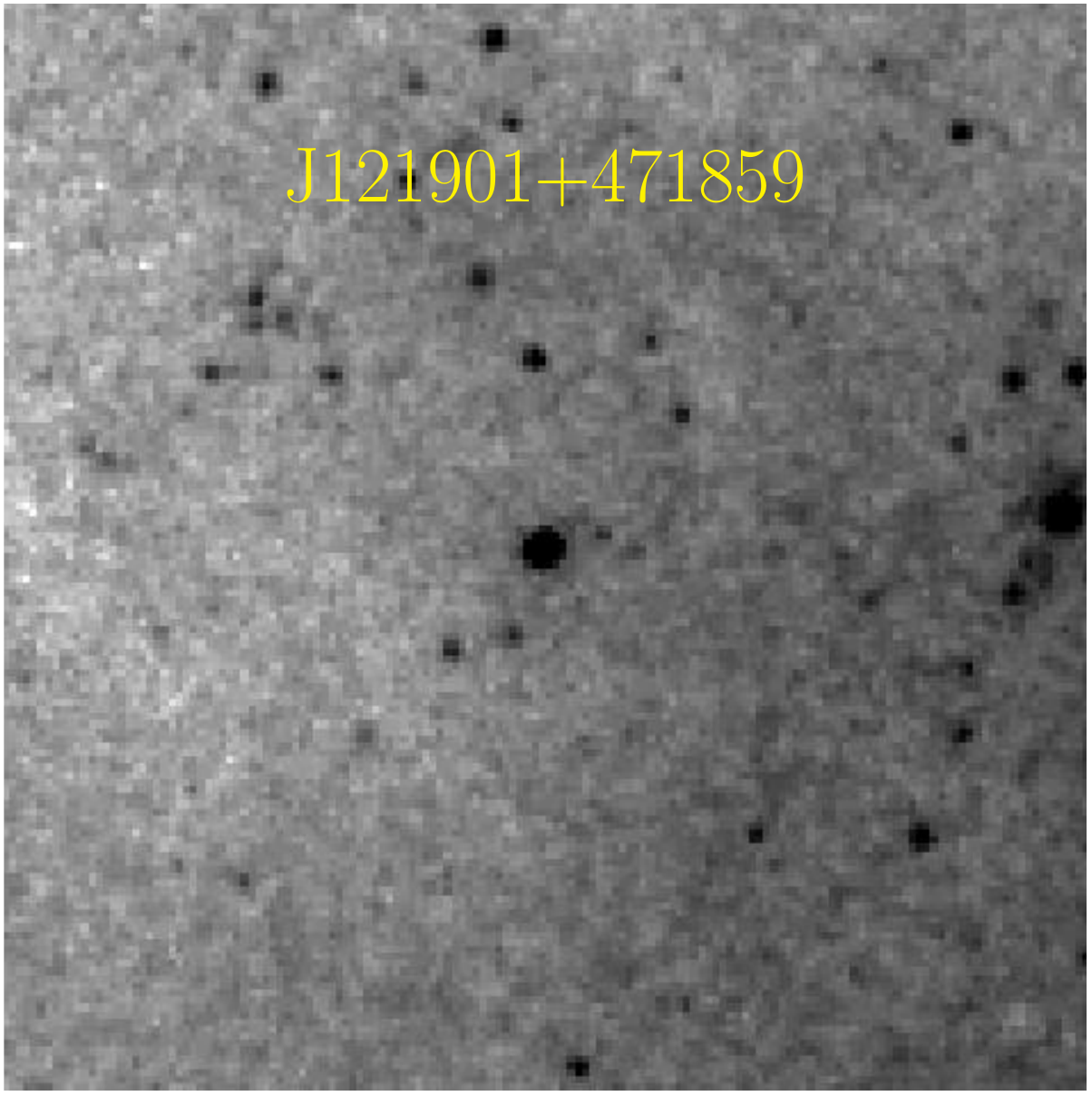}
\end{tabular}
\begin{tabular}{ll}
\hspace*{0.3cm}\raisebox{-0.57cm}{\includegraphics[scale=0.5,angle=0.]{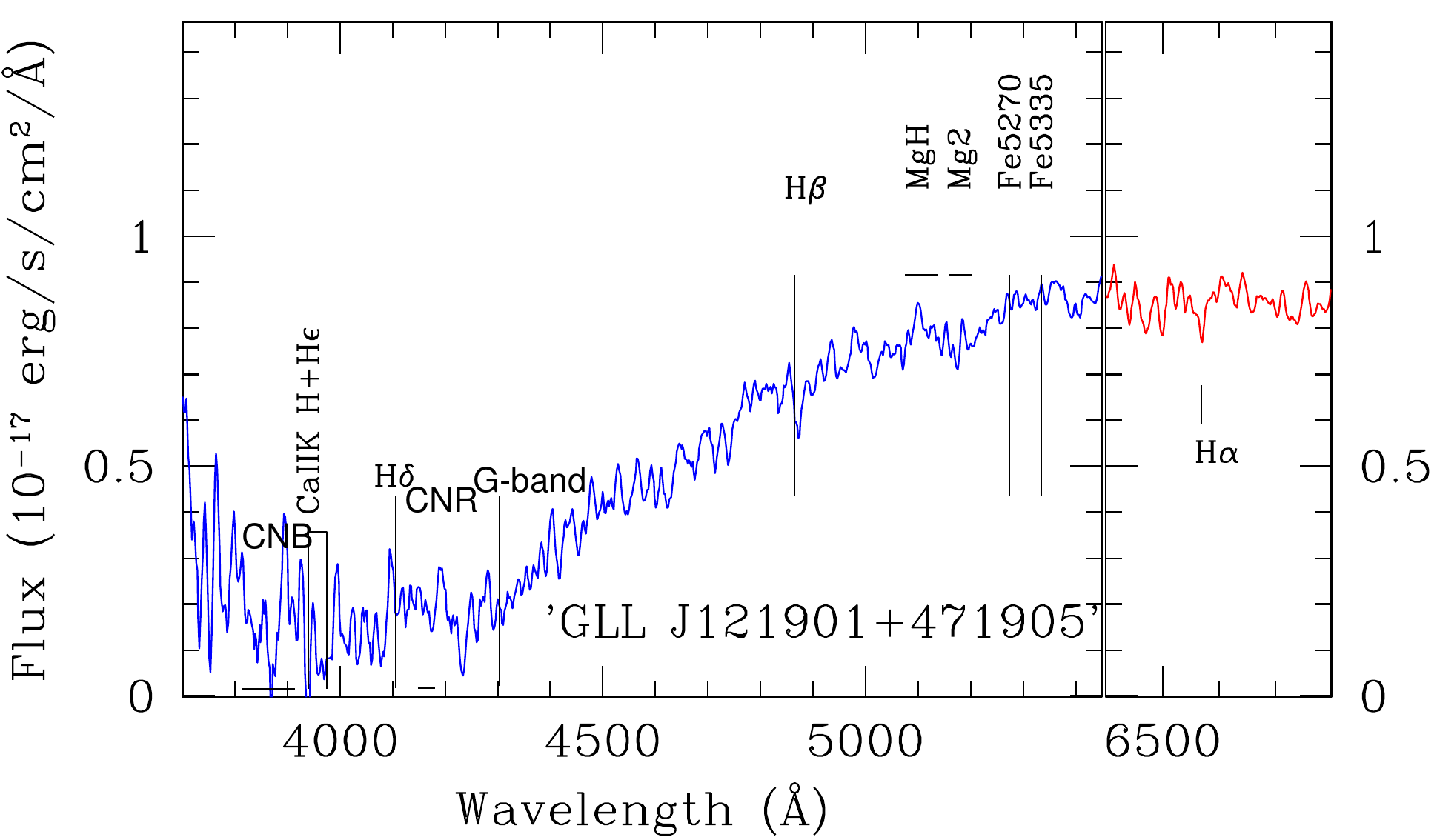}}
&
\hspace*{0.3cm}\includegraphics[scale=0.413]{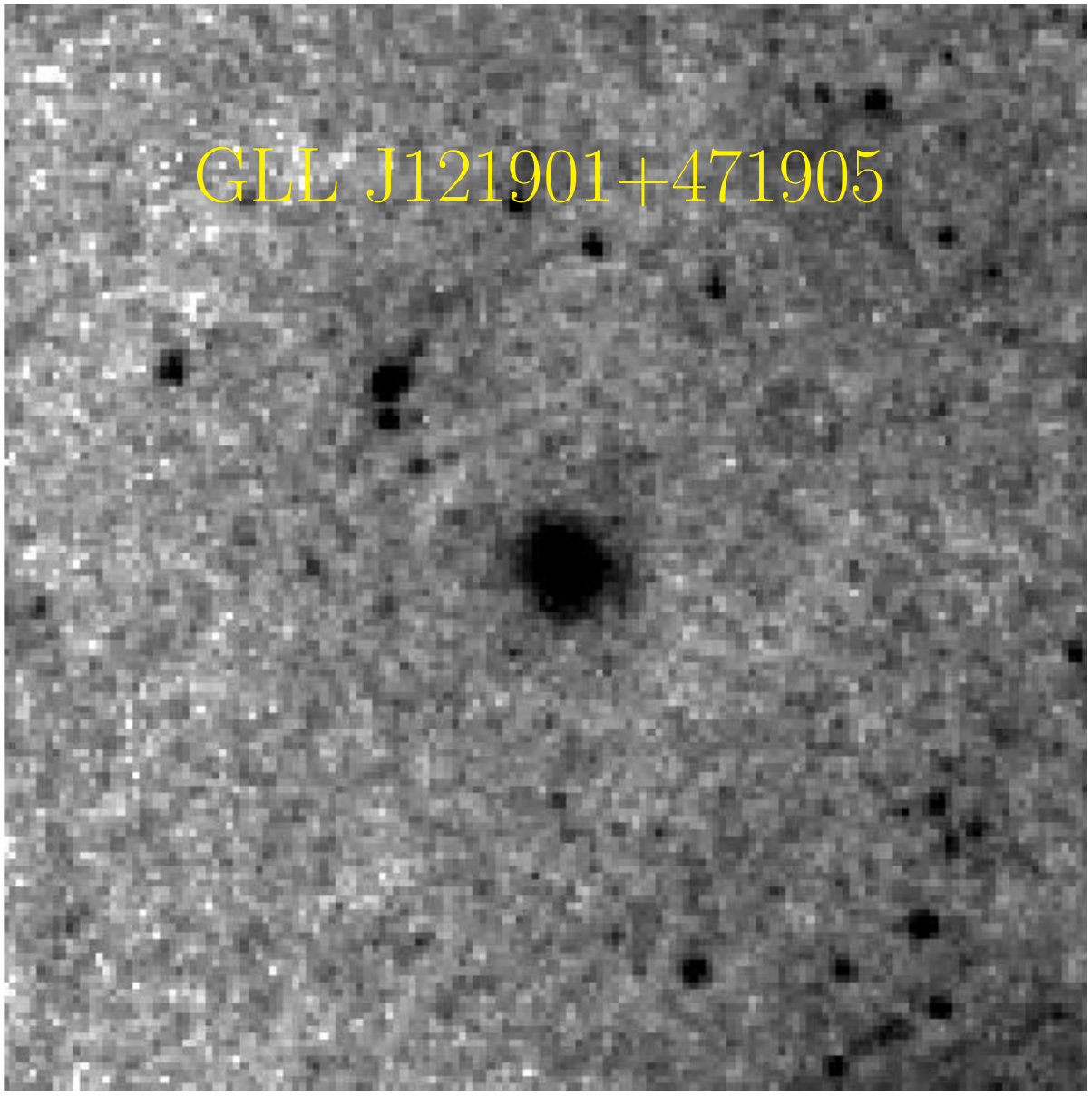}
\end{tabular}
\begin{tabular}{ll}
\hspace*{0.3cm}\raisebox{-0.57cm}{\includegraphics[scale=0.5,angle=0.]{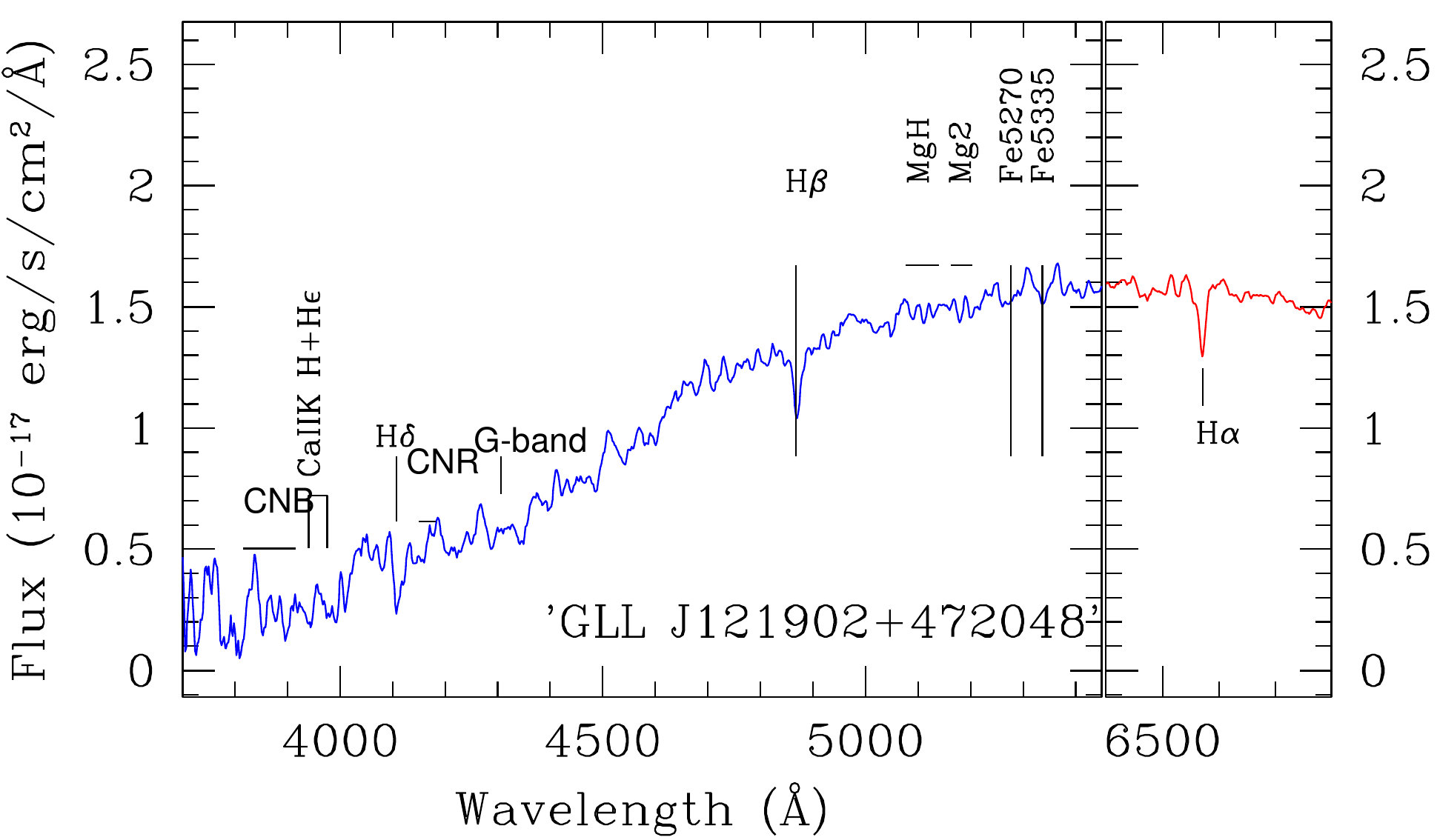}}
&
\hspace*{0.3cm}\includegraphics[scale=0.413]{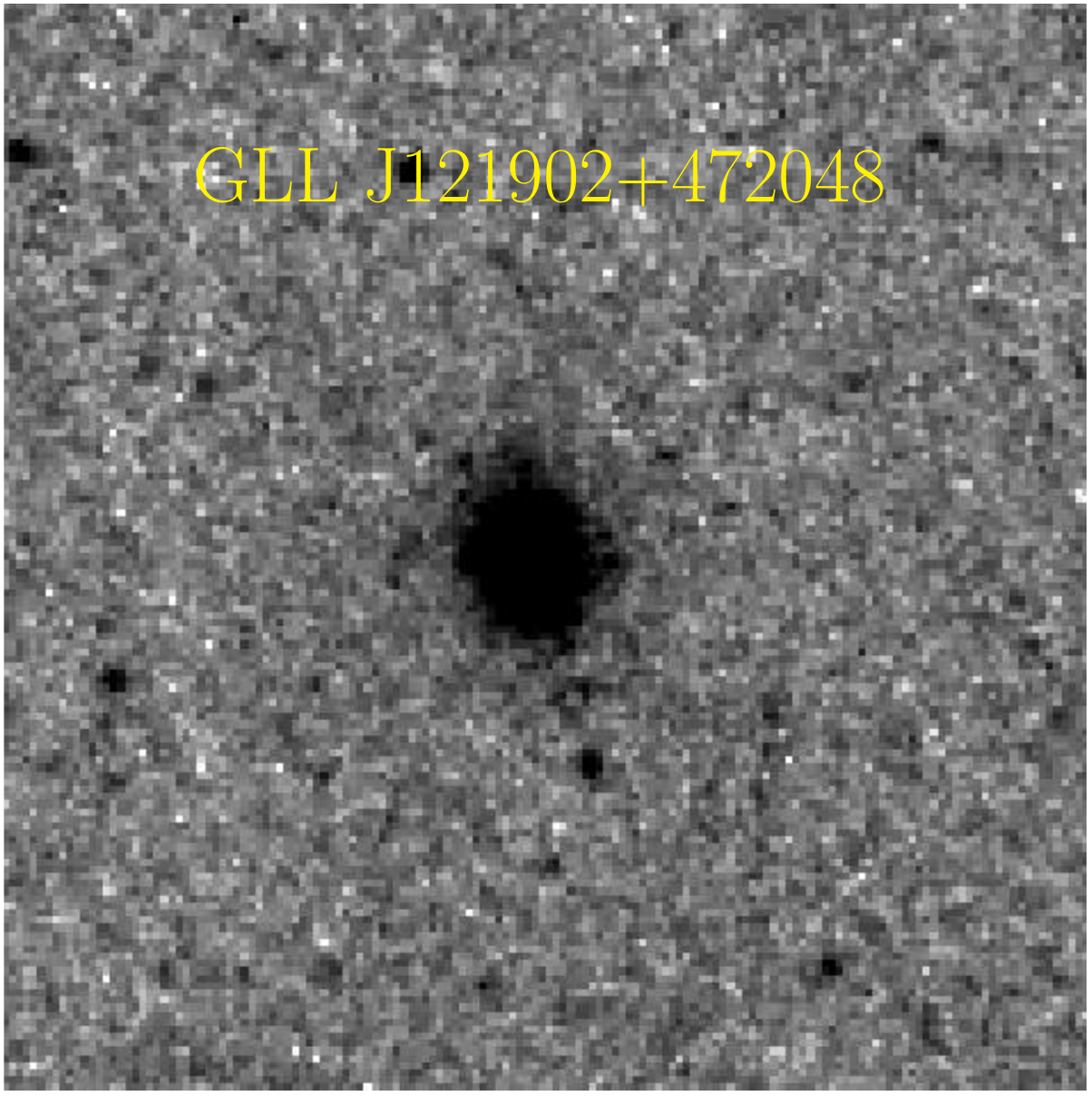}
\end{tabular}
\caption{{\it (Continued--)} Spectra ({\it left}) and grayscales ({\it right}) of confirmed GC candidates.
\label{fig:specim_conf}}
\end{figure}

\clearpage

\setcounter{figure}{3}
\begin{figure}
%\ContinuedFloat
\begin{tabular}{ll}
\hspace*{0.3cm}\raisebox{-0.57cm}{\includegraphics[scale=0.5,angle=0.]{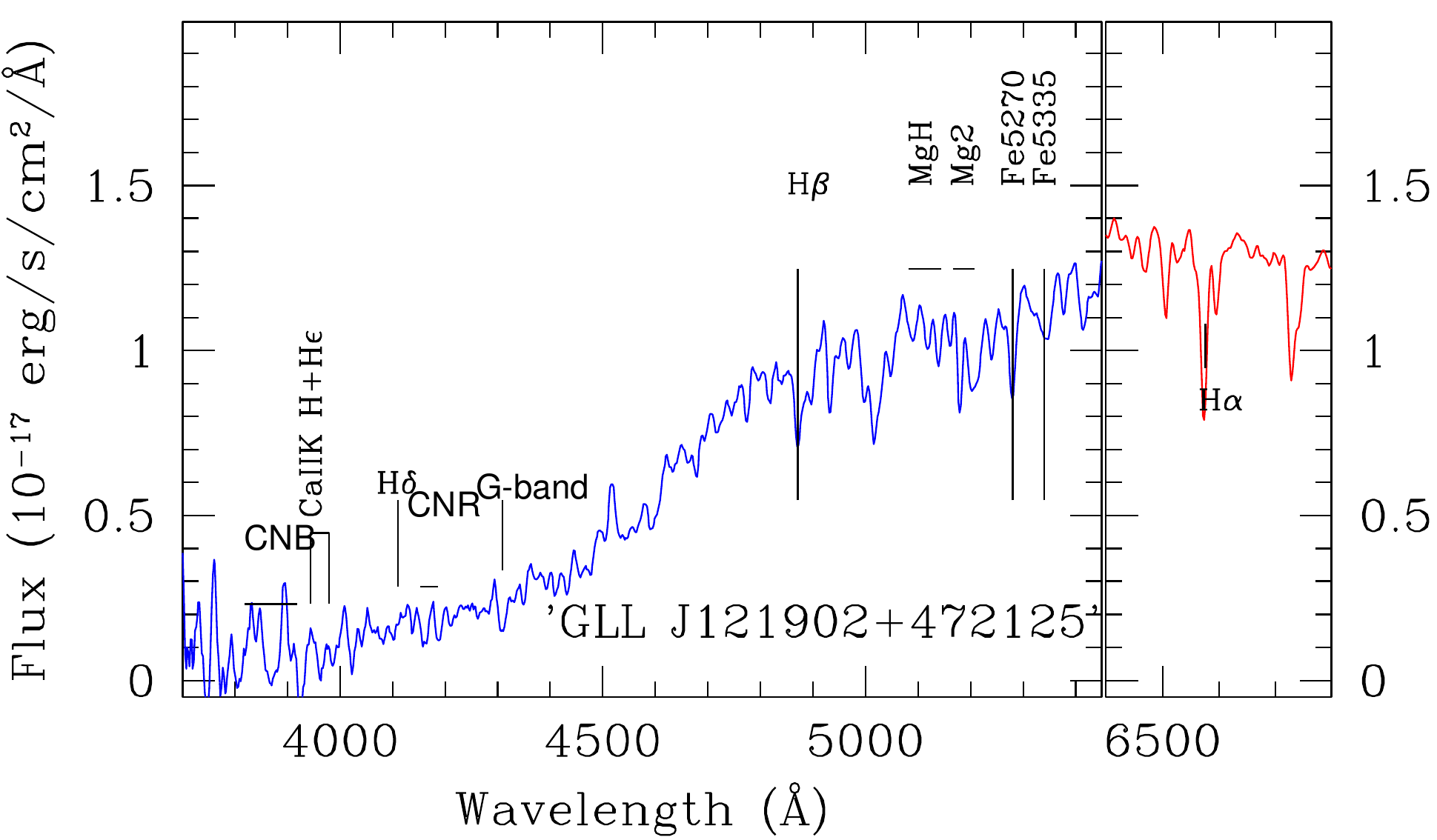}}
&
\hspace*{0.3cm}\includegraphics[scale=0.413]{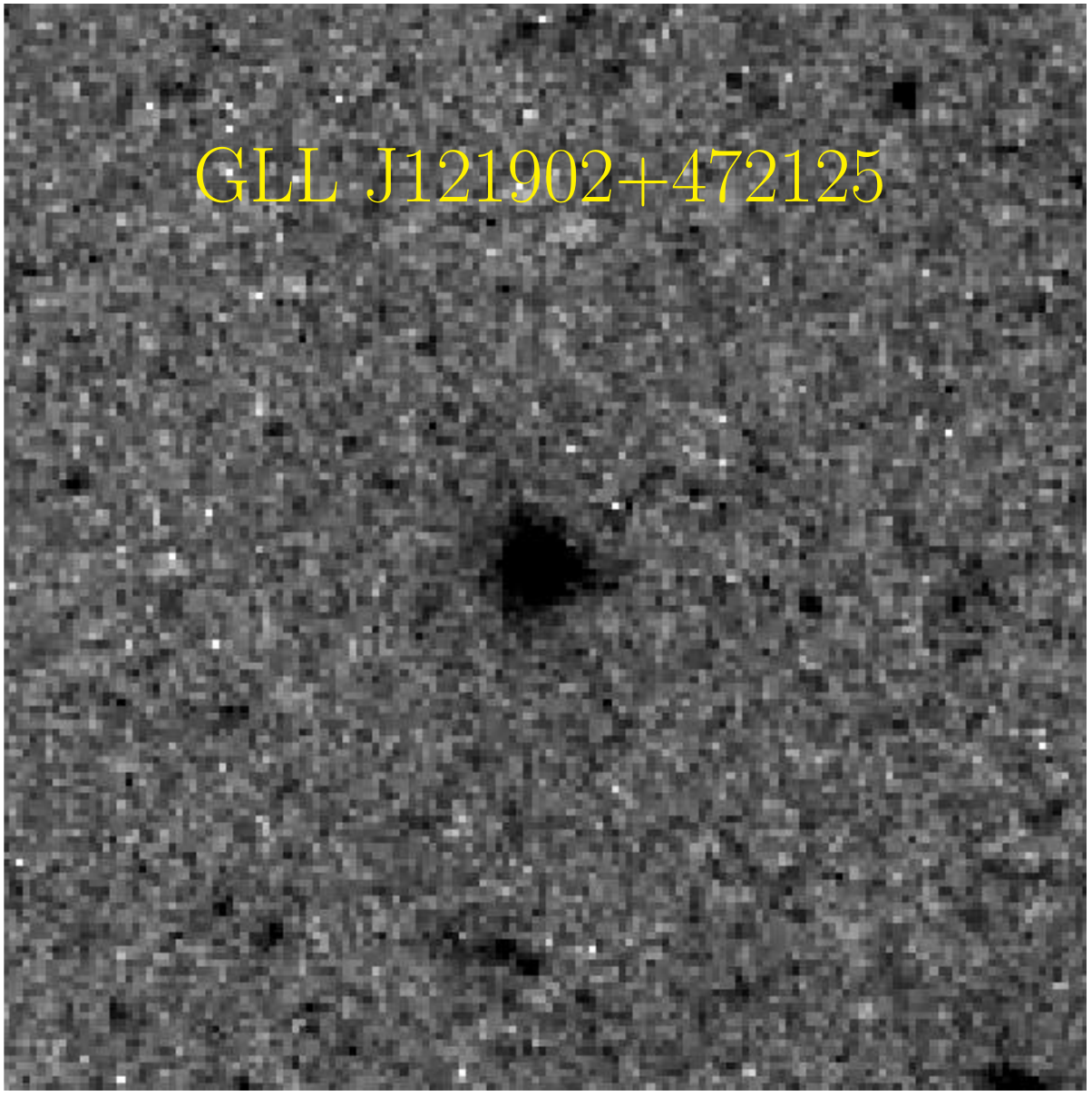}
\end{tabular}
\begin{tabular}{ll}
\hspace*{0.3cm}\raisebox{-0.57cm}{\includegraphics[scale=0.5,angle=0.]{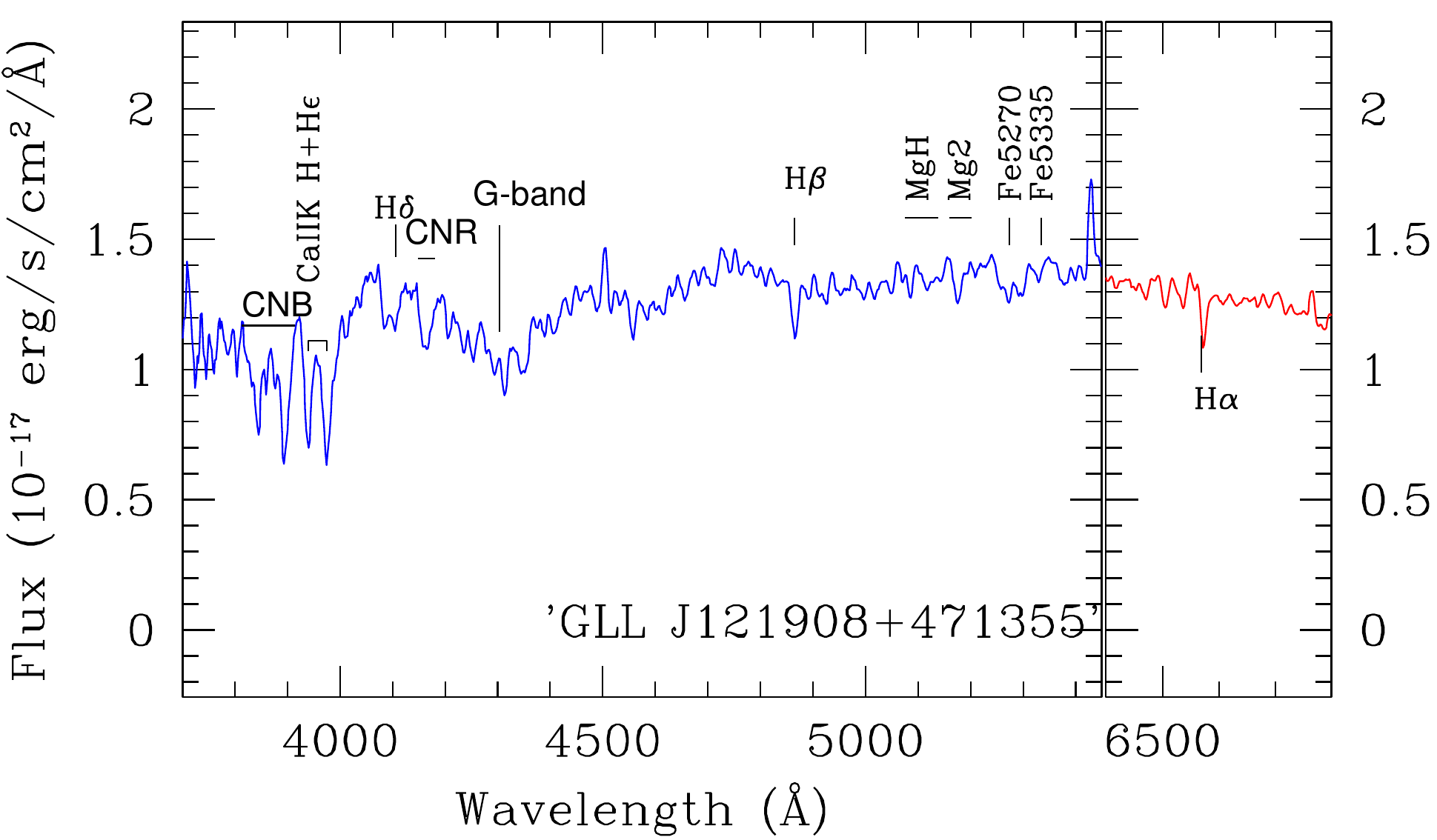}}
&
\hspace*{0.3cm}\includegraphics[scale=0.413]{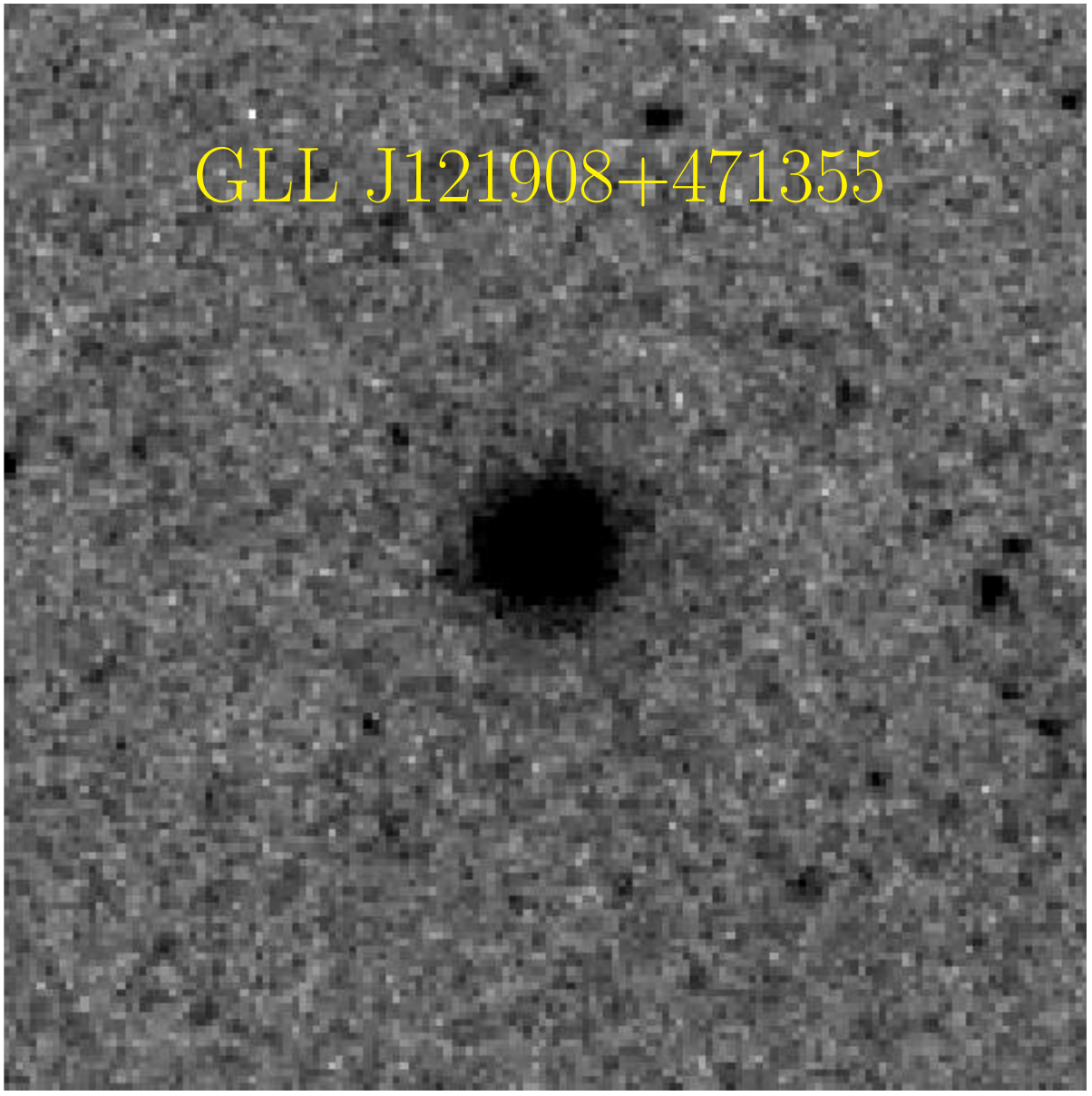}
\end{tabular}
\begin{tabular}{ll}
\hspace*{0.3cm}\raisebox{-0.57cm}{\includegraphics[scale=0.5,angle=0.]{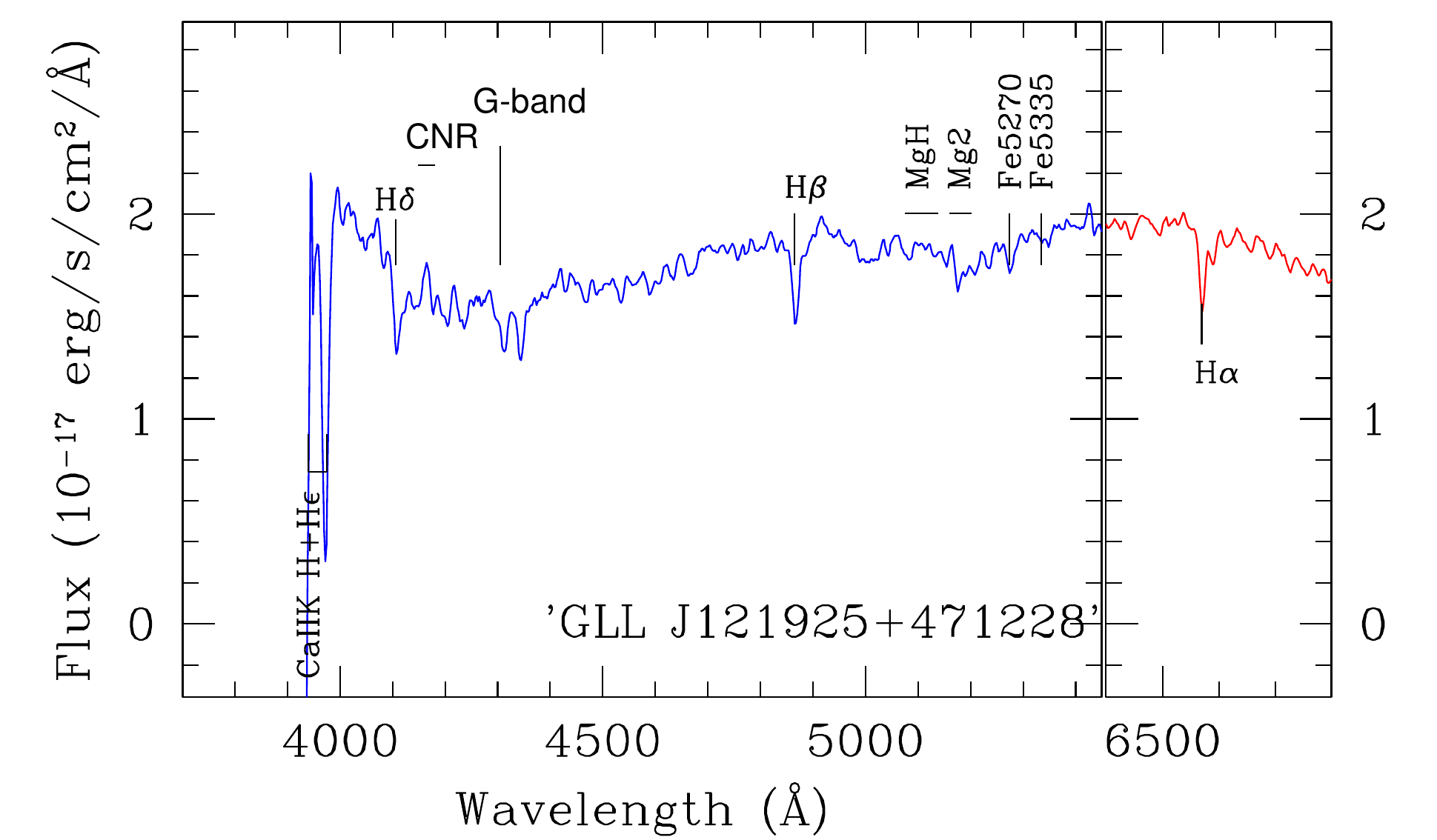}}
&
\hspace*{0.3cm}\includegraphics[scale=0.413]{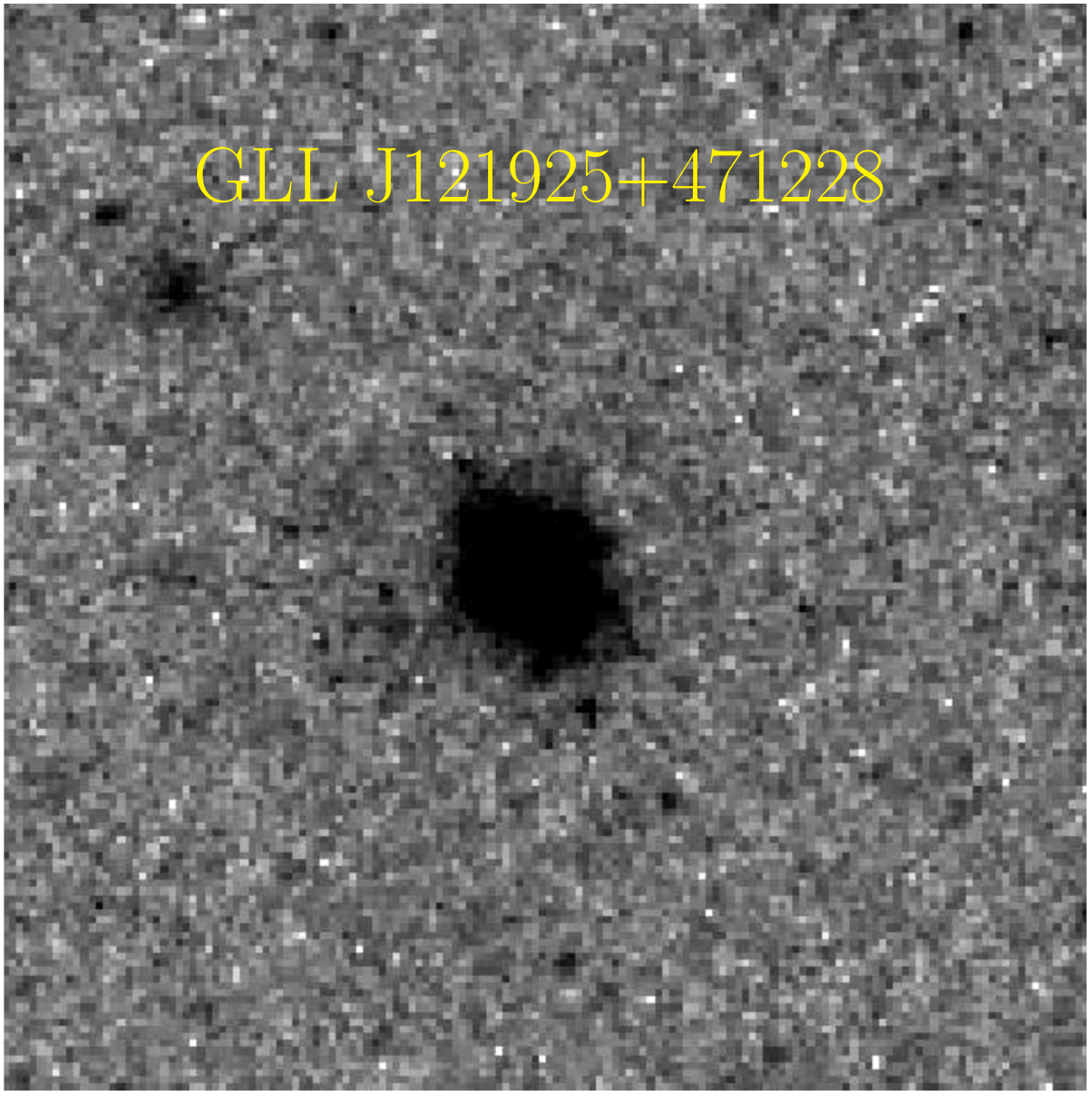}
\end{tabular}
\caption{{\it (Continued--)} Spectra ({\it left}) and grayscales ({\it right}) of confirmed GC candidates.
\label{fig:specim_conf}}
\end{figure}

\clearpage

\setcounter{figure}{3}
\begin{figure}
%\ContinuedFloat
\begin{tabular}{ll}
\hspace*{0.3cm}\raisebox{-0.57cm}{\includegraphics[scale=0.5,angle=0.]{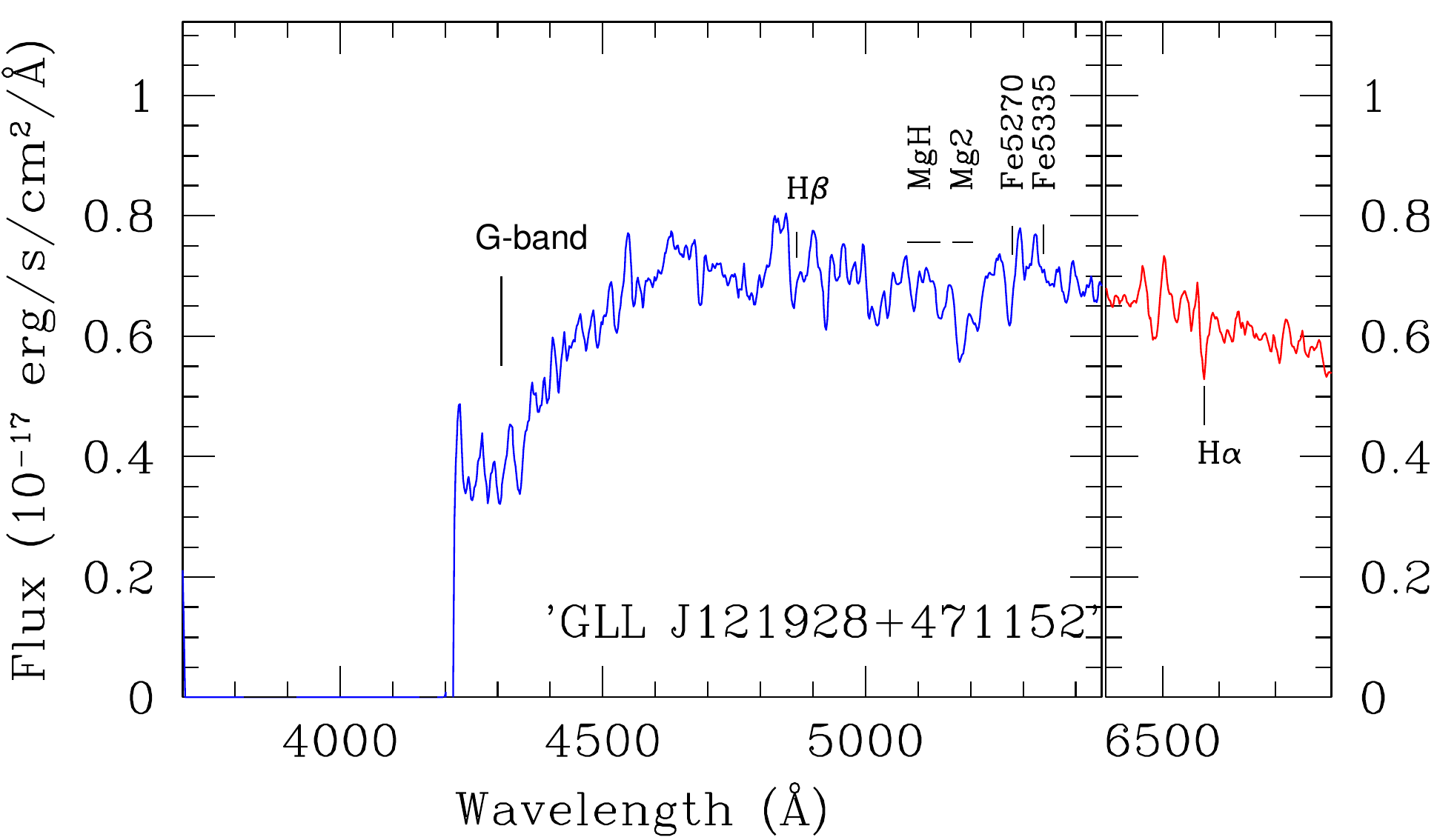}}
&
\hspace*{0.3cm}\includegraphics[scale=0.413]{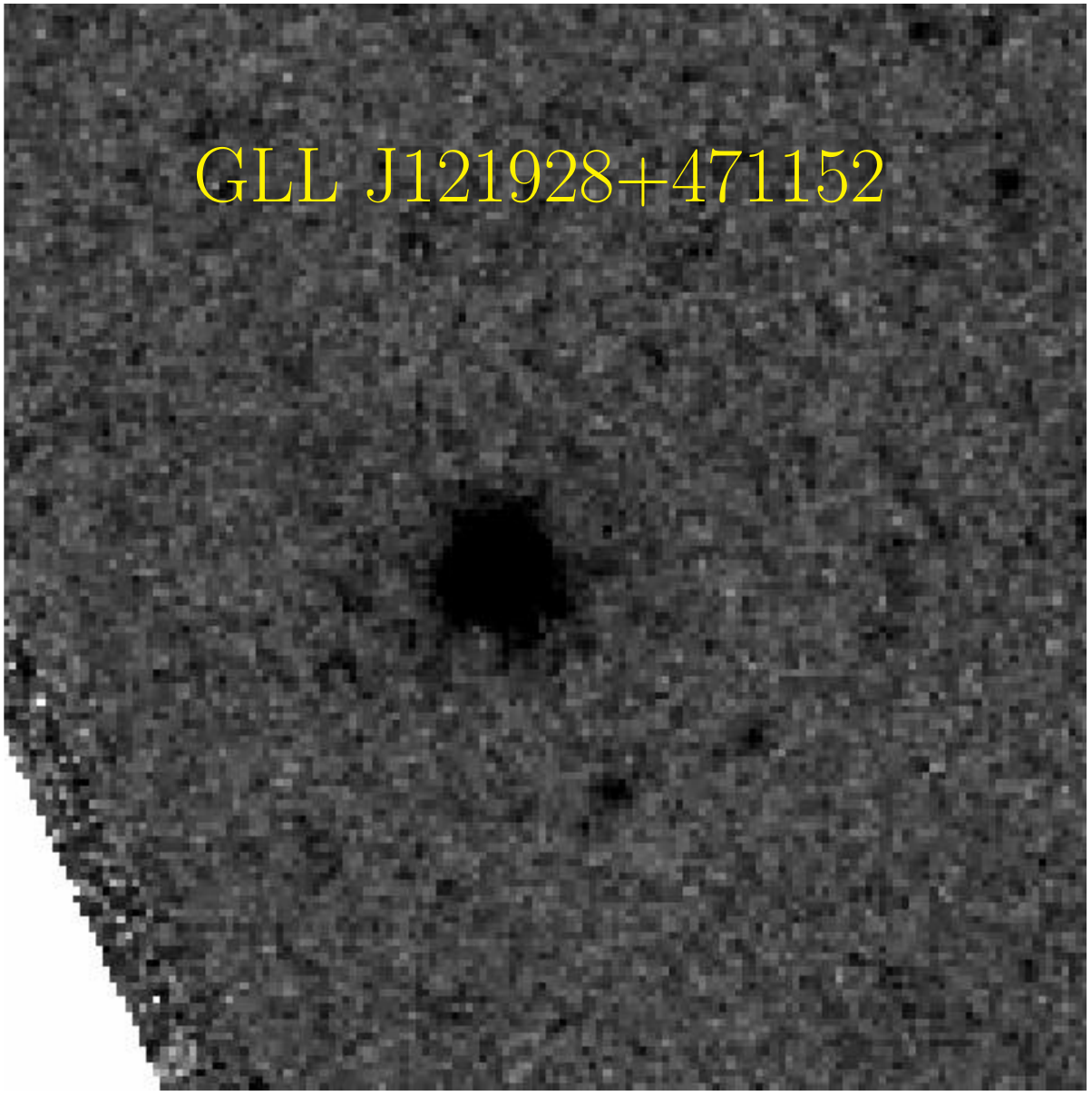}
\end{tabular}
\begin{tabular}{ll}
\hspace*{0.3cm}\raisebox{-0.57cm}{\includegraphics[scale=0.5,angle=0.]{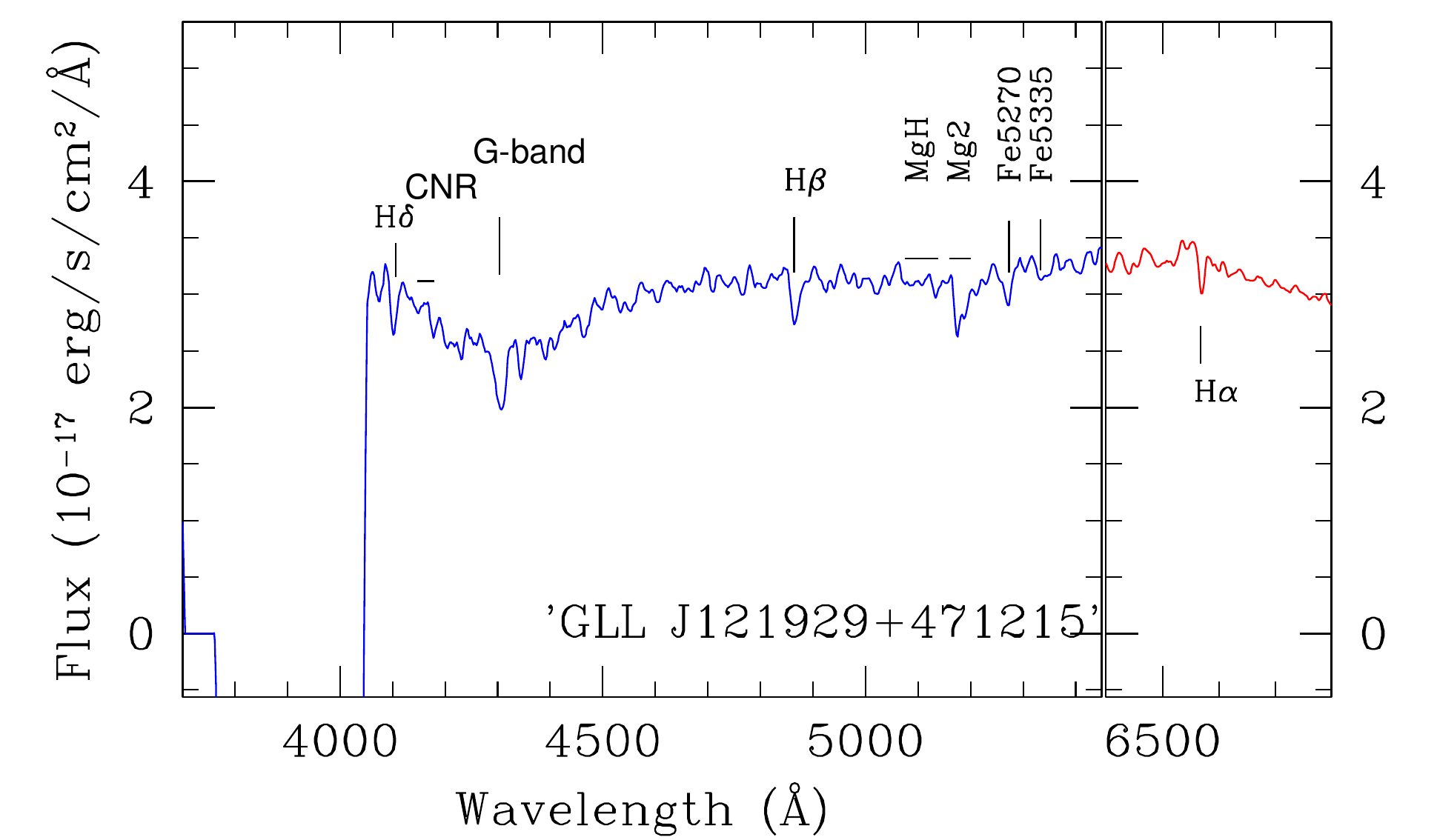}}
&
\hspace*{0.3cm}\includegraphics[scale=0.413]{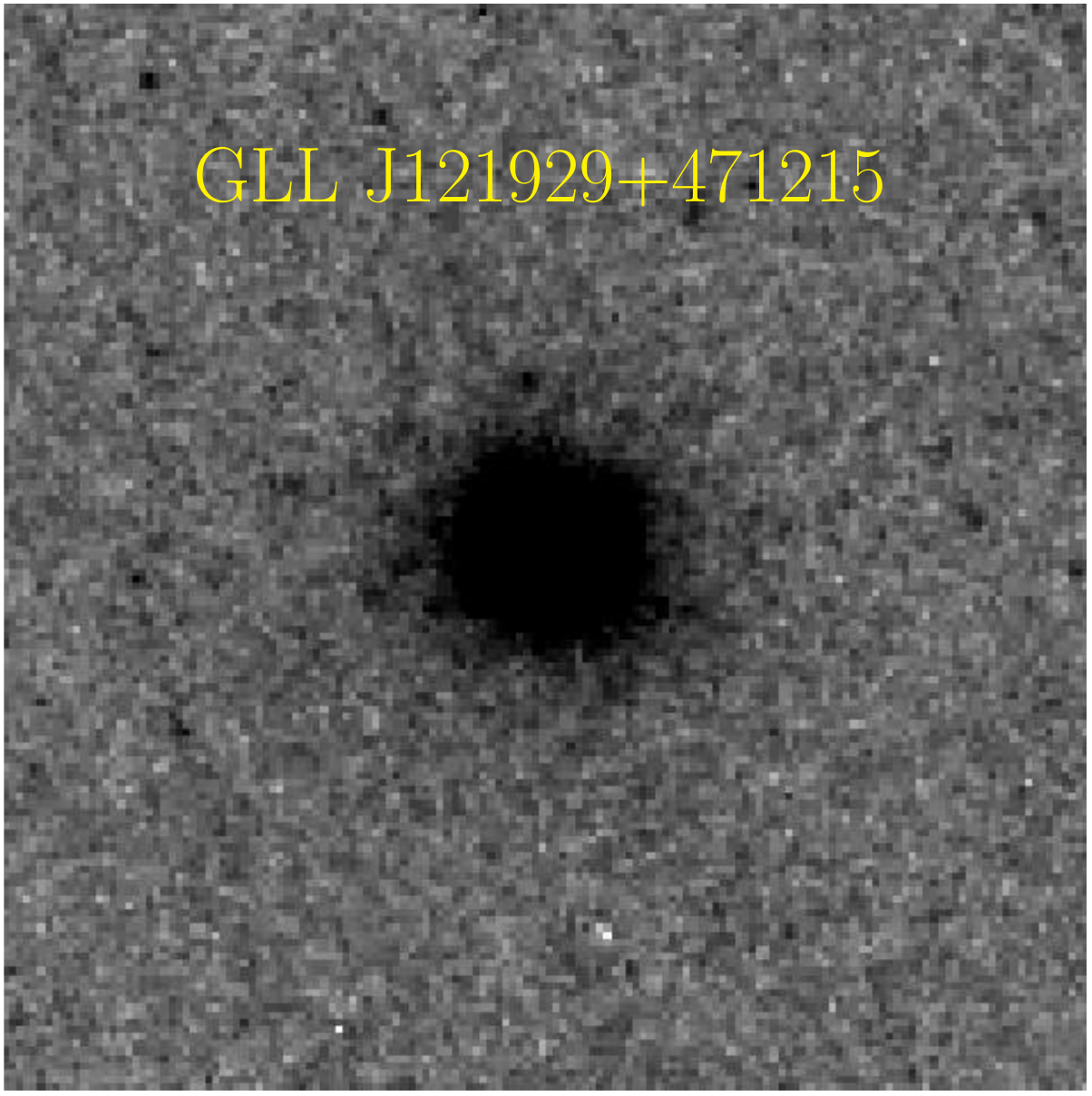}
\end{tabular}
\begin{tabular}{ll}
\hspace*{0.3cm}\raisebox{-0.57cm}{\includegraphics[scale=0.5,angle=0.]{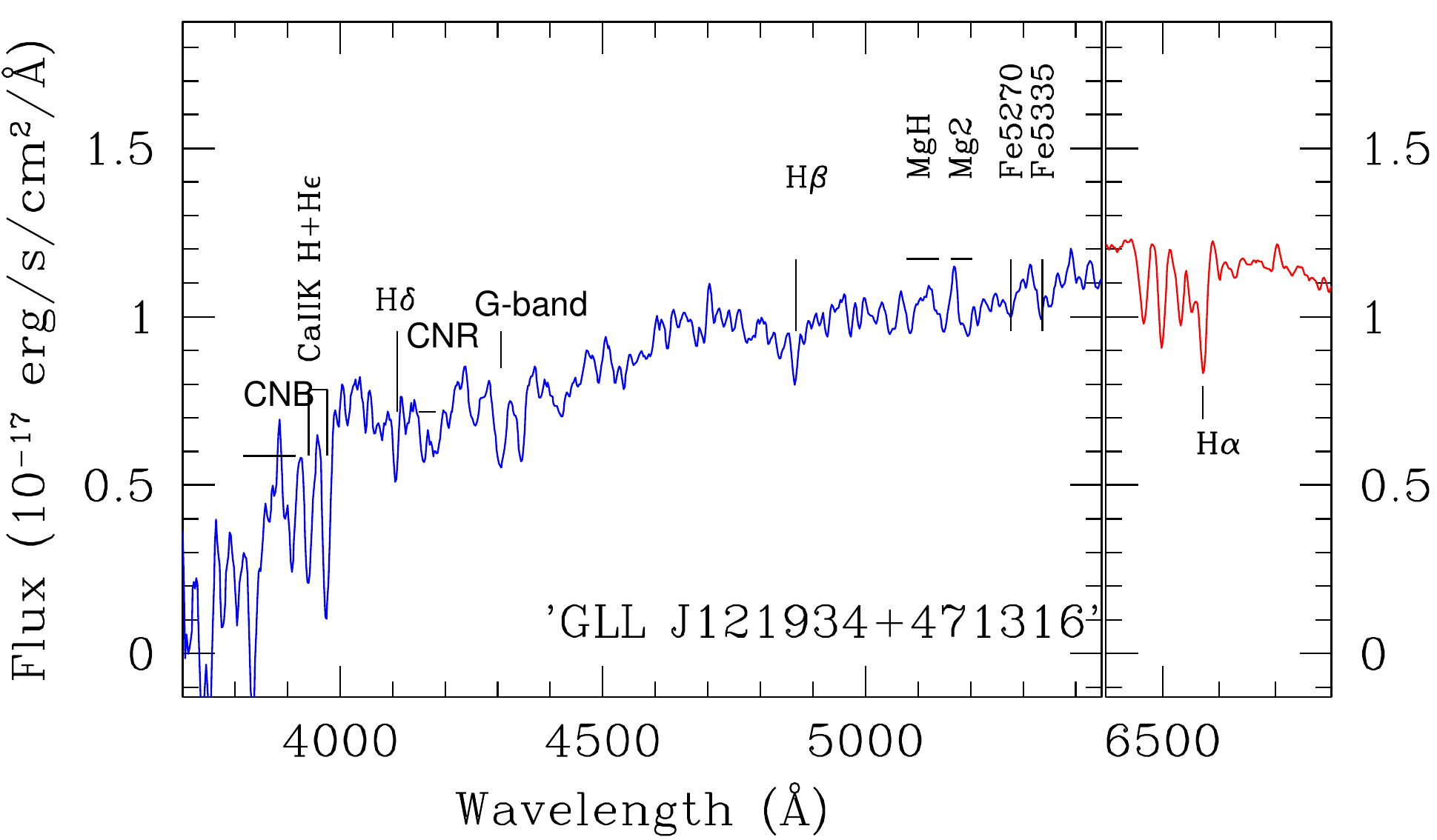}}
&
\hspace*{0.3cm}\includegraphics[scale=0.413]{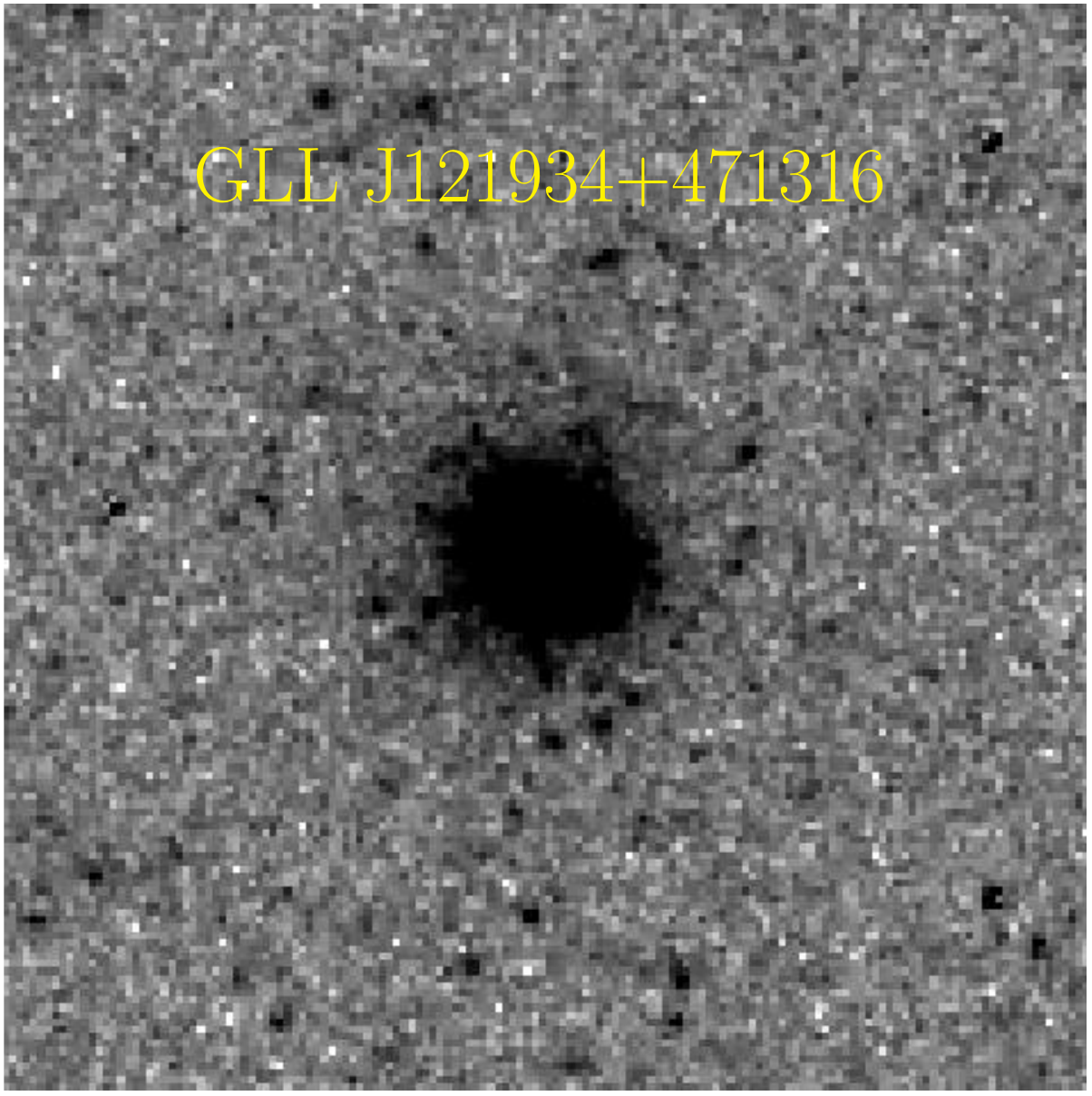}
\end{tabular}
\caption{{\it (Continued--)} Spectra ({\it left}) and grayscales ({\it right}) of confirmed GC candidates.
\label{fig:specim_conf}}
\end{figure}

\clearpage

\setcounter{figure}{4}
\begin{figure}
\begin{tabular}{ll}
\hspace*{0.3cm}\raisebox{-0.57cm}{\includegraphics[scale=0.5,angle=0.]{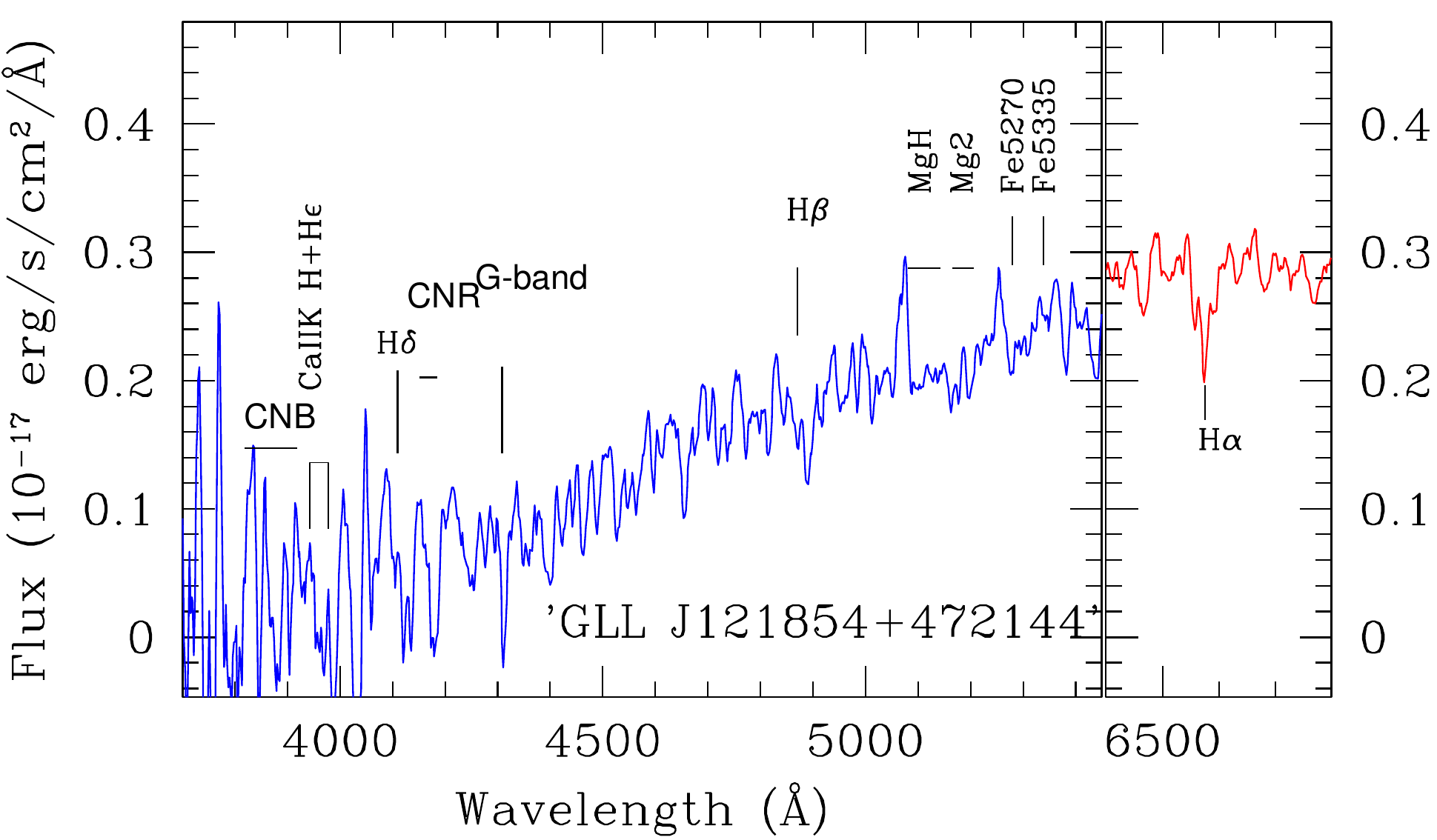}}
&
\hspace*{0.3cm}\includegraphics[scale=0.413]{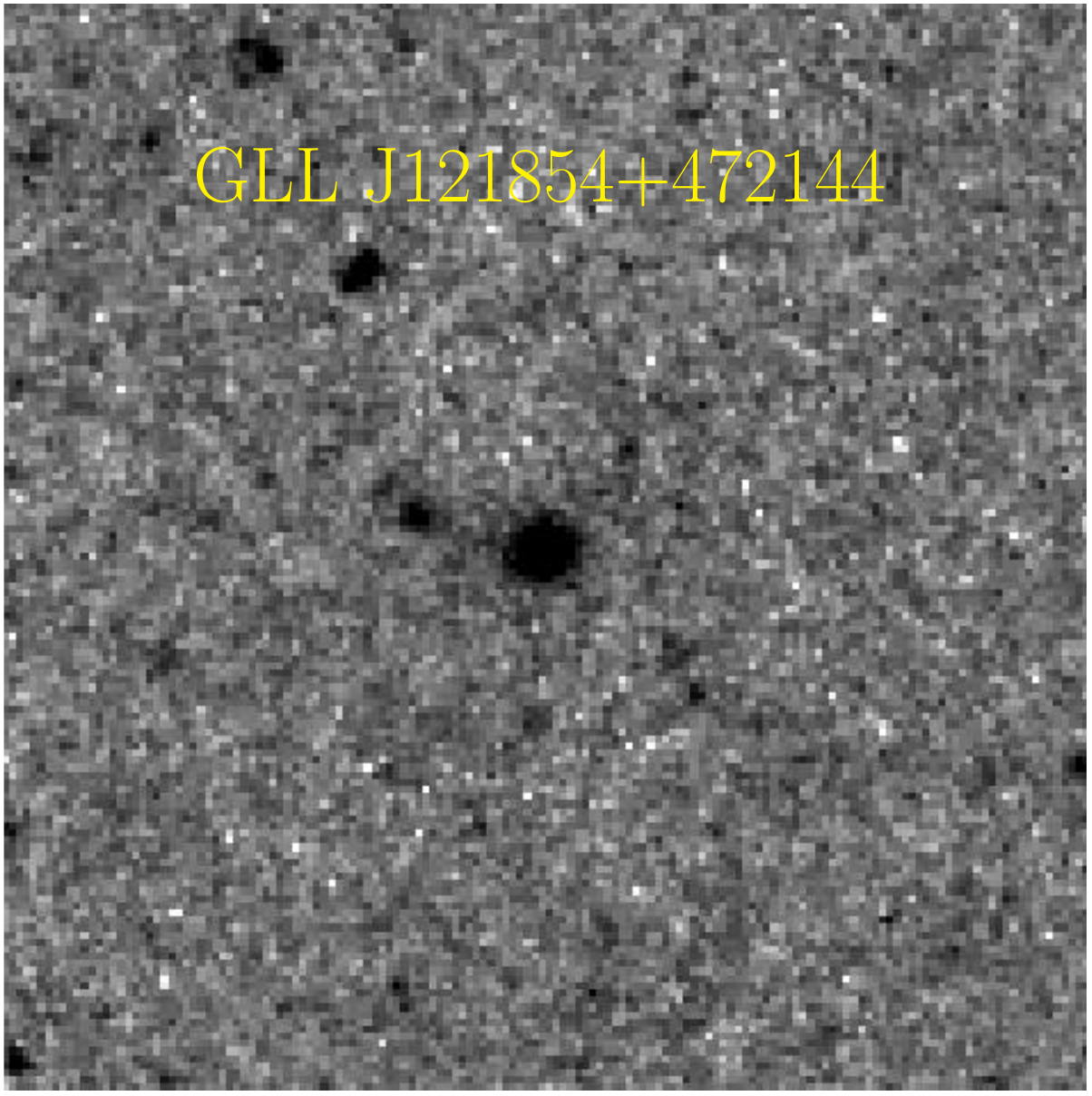}
\end{tabular}
\begin{tabular}{ll}
\hspace*{0.3cm}\raisebox{-0.57cm}{\includegraphics[scale=0.5,angle=0.]{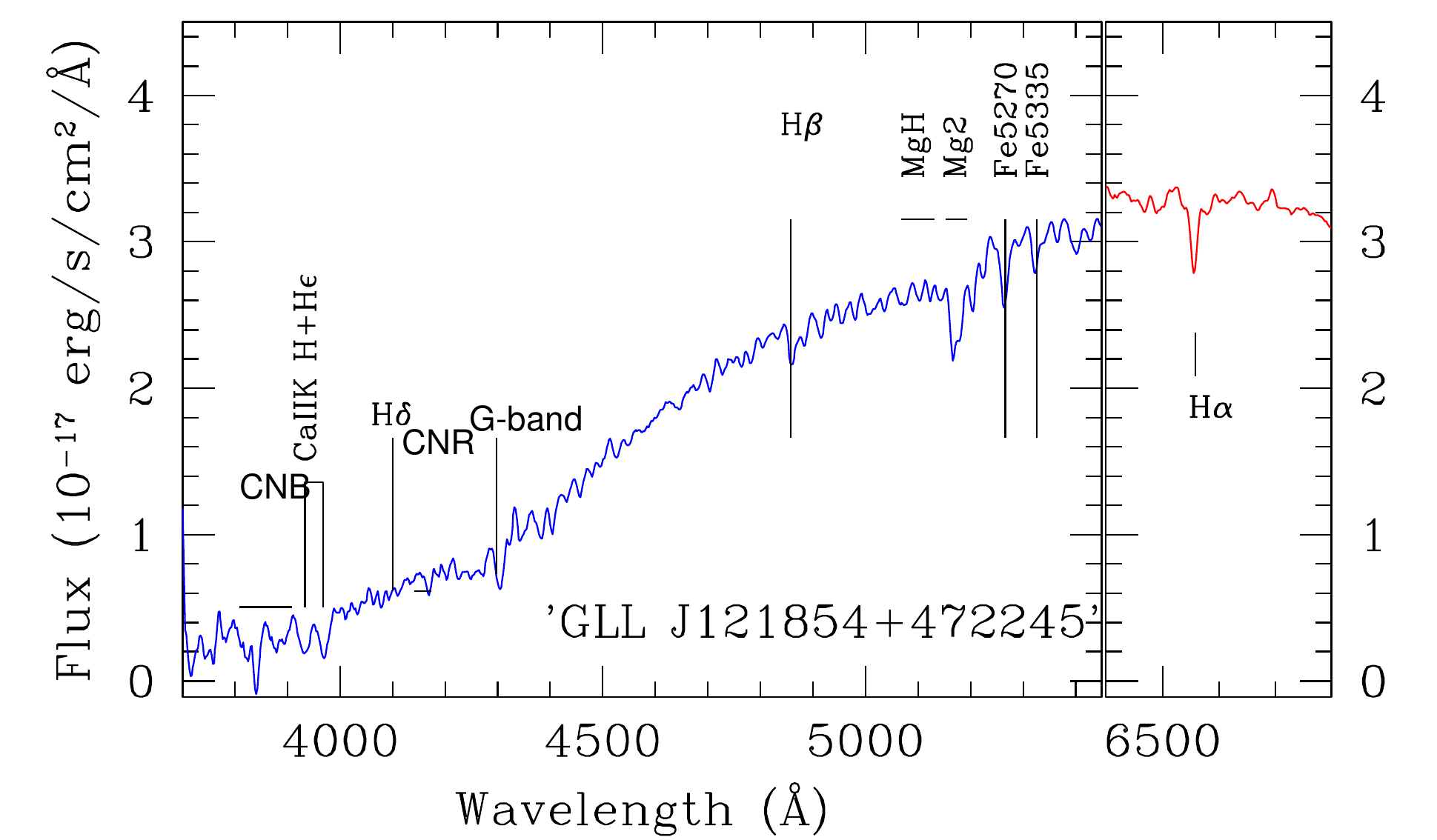}}
&
\hspace*{0.3cm}\includegraphics[scale=0.413]{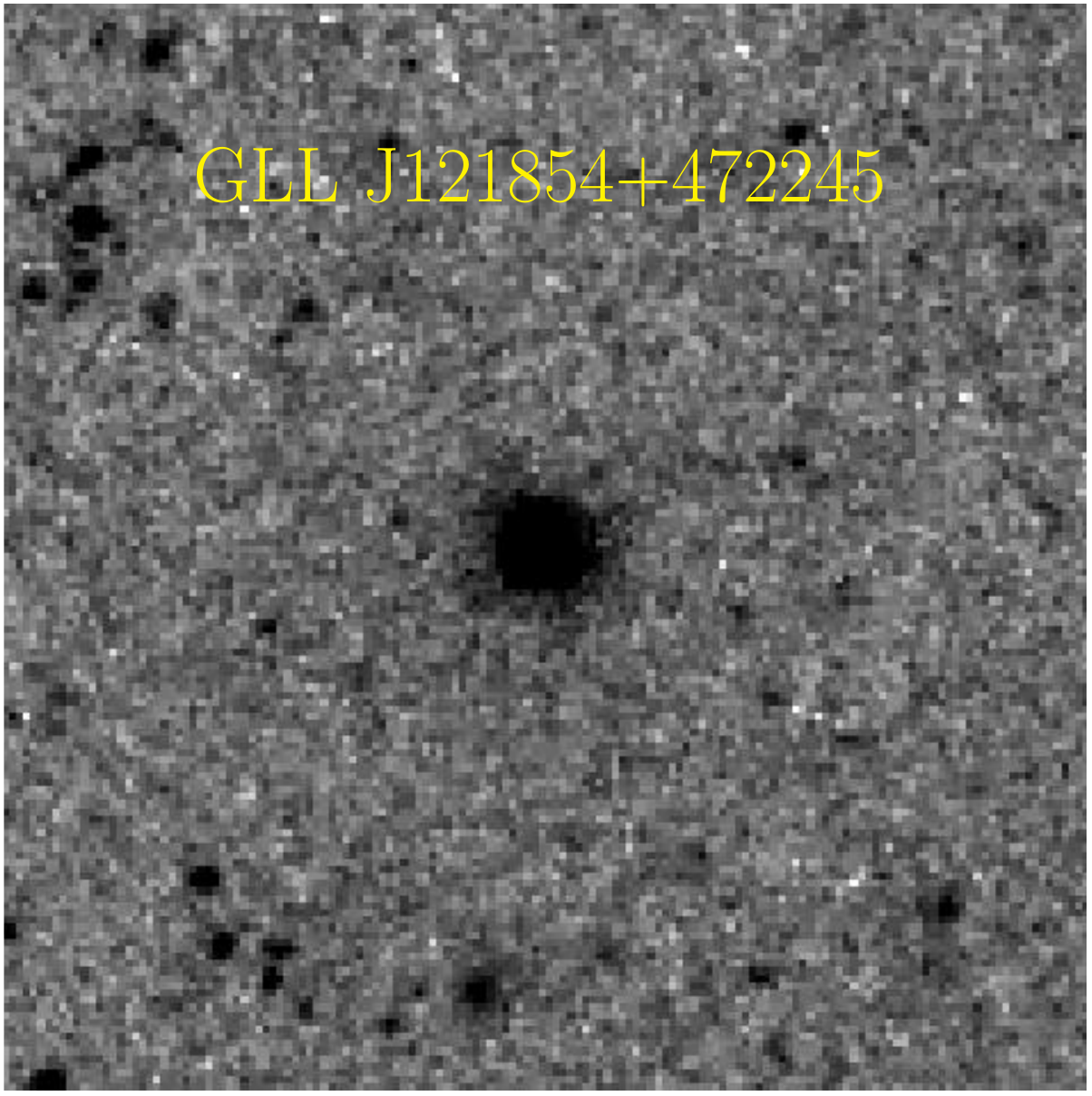}
\end{tabular}
\hspace*{0.5cm}\raisebox{-0.57cm}{\includegraphics[scale=0.5,angle=0]{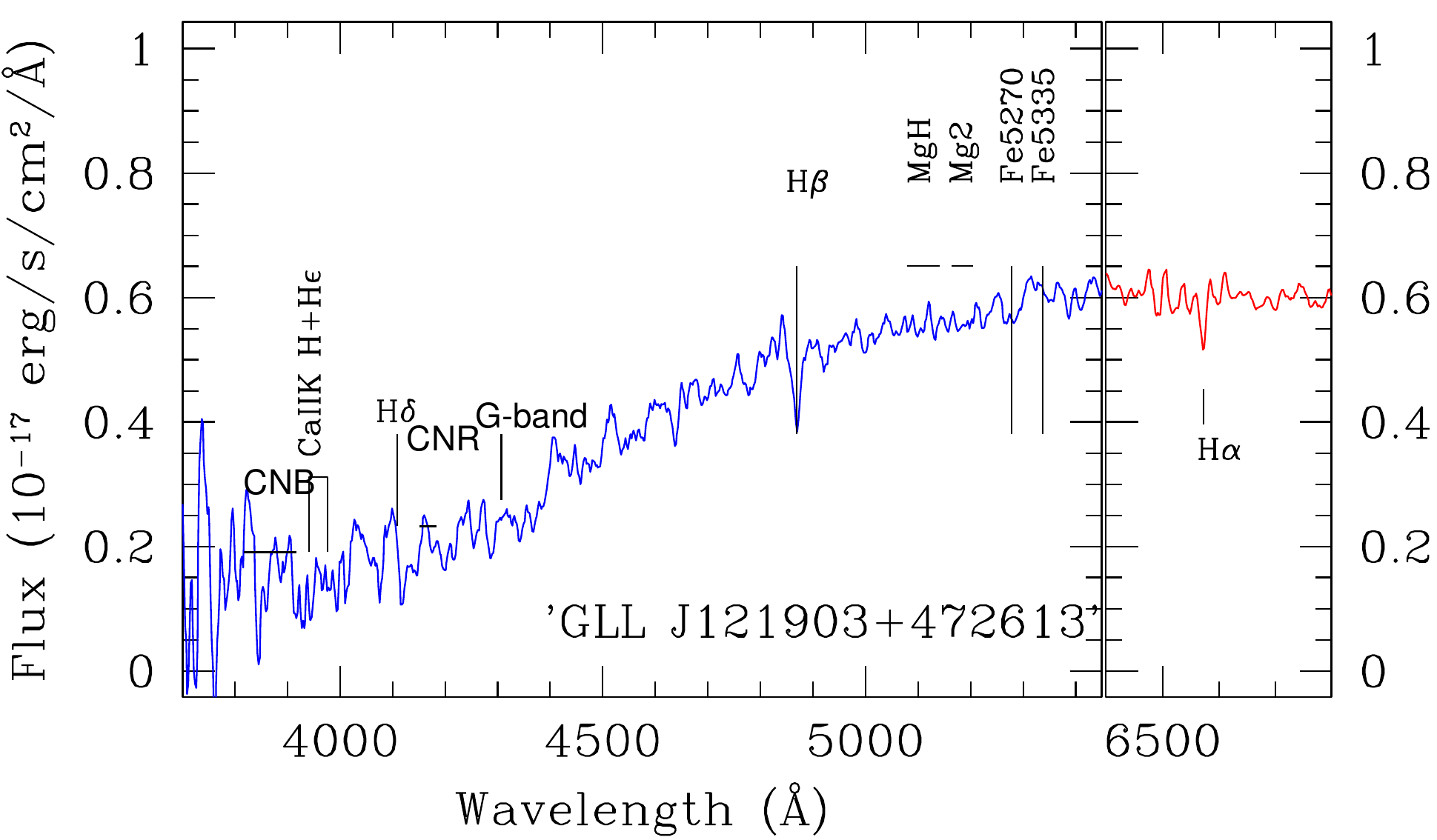}}
\caption{{\it (Continued--)} Spectra ({\it left}) and grayscales ({\it right}) of rejected GC candidates.
{The spectrum of GLL J121854+472144 correlates best with {\it fm32temp}, a non-GC template.
The spectrum of GLL J121854+472245 correlates best with {\it fm32temp}, a non-GC template, and displays a negative radial velocity; the source is not resolved in the ACS images. The spectrum of GLL J121903+472613 does correlate with {\it fglotemp}, a GC template, but only if restricted to wavelengths bluer than 5500$\AA$; there is no {\sl HST} image of the source.} 
\label{fig:specim_rej}}
\end{figure}

\clearpage

\setcounter{figure}{4}
\begin{figure}
\begin{tabular}{ll}
\hspace*{0.3cm}\raisebox{-0.57cm}{\includegraphics[scale=0.5,angle=0.]{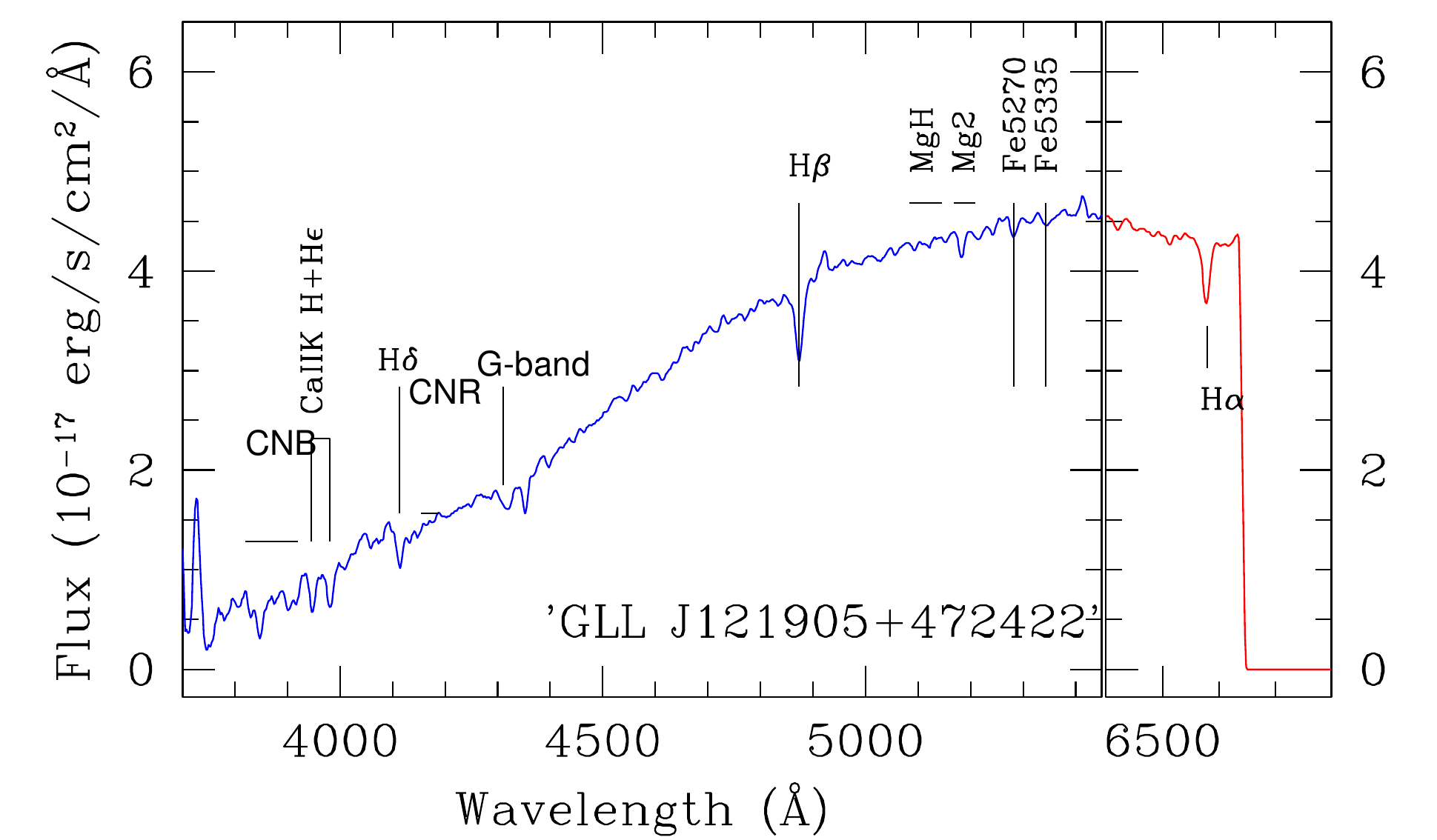}}
&
\hspace*{0.3cm}\includegraphics[scale=0.413]{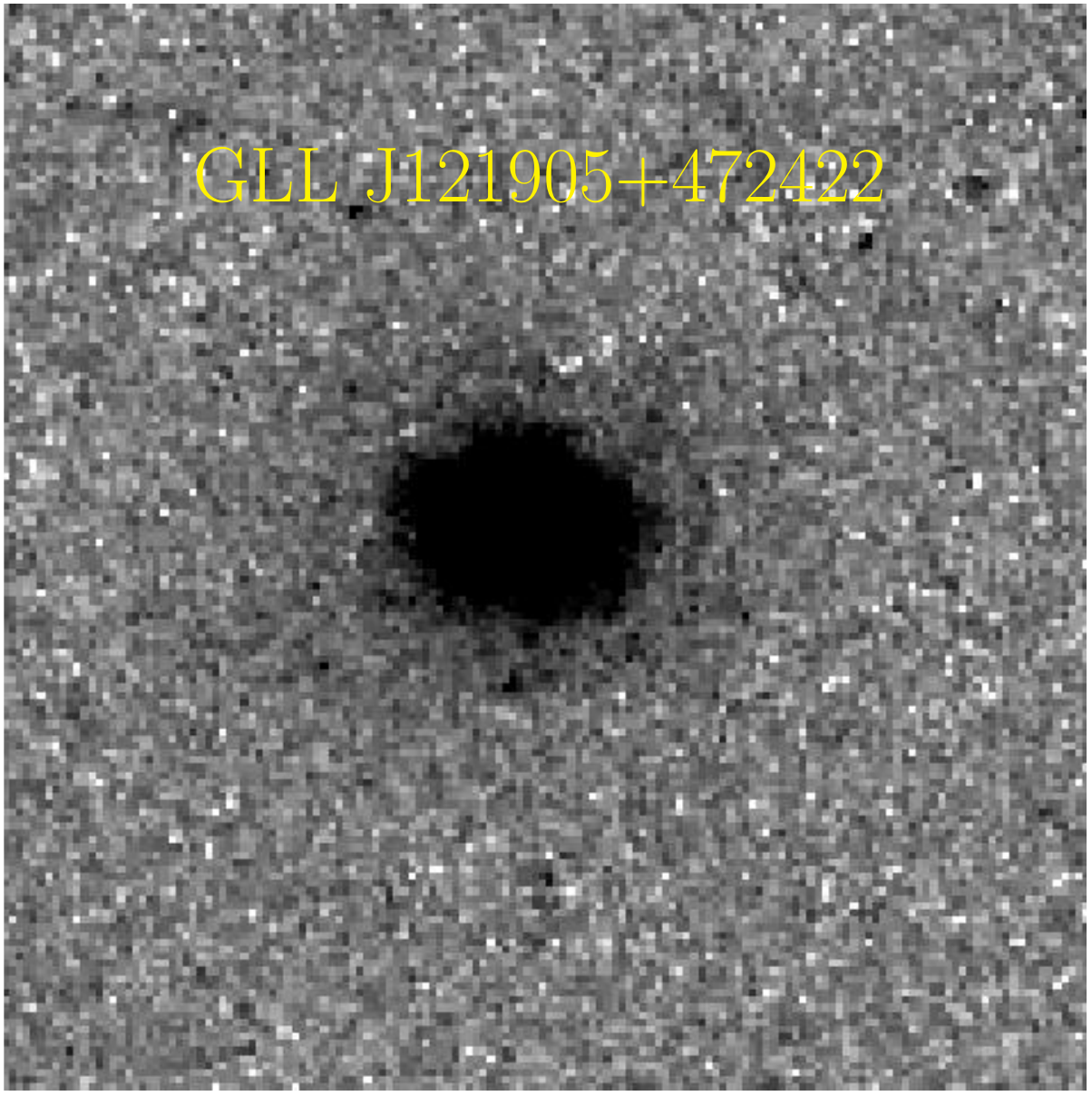}
\end{tabular}
\begin{tabular}{ll}
\hspace*{0.3cm}\raisebox{-0.57cm}{\includegraphics[scale=0.5,angle=0.]{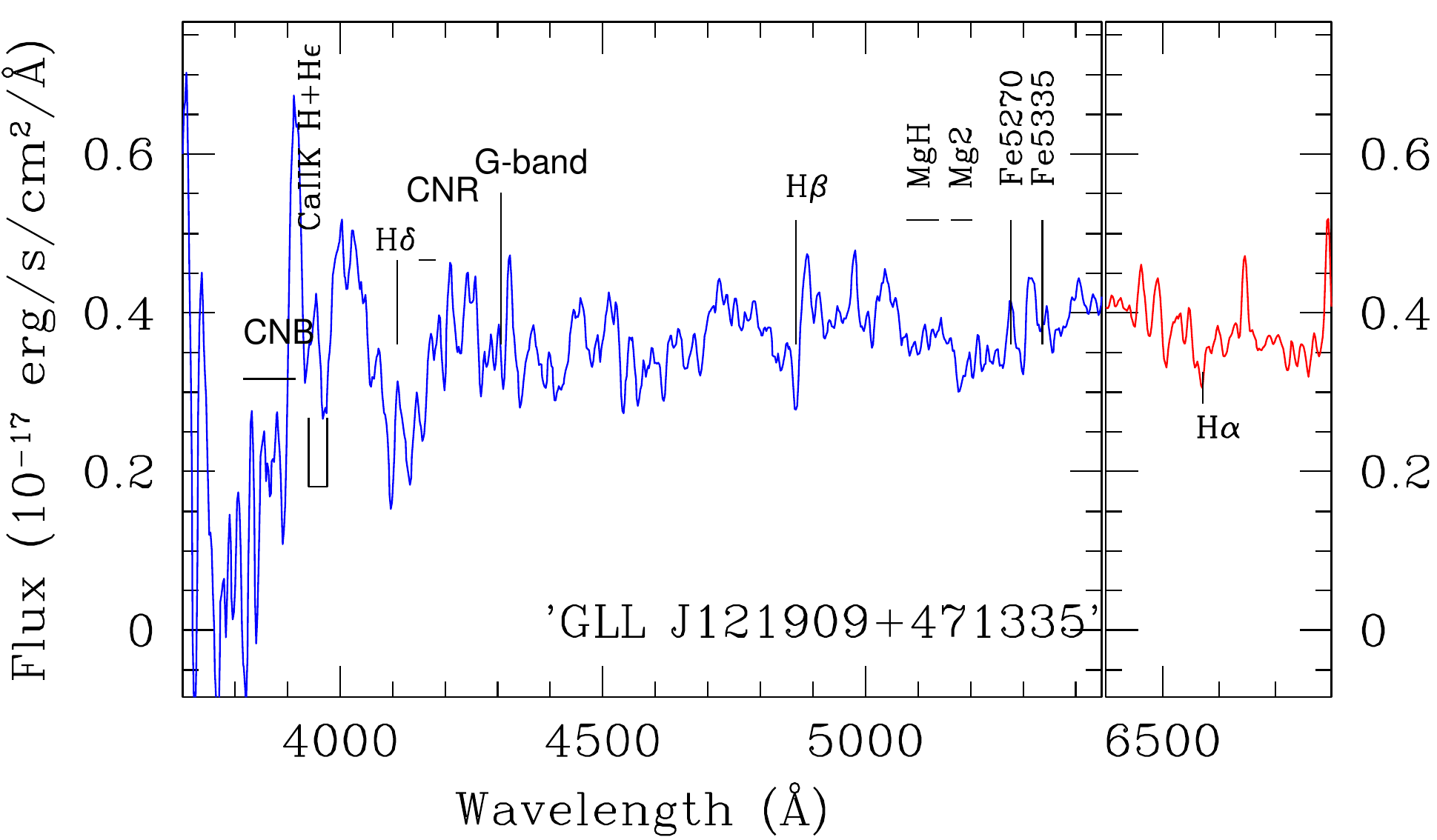}}
&
\hspace*{0.3cm}\includegraphics[scale=0.413]{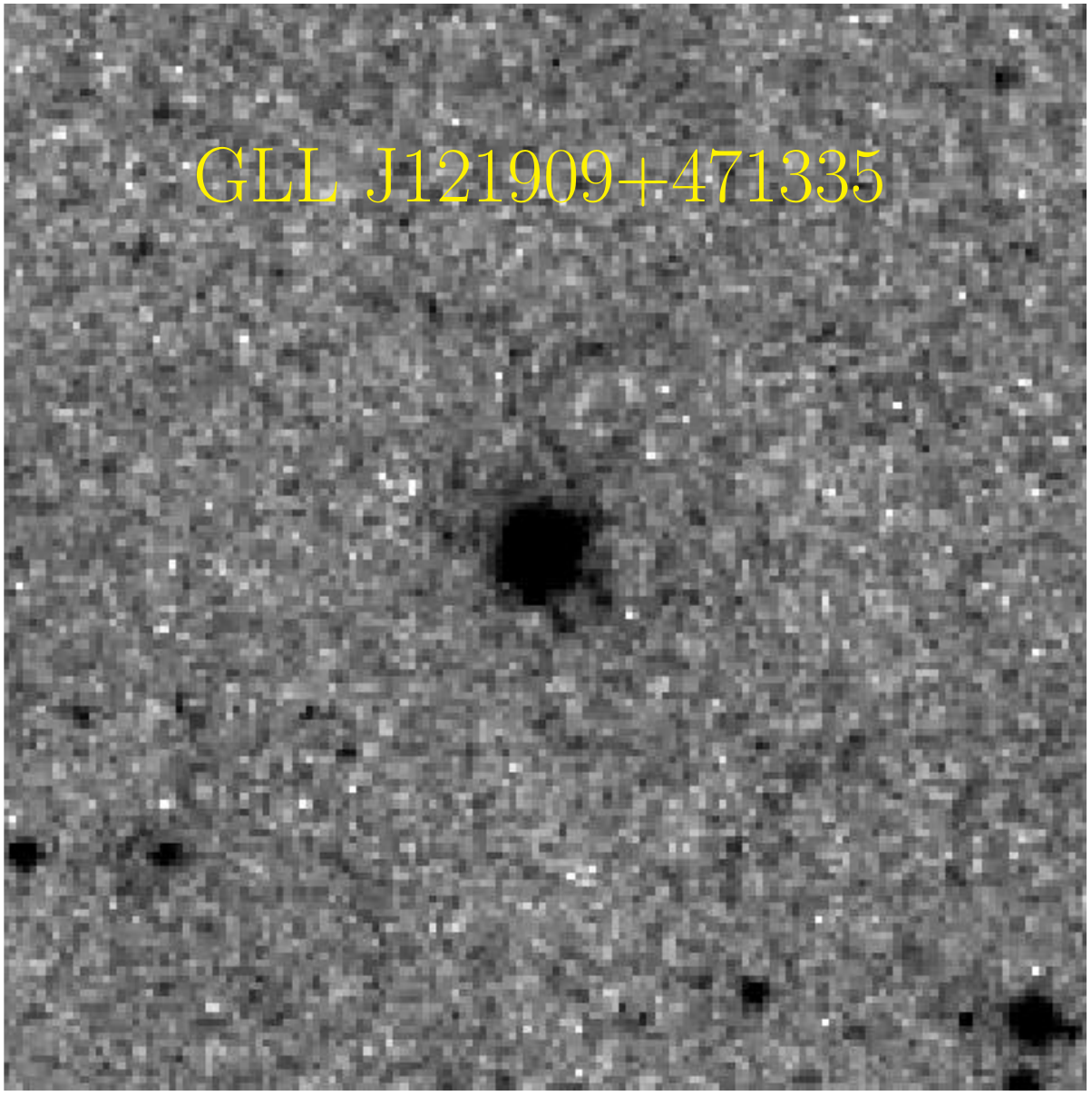}
\end{tabular}
\begin{tabular}{ll}
\hspace*{0.3cm}\raisebox{-0.57cm}{\includegraphics[scale=0.5,angle=0.]{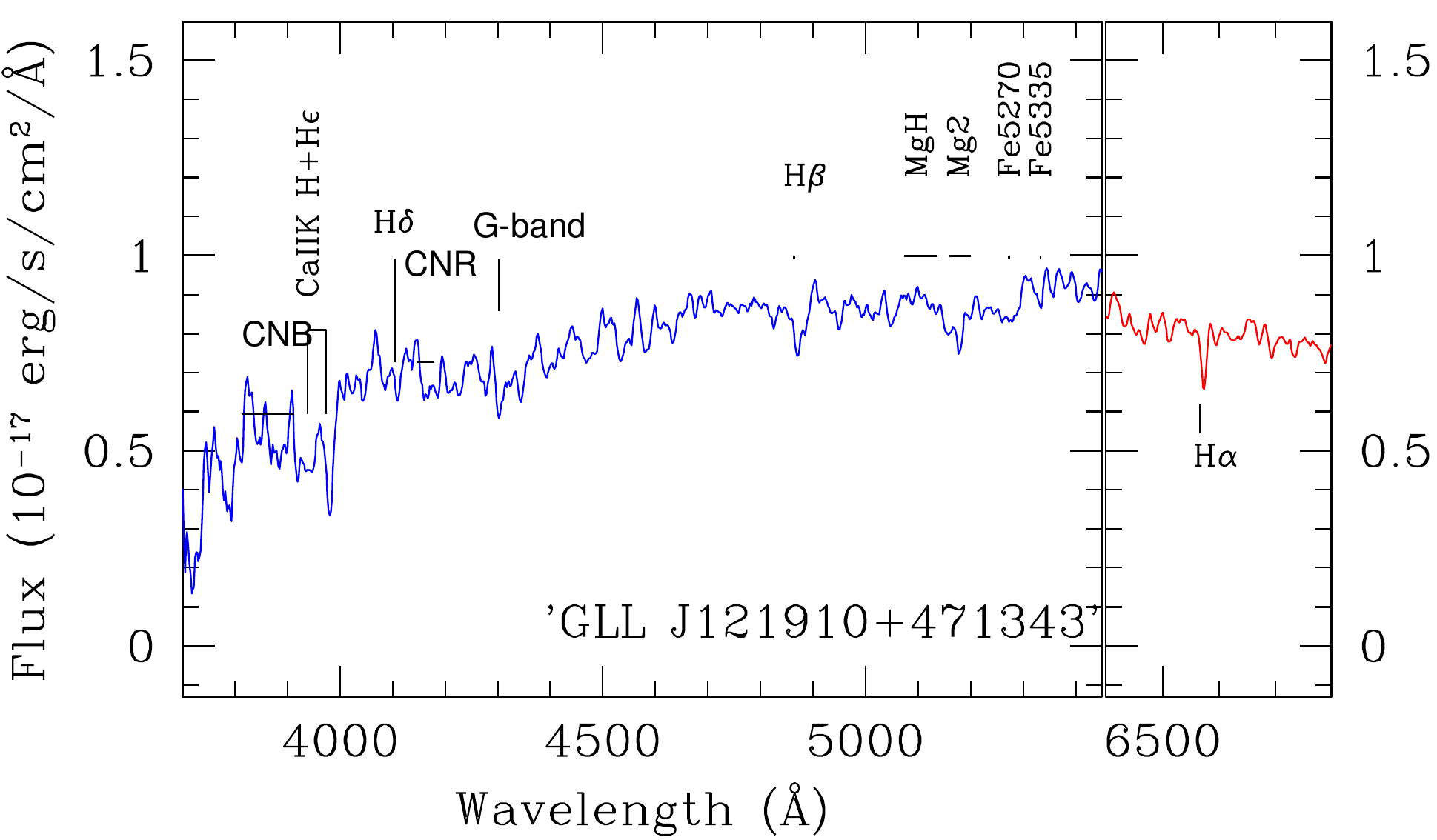}}
&
\hspace*{0.3cm}\includegraphics[scale=0.413]{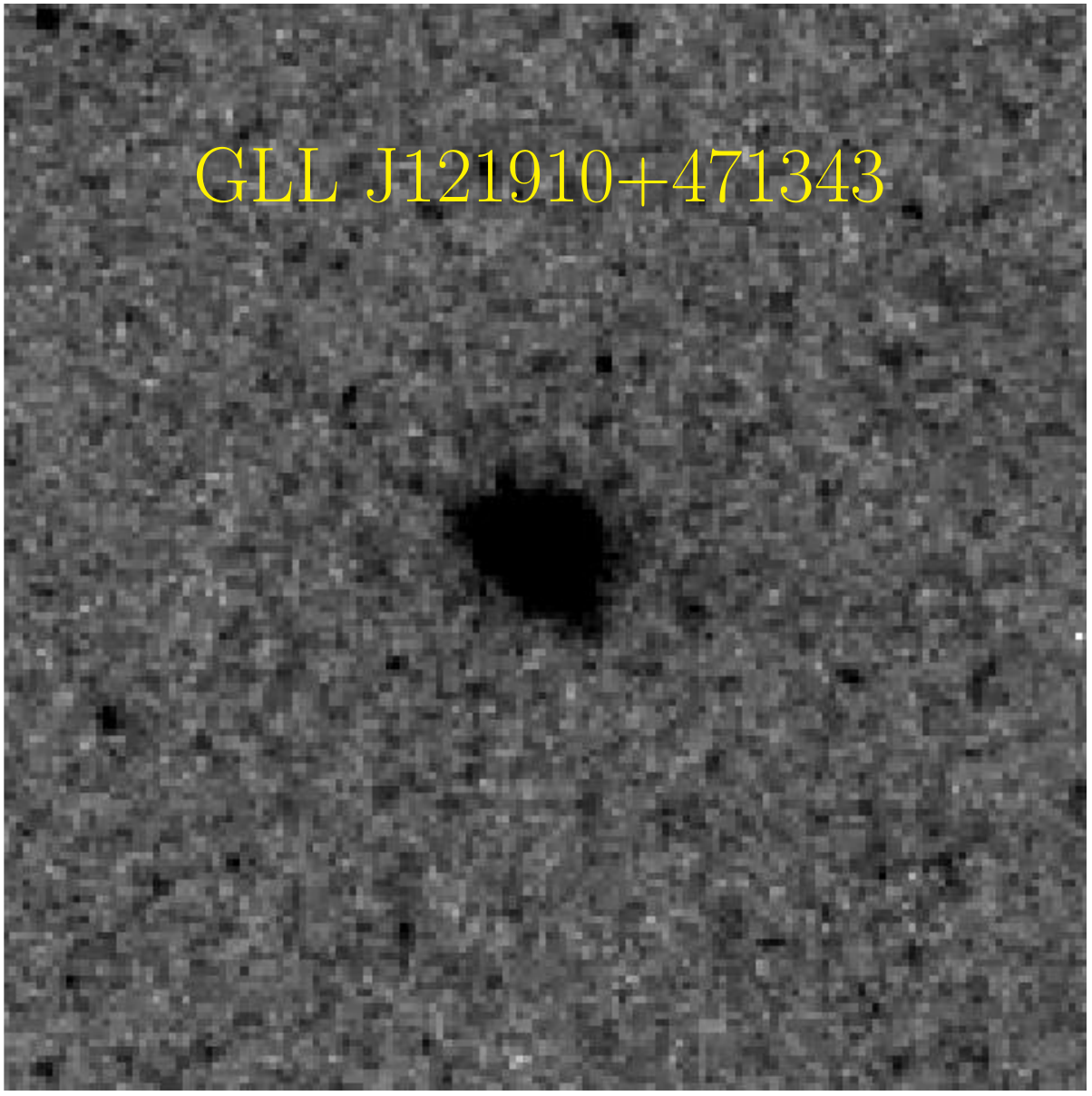}
\end{tabular}
\caption{{\it (Continued--)} Spectra ({\it left}) and grayscales ({\it right}) of rejected GC candidates.
{The spectrum of GLL J121905+472422 correlates best with {\it fm32temp}, a non-GC template. The spectrum of GLL J121909+471335 correlates best with {\it eltemp}, a non-GC template. GLL J121910+471343 looks quite elongated in {\sl HST} images.}
\label{fig:specim_rej}}
\end{figure}

\clearpage

\setcounter{figure}{5}
\begin{figure}
\begin{tabular}{ll}
\hspace*{0.3cm}\raisebox{-0.57cm}{\includegraphics[scale=0.5,angle=0.]{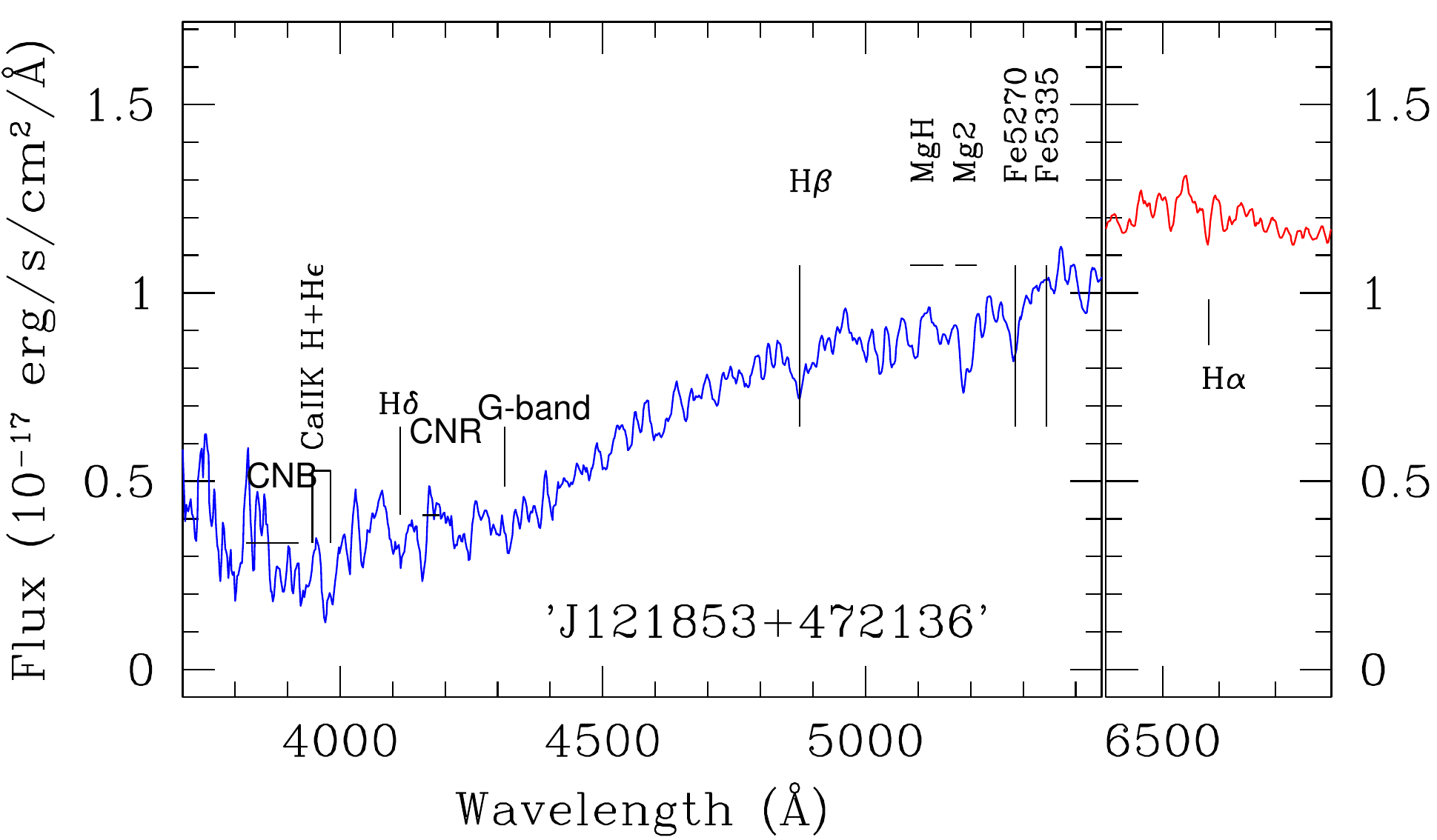}}
&
\hspace*{0.3cm}\includegraphics[scale=0.413]{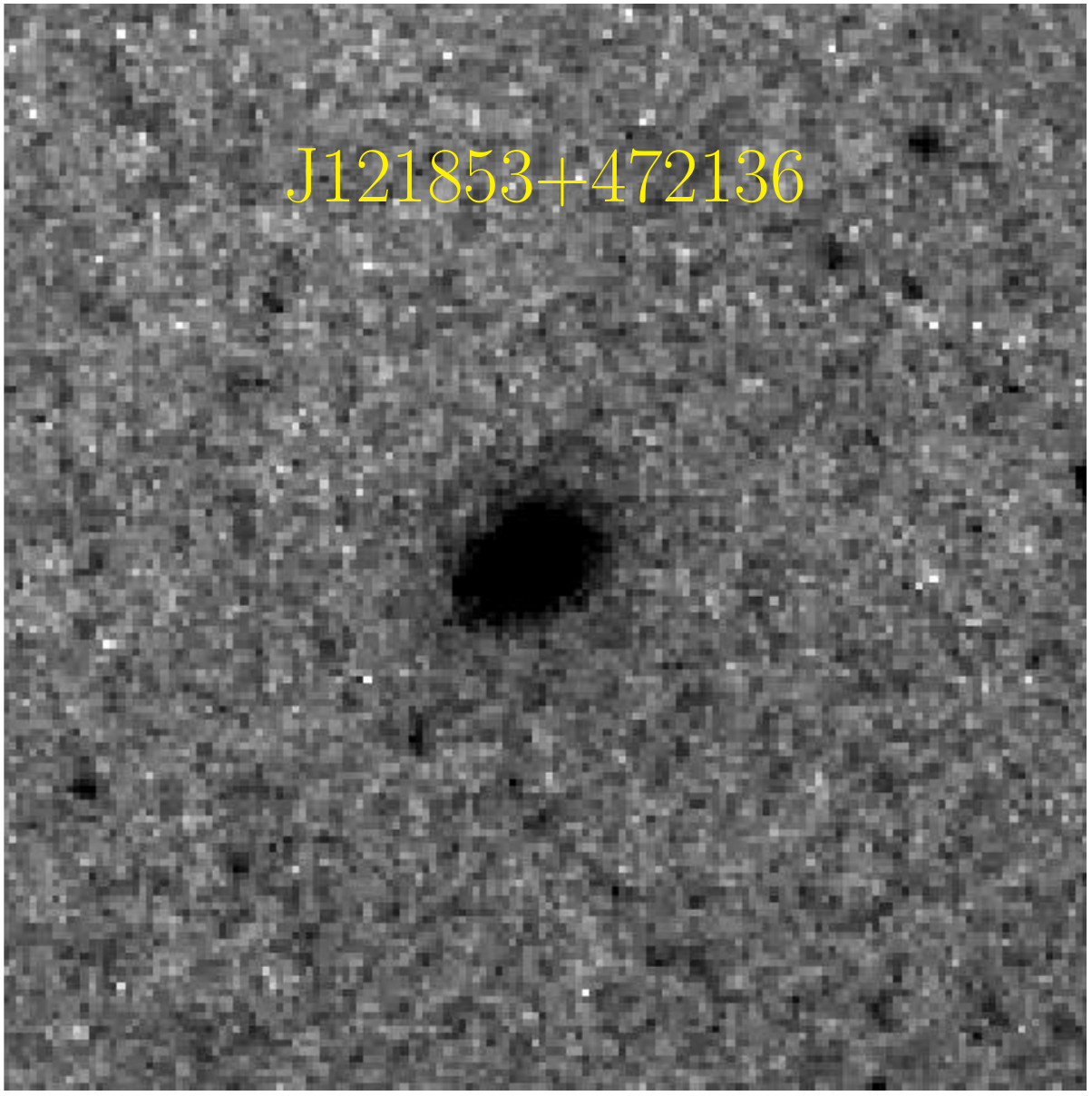}
\end{tabular}
\begin{tabular}{ll}
\hspace*{0.3cm}\raisebox{-0.57cm}{\includegraphics[scale=0.5,angle=0.]{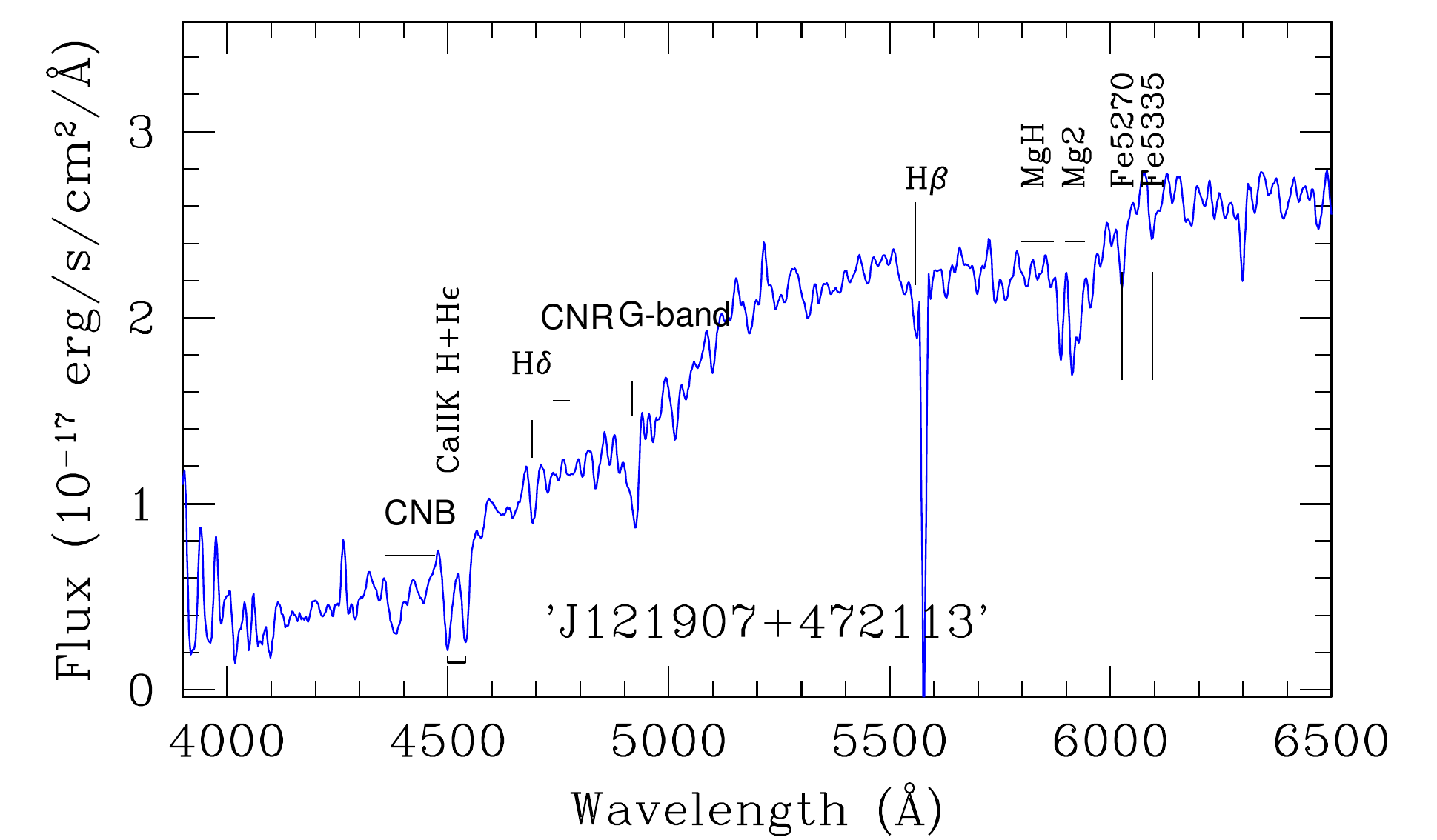}}
&
\hspace*{0.3cm}\includegraphics[scale=0.413]{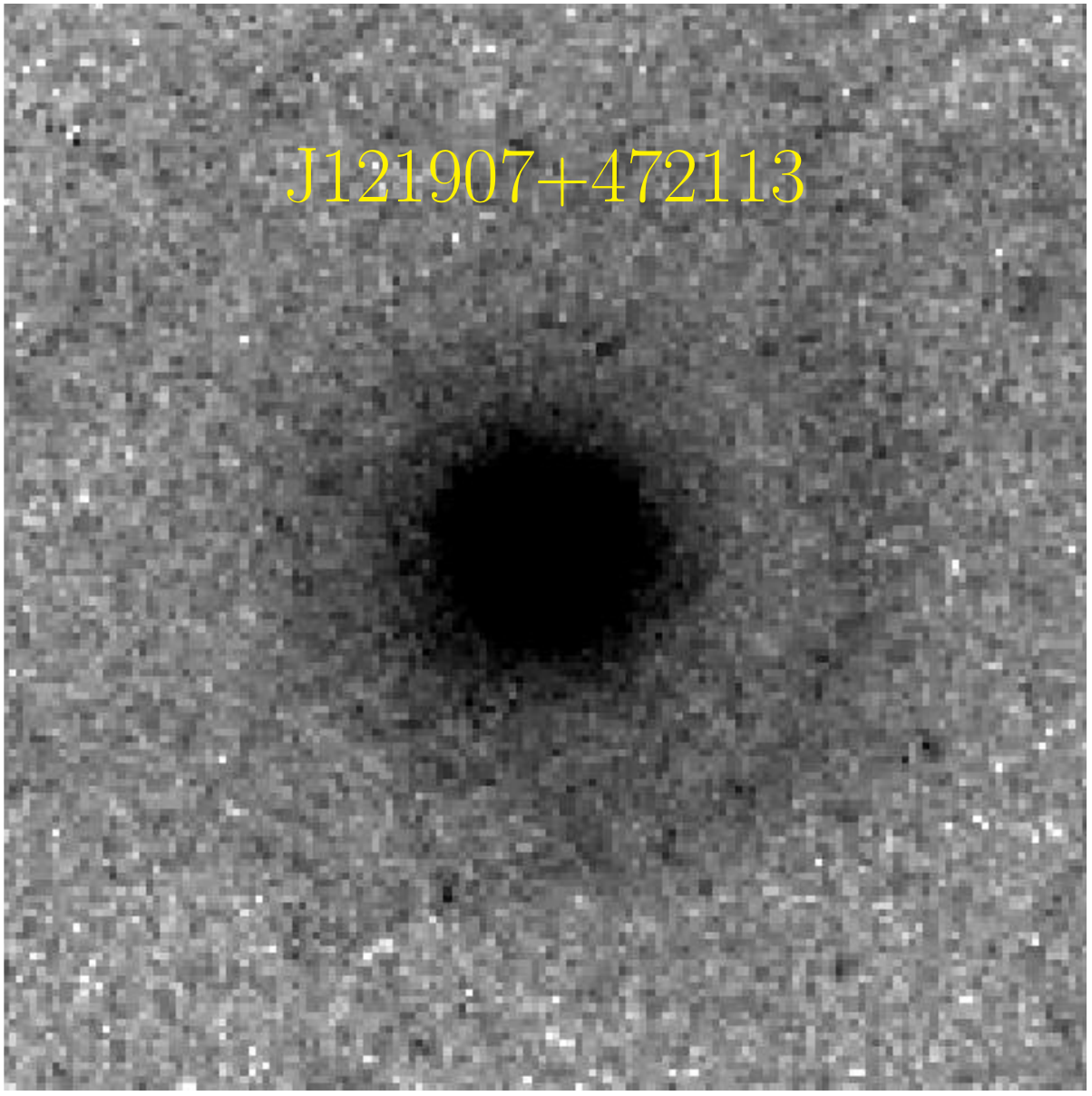}
\end{tabular}
\begin{tabular}{ll}
\hspace*{0.3cm}\raisebox{-0.57cm}{\includegraphics[scale=0.5,angle=0.]{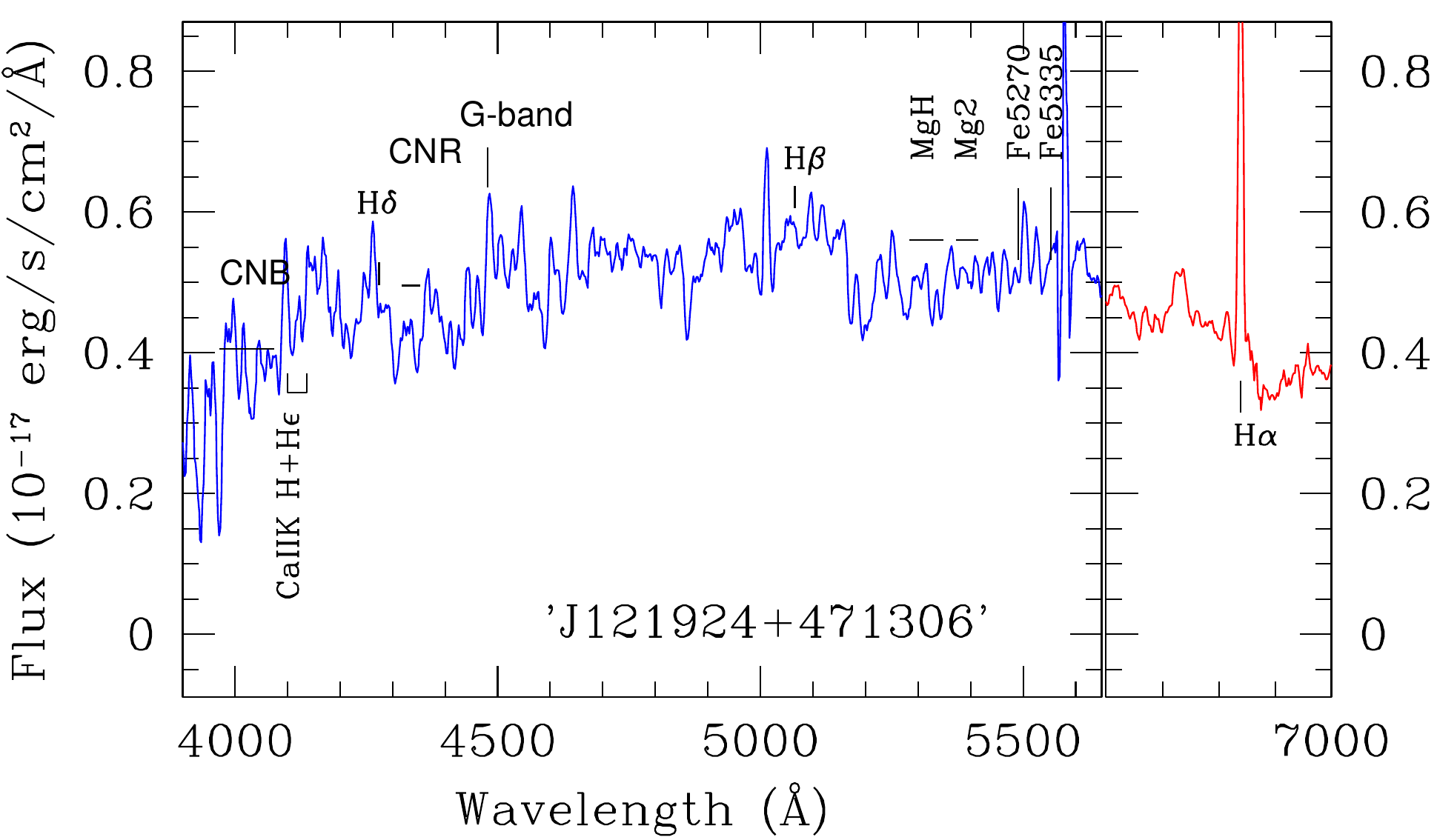}}
&
\hspace*{0.3cm}\includegraphics[scale=0.413]{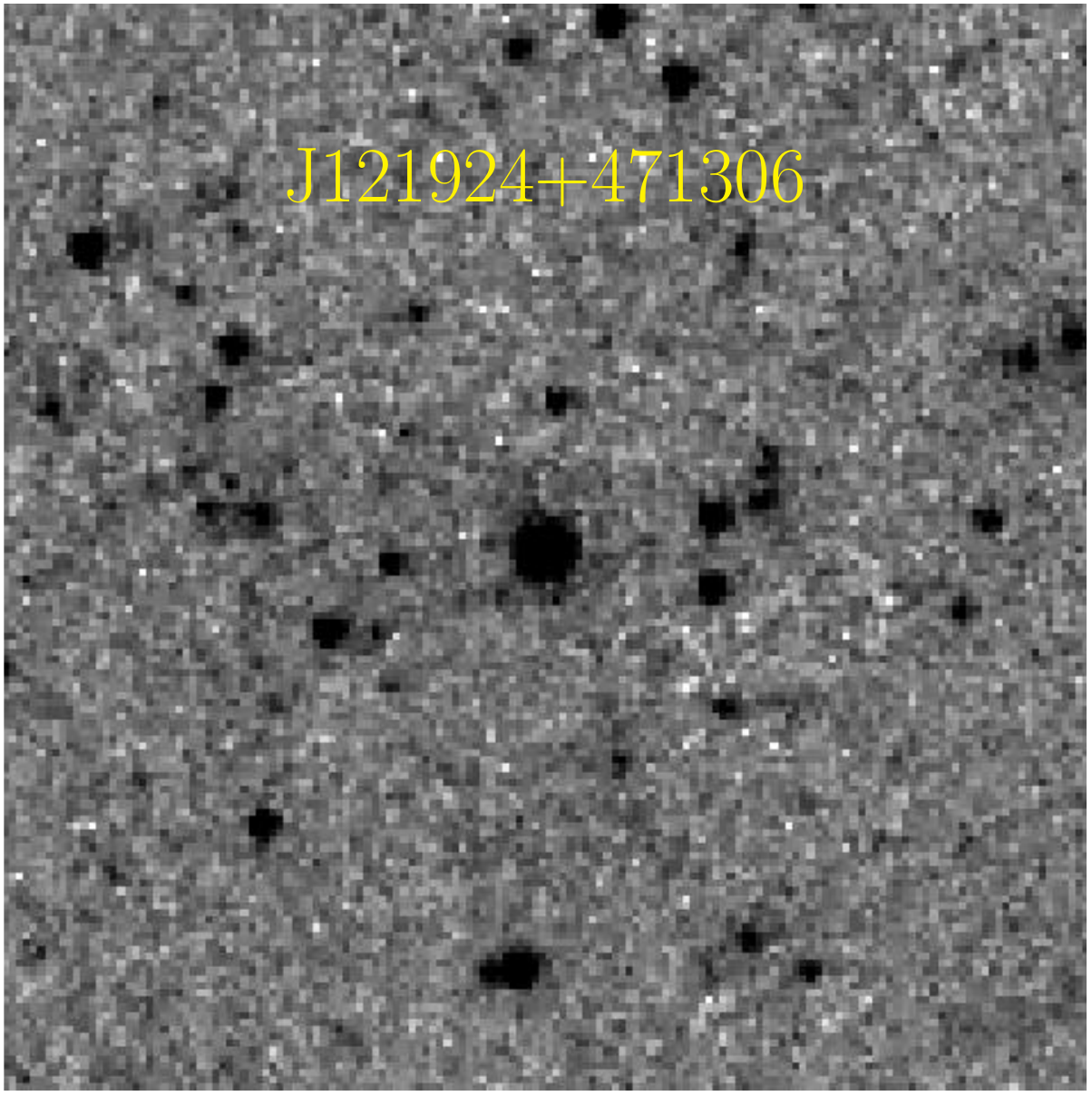}
\end{tabular}
\caption{{\it (Continued--)} Spectra ({\it left}) and grayscales ({\it right}) of non-candidates.
\label{fig:specim_non}}
\end{figure}

\clearpage

\setcounter{figure}{5}
\begin{figure}
%\ContinuedFloat
\begin{tabular}{ll}
\hspace*{0.3cm}\raisebox{-0.57cm}{\includegraphics[scale=0.5,angle=0.]{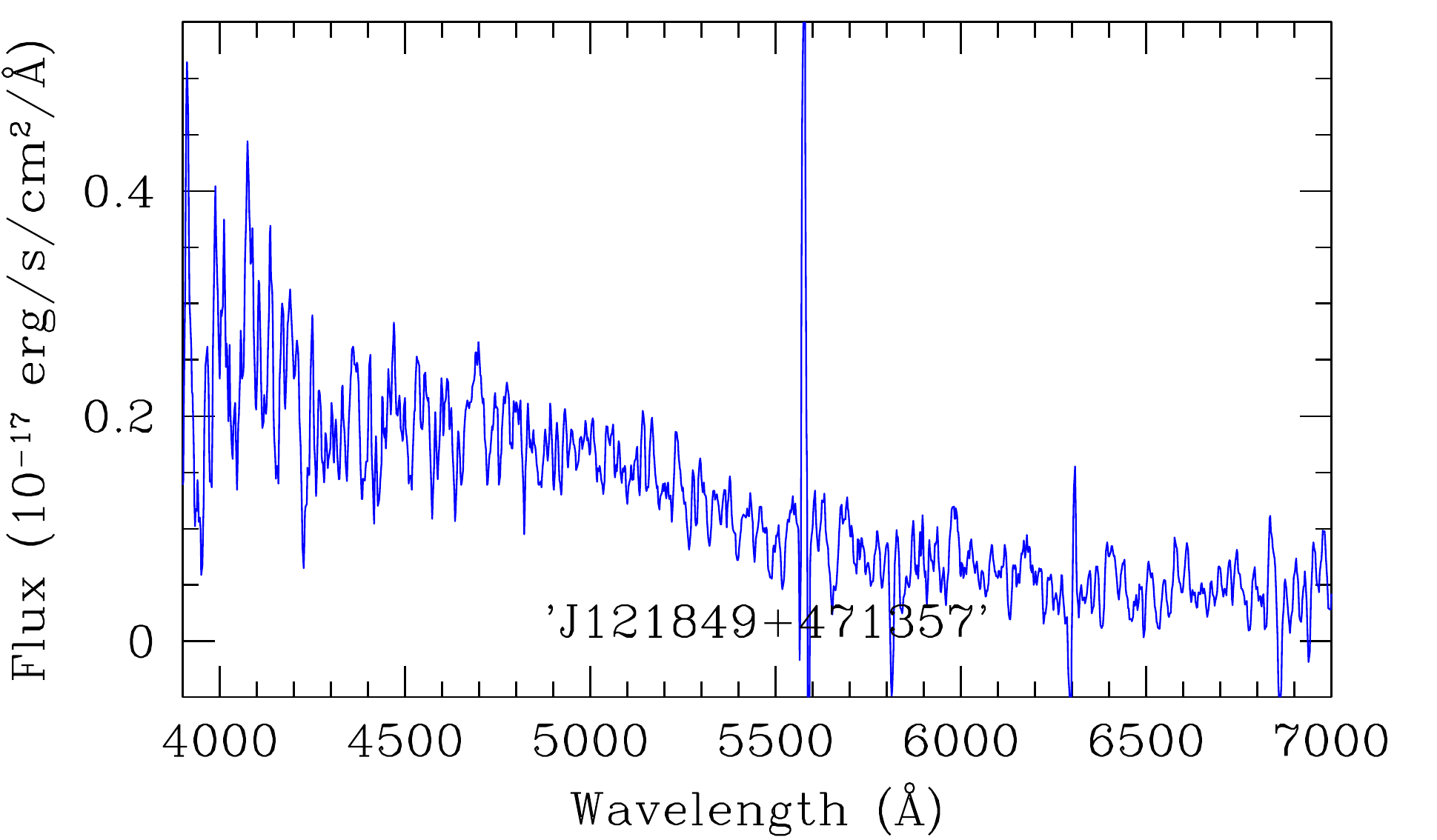}}
&
\hspace*{0.3cm}\includegraphics[scale=0.413]{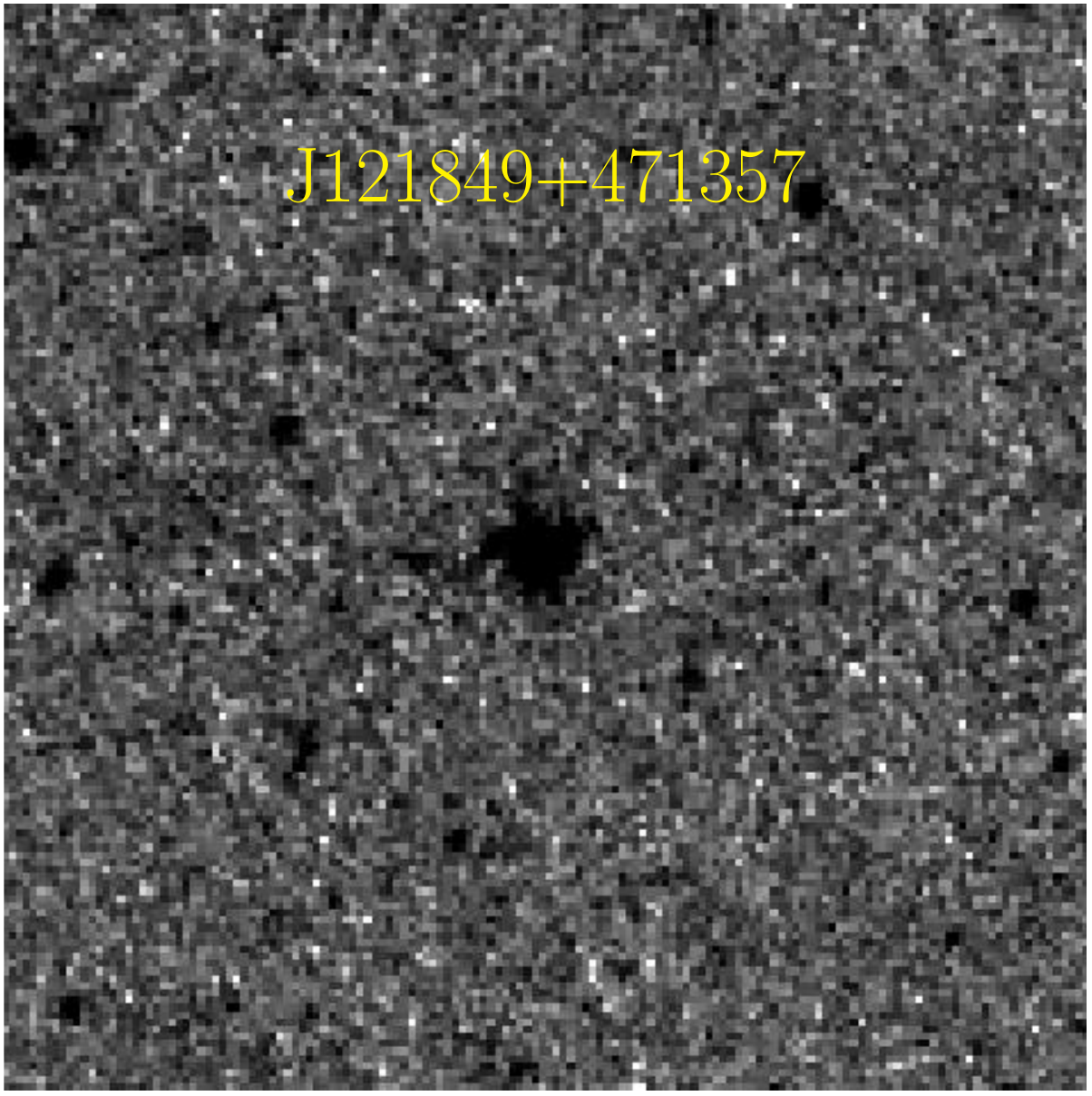}
\end{tabular}
\begin{tabular}{ll}
\hspace*{0.3cm}\raisebox{-0.57cm}{\includegraphics[scale=0.5,angle=0.]{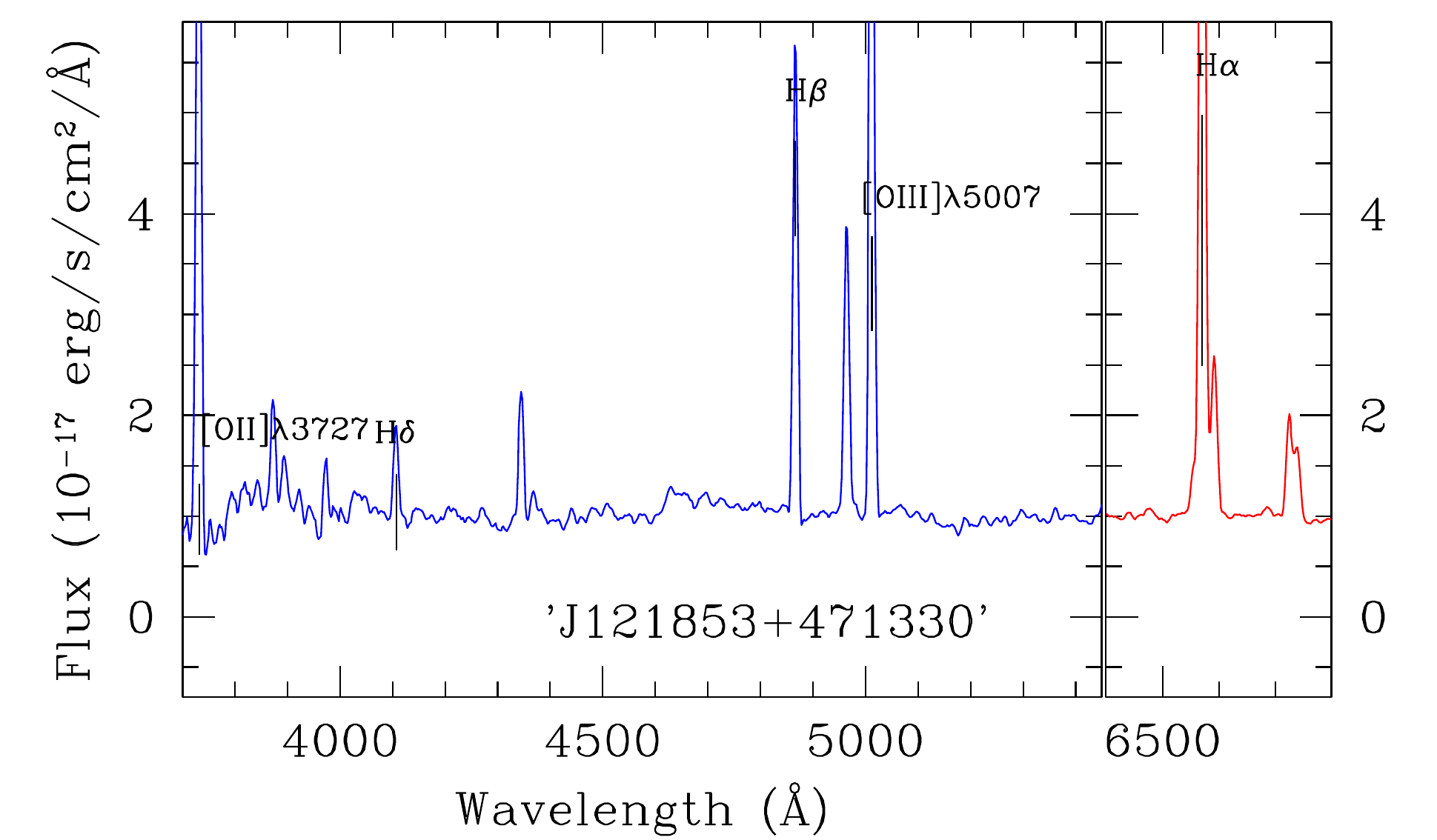}}
&
\hspace*{0.3cm}\includegraphics[scale=0.413]{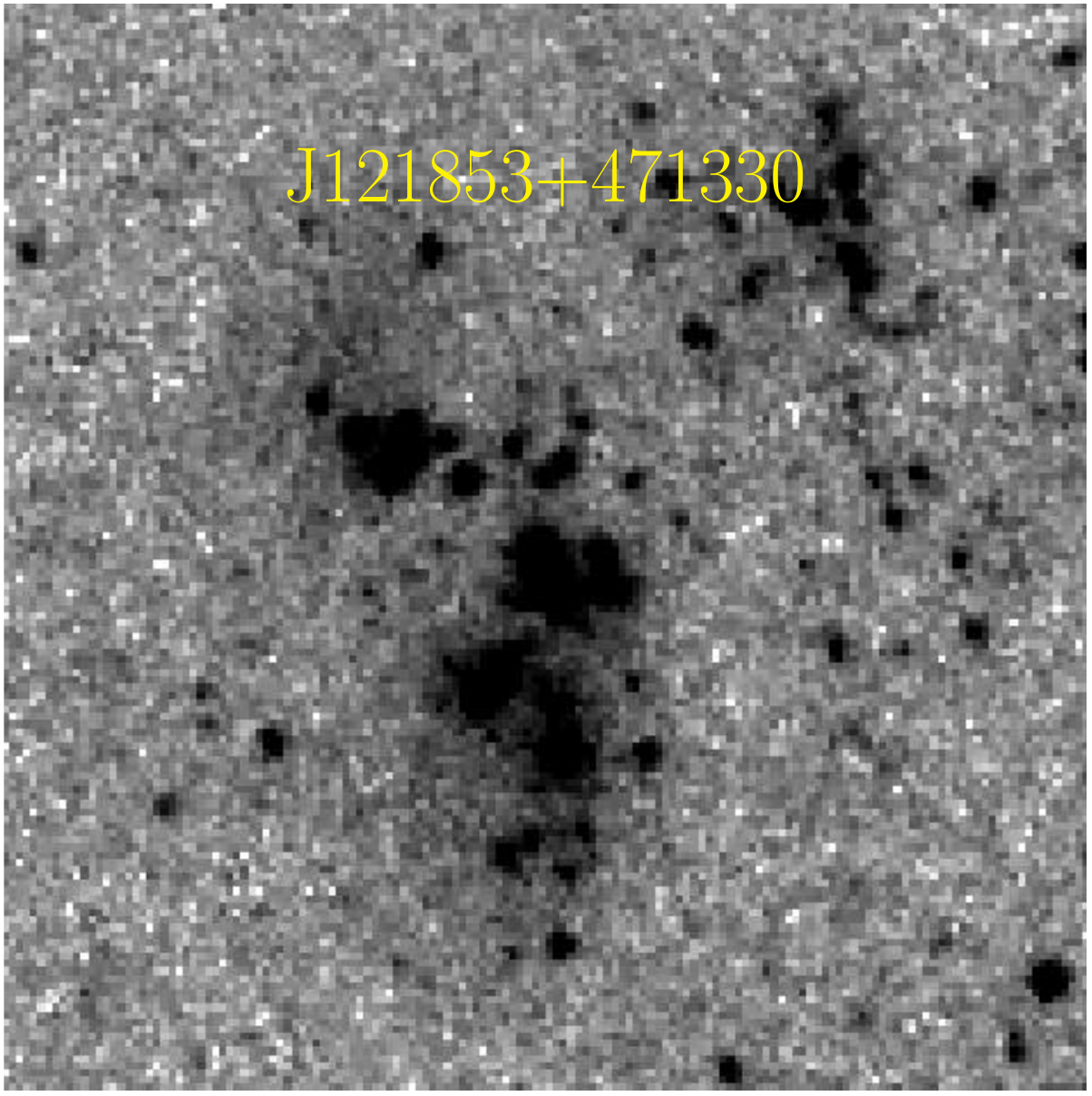}
\end{tabular}
\begin{tabular}{ll}
\hspace*{0.3cm}\raisebox{-0.57cm}{\includegraphics[scale=0.5,angle=0.]{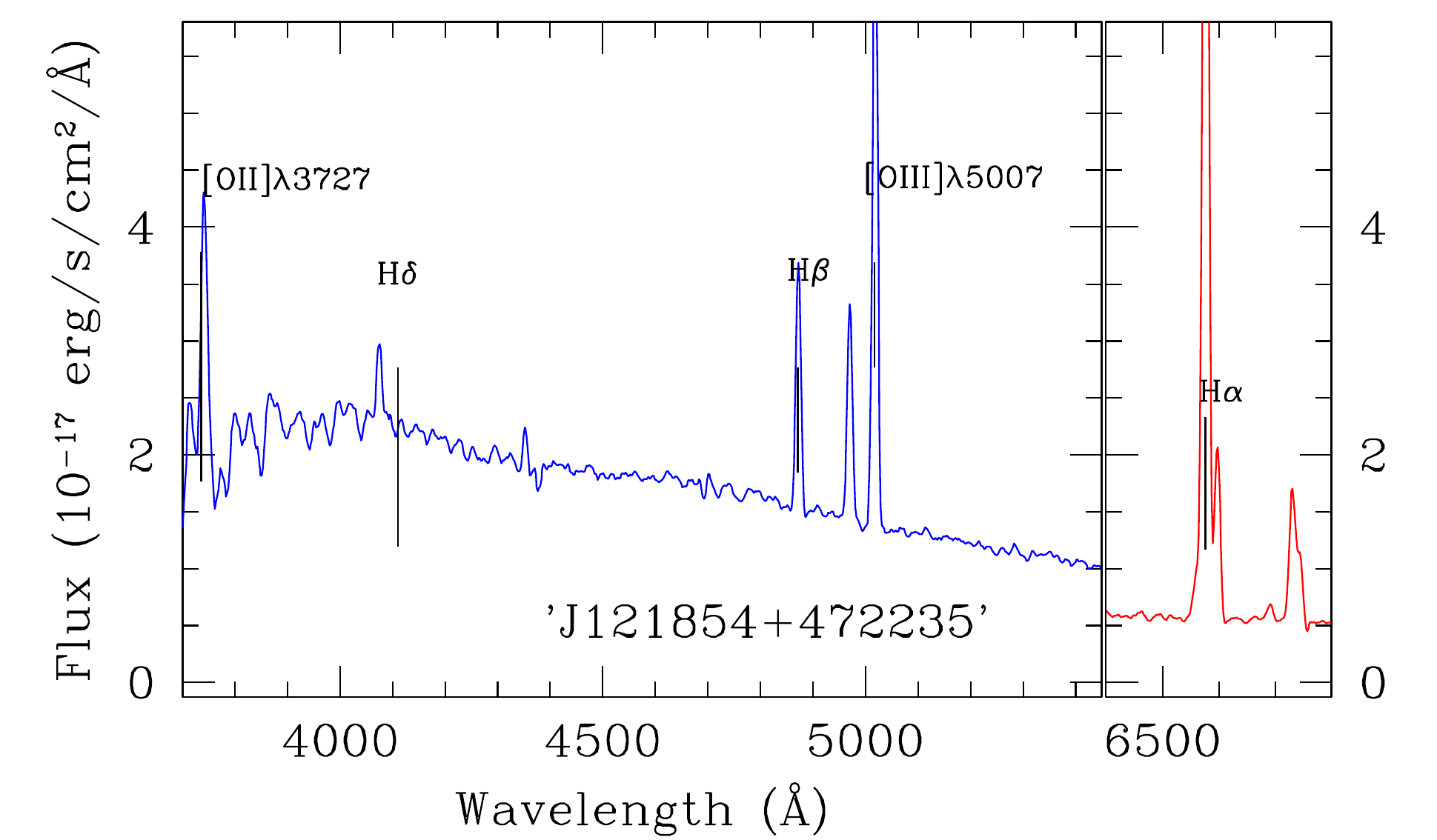}}
&
\hspace*{0.3cm}\includegraphics[scale=0.413]{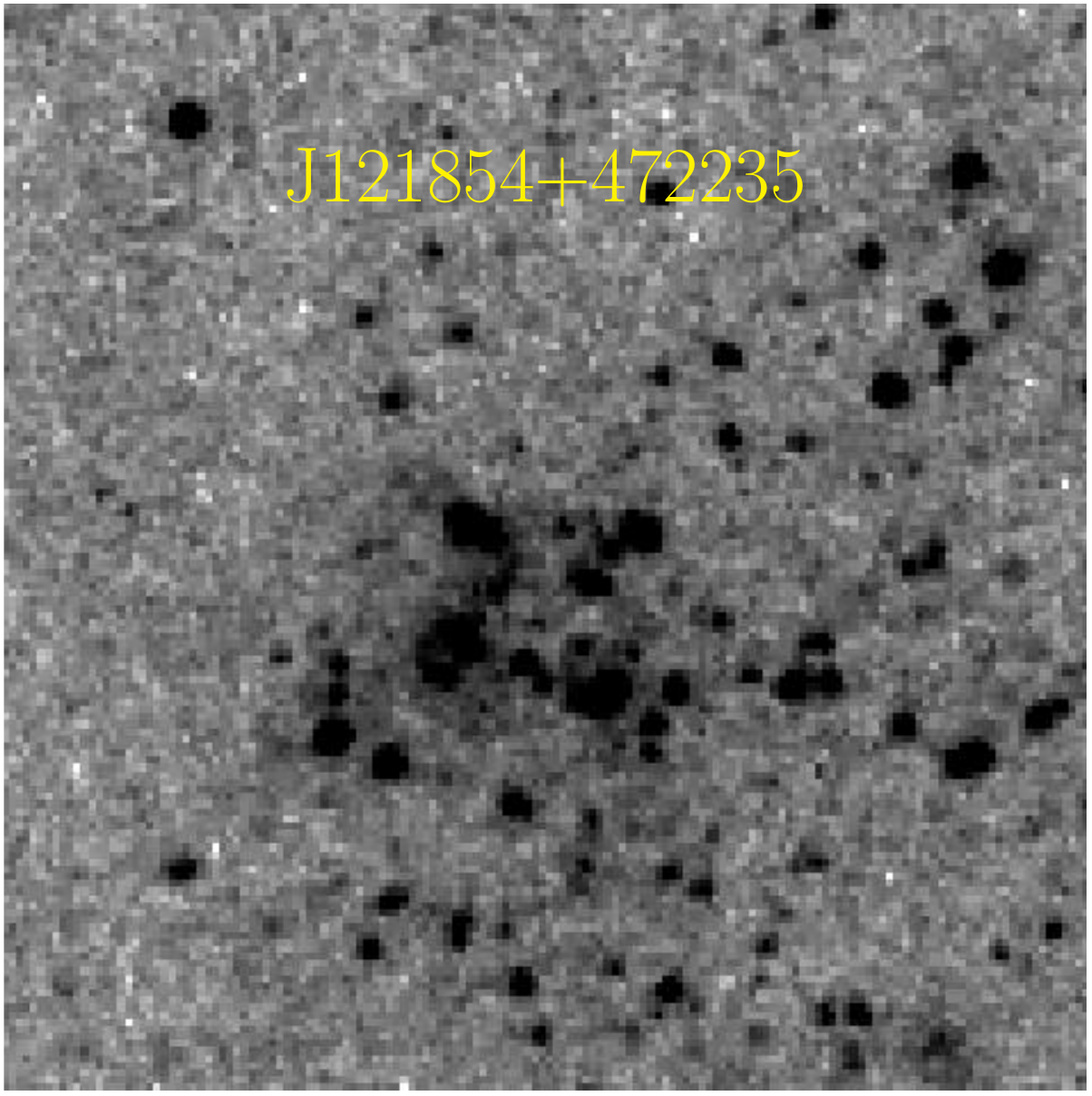}
\end{tabular}
\caption{{\it (Continued--)} Spectra ({\it left}) and grayscales ({\it right}) of non-candidates.
\label{fig:specim_non}}
\end{figure}

\clearpage

\setcounter{figure}{5}
\begin{figure}
%\ContinuedFloat
\begin{tabular}{ll}
\hspace*{0.3cm}\raisebox{-0.57cm}{\includegraphics[scale=0.5,angle=0.]{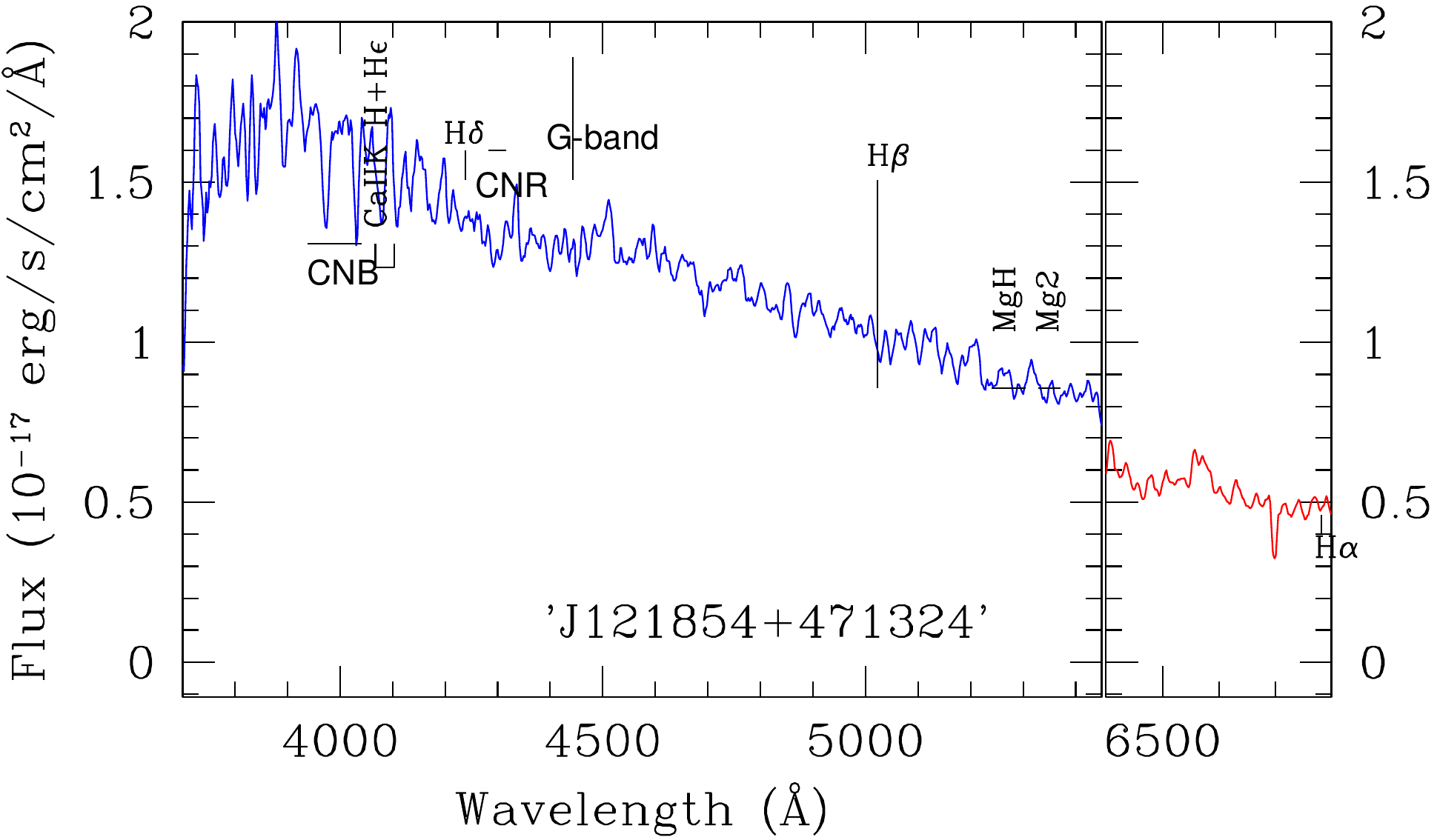}}
&
\hspace*{0.3cm}\includegraphics[scale=0.413]{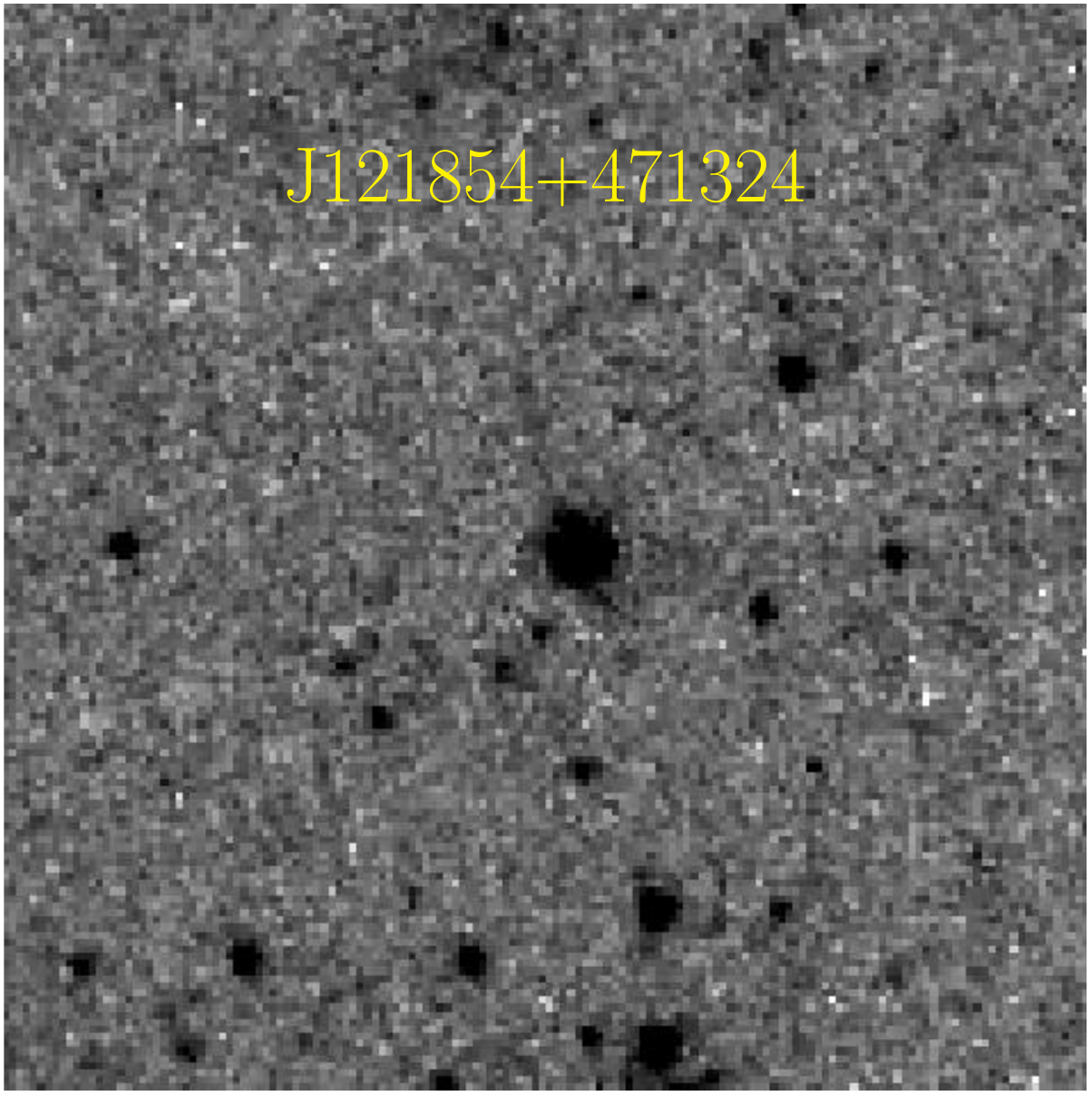}
\end{tabular}
\begin{tabular}{ll}
\hspace*{0.3cm}\raisebox{-0.57cm}{\includegraphics[scale=0.5,angle=0.]{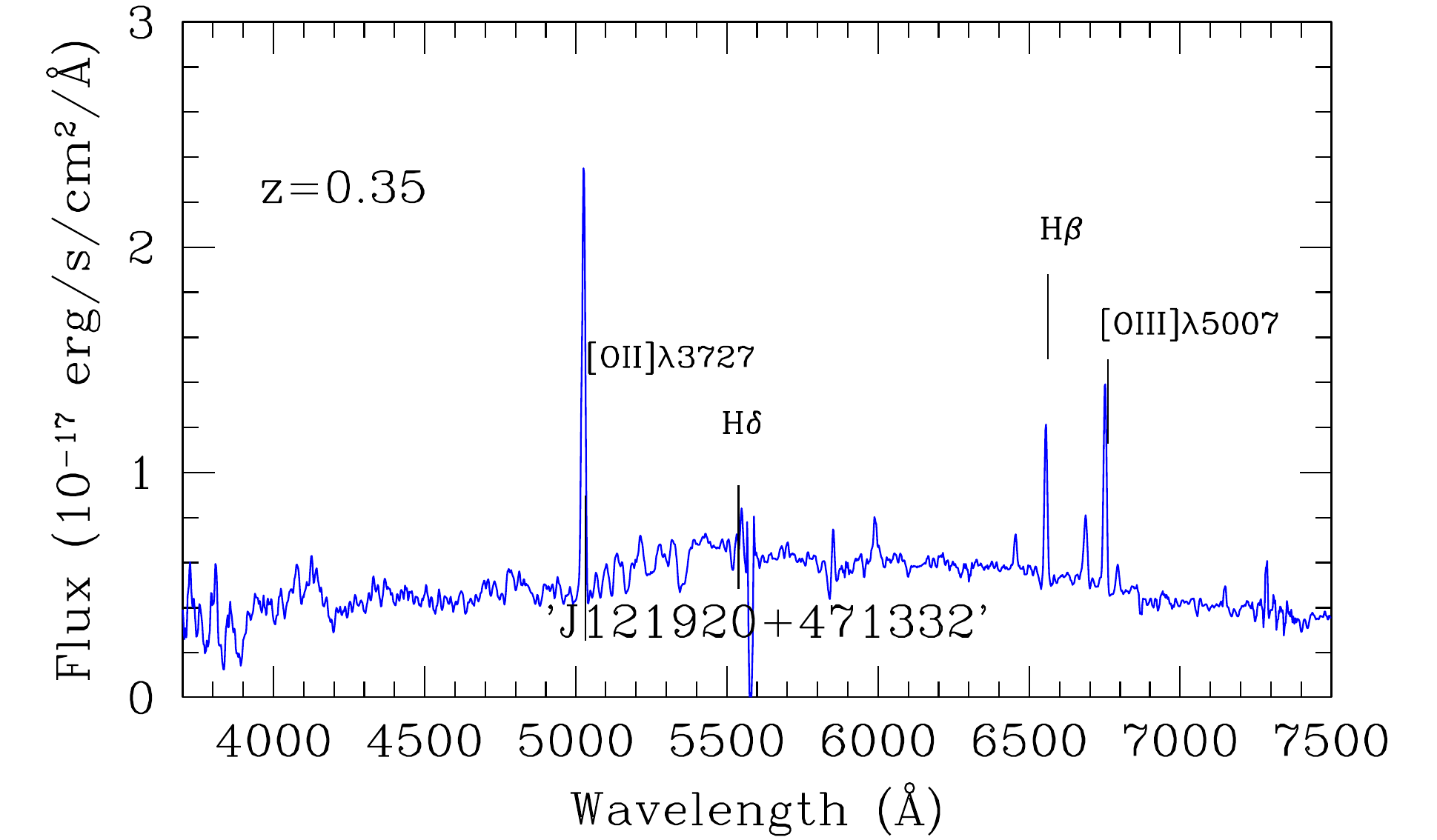}}
&
\hspace*{0.3cm}\includegraphics[scale=0.413]{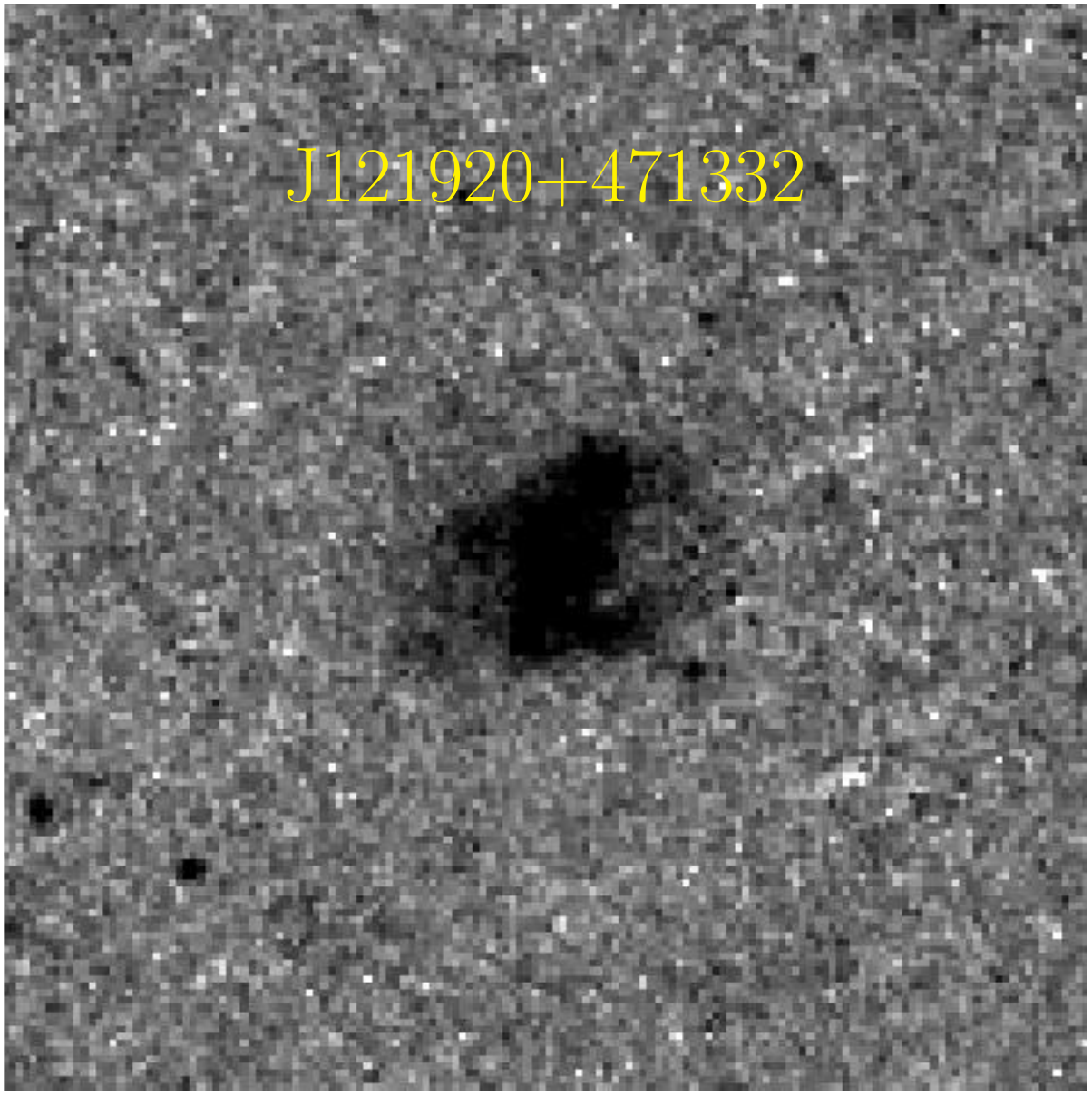}
\end{tabular}
\begin{tabular}{ll}
\hspace*{0.3cm}\raisebox{-0.57cm}{\includegraphics[scale=0.5,angle=0.]{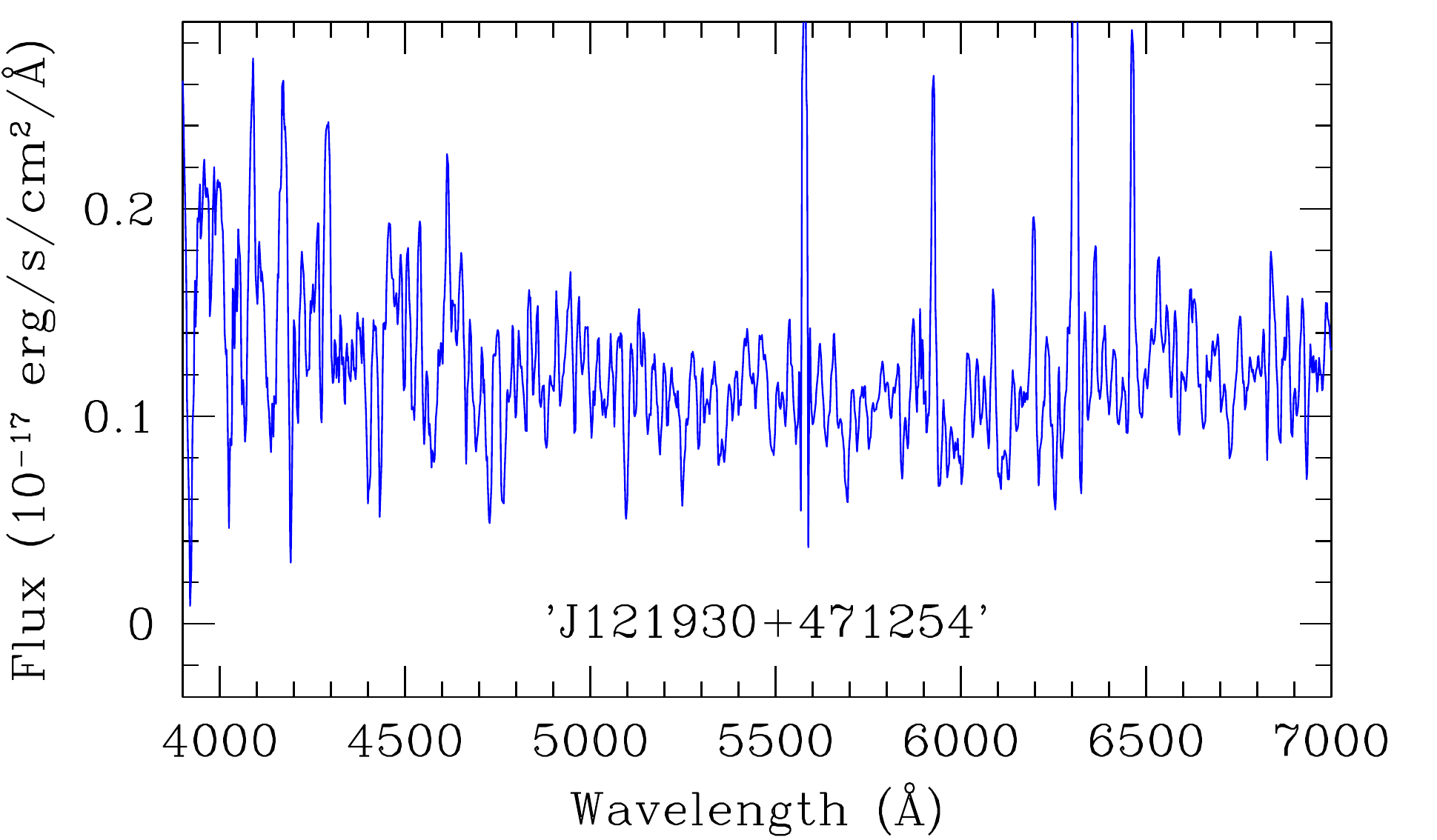}}
&
\hspace*{0.3cm}\includegraphics[scale=0.413]{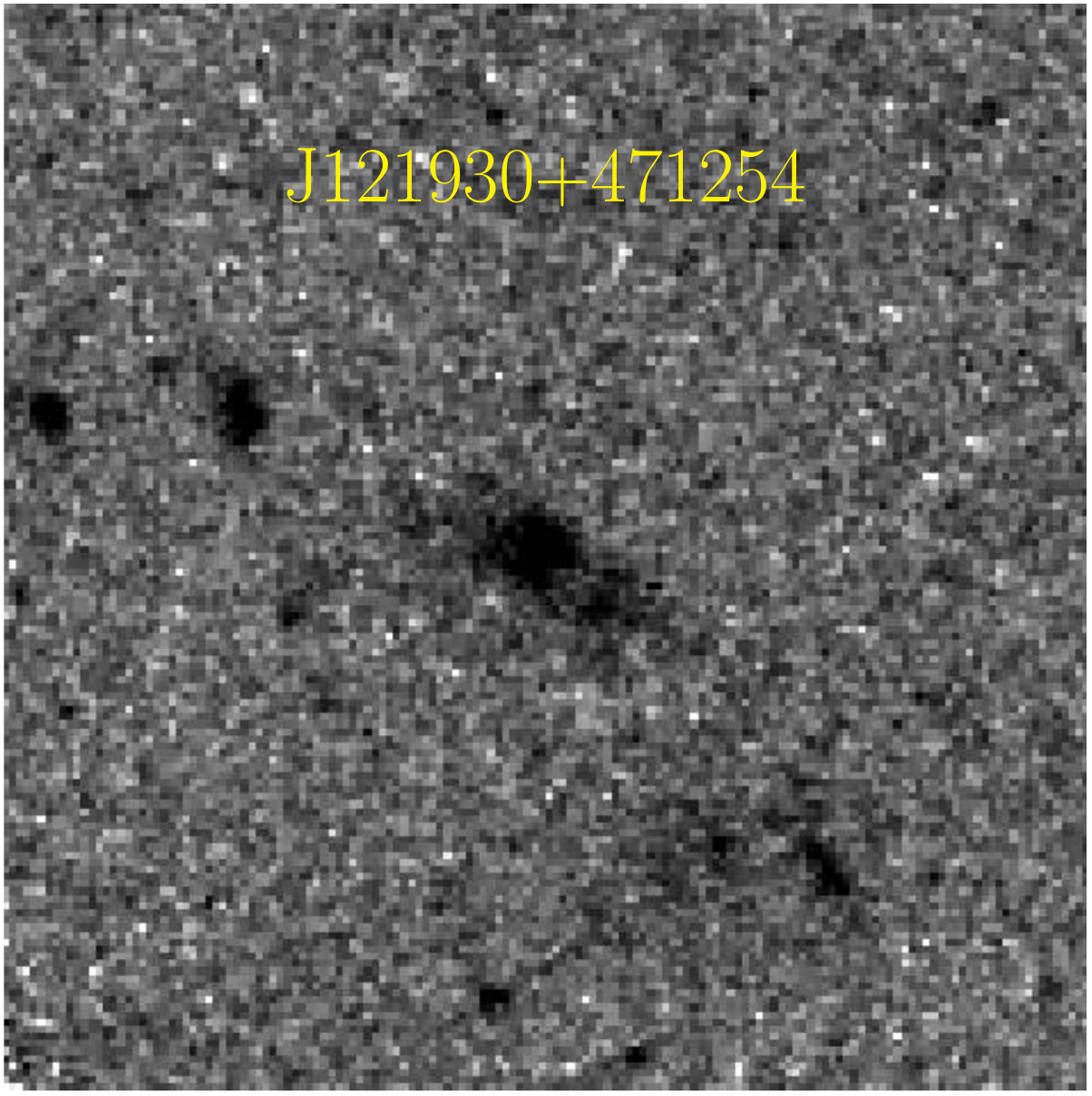}
\end{tabular}
\caption{{\it (Continued--)} Spectra ({\it left}) and grayscales ({\it right}) of non-candidates.
\label{fig:specim_non}}
\end{figure}




\clearpage

\begin{thebibliography}{}

\bibitem[Agertz et al.(2009)]{ager09} Agertz, O., Teyssier, R., \& Moore, B.\ 2009, \mnras, 397, L64 

\bibitem[Bertin \& Arnouts(1996)]{bert96} Bertin, E., \& Arnouts, S.\ 1996, \aaps, 117, 393

\bibitem[Bruzual \& Charlot(2003)]{bruz03} Bruzual, G., \& Charlot, S.\ 2003, \mnras, 344, 1000

\bibitem[Burkert \& Tremaine(2010)]{burk10} Burkert, A., \& Tremaine, S.\ 2010, \apj, 720, 516

\bibitem[Cappellari \& Emsellem(2004)]{capp04} Cappellari, M., \& Emsellem, E.\ 2004, \pasp, 116, 138

\bibitem[Desai et al.(2012)]{desa12} Desai, S., Armstrong, R., Mohr, J.~J., et al.\ 2012, \apj, 757, 83

\bibitem[Dekel et al.(2009)]{deke09} Dekel, A., Sari, R., \& Ceverino, D.\ 2009, \apj, 703, 785 

\bibitem[Dessauges-Zavadsky \& Adamo(2018)]{dess18} Dessauges-Zavadsky, M., \& Adamo, A.\ 2018, \mnras, 479, L118 

\bibitem[Dessauges-Zavadsky et al.(2015)]{dess15} Dessauges-Zavadsky, M., Zamojski, M., Schaerer, D., et al.\ 2015, \aap, 577, A50 

\bibitem[de Vaucouleurs et al.(1991)]{rc3} de Vaucouleurs, G., de Vaucouleurs, A., Corwin, H.~G., Jr., et al.\ 1991, Third Reference Catalogue of Bright Galaxies.~Volume I: Explanations and references.~ Volume II: Data for galaxies between 0$^{h}$ and 12$^{h}$.~ Volume III: Data for galaxies between 12$^{h}$ and 24$^{h}$., by de Vaucouleurs, G.; de Vaucouleurs, A.; Corwin, H.~G., Jr.; Buta, R.~J.; Paturel, G.; Fouqu{\'e}, P..~Springer, New York, NY (USA), 1991, 2091 p.

\bibitem[Ellis \& Bland-Hawthorn(2007)]{elli07} Ellis, S.~C., \& Bland-Hawthorn, J.\ 2007, \mnras, 377, 815

\hspace*{-0.38cm}\hypertarget{fabe79}{Faber, S.~M., \& Gallagher, J.~S.\ 1979, \araa, 17, 135}

%\bibitem[Ferrarese(2002)]{ferr02} Ferrarese, L.\ 2002, \apj, 578, 90 

\bibitem[F{\"o}rster Schreiber et al.(2009)]{foer09} F{\"o}rster Schreiber, N.~M., Genzel, R., Bouch{\'e}, N., et al.\ 2009, \apj, 706, 1364 

\bibitem[Giovanelli et al.(1994)]{giov94} Giovanelli, R., Haynes, M.~P., Salzer, J.~J., et al.\ 1994, \aj, 107, 2036 

\bibitem[Girardi et al.(2000)]{gira00} Girardi, L., Bressan, A., Bertelli, G., \& Chiosi, C.\ 2000, \aaps, 141, 371 

\bibitem[Georgiev et al.(2010)]{geor10} Georgiev, I.~Y., Puzia, T.~H., Goudfrooij, P., \& Hilker, M.\ 2010, \mnras, 406, 1967 

\bibitem[G{\'o}mez-Gonz{\'a}lez et al.(2016)]{gome16} G{\'o}mez-Gonz{\'a}lez, V.~M.~A., Mayya, Y.~D., \& Rosa-Gonz{\'a}lez, D.\ 2016, \mnras, 460, 155

\bibitem[Gonz{\'a}lez-L{\'o}pezlira et al.(2017)]{gonz17} Gonz{\'a}lez-L{\'o}pezlira, R.~A., Lomel{\'{\i}}-N{\'u}{\~n}ez, L., {\'A}lamo-Mart{\'{\i}}nez, K., et al.\ 2017, \apj, 835, 184 

{\bibitem[Harris 1996 (2010 edition) ]{harr96} Harris, W.~E.\ 1996, \aj, 112, 1487}

\bibitem[Harris et al.(2017)]{harr17} Harris, W.~E., Blakeslee, J.~P., \& Harris, G.~L.~H.\ 2017, \apj, 836, 67 

\bibitem[Harris \& Harris(2011)]{harr11} Harris, G.~L.~H., \& Harris, W.~E.\ 2011, \mnras, 410, 2347

\bibitem[Harris et al.(2013)]{harr13} Harris, W.~E., Harris, G.~L.~H., \& Alessi, M.\ 2013, \apj, 772, 82

\bibitem[Harris et al.(2014)]{harr14} Harris, G.~L.~H., Poole, G.~B., \& Harris, W.~E.\ 2014, \mnras, 438, 2117

\bibitem[Harrison et al.(2017)]{charr17} Harrison, C.~M., Johnson, H.~L., Swinbank, A.~M., et al.\ 2017, \mnras, 467, 1965 

\bibitem[Heald et al.(2011)]{heal11} Heald, G., J{\'o}zsa, G., Serra, P., et al.\ 2011, \aap, 526, A118 

\bibitem[Humphreys et al.(2013)]{hump13} Humphreys, E.~M.~L., Reid, M.~J., Moran, J.~M., Greenhill, L.~J., \& Argon, A.~L.\ 2013, \apj, 775, 13

\bibitem[Kissler-Patig et al.(1999)]{kiss99} Kissler-Patig, M., Ashman, K.~M., Zepf, S.~E., \& Freeman, K.~C.\ 1999, \aj, 118, 197

\bibitem[Kruijssen(2015)]{krui15} Kruijssen, J.~M.~D.\ 2015, \mnras, 454, 1658 

\bibitem[Kurtz \& Mink(1998)]{kurt98} Kurtz, M.~J., \& Mink, D.~J.\ 1998, \pasp, 110, 934 

\bibitem[Kron(1980)]{kron80} Kron, R.~G.\ 1980, \apjs, 43, 305

\bibitem[L{\"a}sker et al.(2016)]{laes16} L{\"a}sker, R., Greene, J.~E., Seth, A., et al.\ 2016, \apj, 825, 3

\bibitem[Larsen(1999)]{lars99} Larsen, S.~S.\ 1999, \aaps, 139, 393

\bibitem[Lee et al.(2008)]{lee08} Lee, M.~G., Hwang, H.~S., Kim, S.~C., et al.\ 2008, \apj, 674, 886 

\bibitem[Lupton(1989)]{lupt89} Lupton, R.~H.\ 1989, \aj, 97, 1350 

\bibitem[Lupton \& Monger(1991)]{lupt91} Lupton, R., \& Monger, P.\ 1991, Unpublished paper, 1991

\bibitem[Mu{\~n}oz et al.(2014)]{muno14} Mu{\~n}oz, R.~P., Puzia, T.~H., Lan{\c c}on, A., et al.\ 2014, \apjs, 210, 4

\bibitem[Nantais \& Huchra(2010)]{nant10} Nantais, J.~B., \& Huchra, J.~P.\ 2010, \aj, 139, 2620 

\bibitem[Patr{\'{\i}}cio et al.(2018)]{patr18} Patr{\'{\i}}cio, V., Richard, J., Carton, D., et al.\ 2018, \mnras, 477, 18 

\bibitem[Pawlowski et al.(2012)]{pawl12} Pawlowski, M.~S., Pflamm-Altenburg, J., \& Kroupa, P.\ 2012, \mnras, 423, 1109

\bibitem[Peng et al.(2010)]{peng10} Peng, C.~Y., Ho, L.~C., Impey, C.~D., \& Rix, H.-W.\ 2010, \aj, 139, 2097

\bibitem[Peng et al.(2008)]{peng08} Peng, E.~W., Jord{\'a}n, A., C{\^o}t{\'e}, P., et al.\ 2008, \apj, 681, 197 

\bibitem[Perrett et al.(2002)]{perr02} Perrett, K.~M., Bridges, T.~J., Hanes, D.~A., et al.\ 2002, \aj, 123, 2490 

\bibitem[Powalka et al.(2016)]{powa16} Powalka, M., Lan{\c c}on, A., Puzia, T.~H., et al.\ 2016, \apjs, 227, 12 

\bibitem[Reid et al.(2014)]{reid14} Reid, M.~J., Menten, K.~M., Brunthaler, A., et al.\ 2014, \apj, 783, 130 

\bibitem[Robin \& Creze(1986)]{robi86} Robin, A., \& Creze, M.\ 1986, \aap, 157, 71

\bibitem[Rodriguez-Gomez et al.(2015)]{rodr15} Rodriguez-Gomez, V., Genel, S., Vogelsberger, M., et al.\ 2015, \mnras, 449, 49 

\bibitem[Rhode(2012)]{rhod12} Rhode, K.~L.\ 2012, \aj, 144, 154

\bibitem[Sadoun \& Colin(2012)]{sado12} Sadoun, R., \& Colin, J.\ 2012, \mnras, 426, L51

%\bibitem[Sani et al.(2011)]{sani11} Sani, E., Marconi, A., Hunt, L.~K., \& Risaliti, G.\ 2011, \mnras, 413, 1479 

\bibitem[Schlafly \& Finkbeiner(2011)]{schl11} Schlafly, E.~F., \& Finkbeiner, D.~P.\ 2011, \apj, 737, 103 

\bibitem[Spitler \& Forbes(2009)]{spit09} Spitler, L.~R., \& Forbes, D.~A.\ 2009, \mnras, 392, L1

\bibitem[Tacconi et al.(2013)]{tacc13} Tacconi, L.~J., Neri, R., Genzel, R., et al.\ 2013, \apj, 768, 74

\bibitem[Terlouw \& Vogelaar(2015)]{terl15} Terlouw, J.~P., \& Vogelaar, M.~G.~R.\ 2015,
        Kapteyn Package, version 2.3, Kapteyn Astronomical Institute, Groningen,
        available from \url{http://www.astro.rug.nl/software/kapteyn/}

\bibitem[Tody(1986)]{tody86} Tody, D.\ 1986, \procspie, 627, 733 

\bibitem[Tody(1993)]{tody93} Tody, D.\ 1993, Astronomical Data Analysis Software and Systems II, 52, 173

\bibitem[Turner et al.(2017)]{turn17} Turner, O.~J., Cirasuolo, M., Harrison, C.~M., et al.\ 2017, \mnras, 471, 1280 


\bibitem[White \& Shawl(1987)]{whit87} White, R.~E., \& Shawl, S.~J.\ 1987, \apj, 317, 246 

\bibitem[Wisnioski et al.(2015)]{wisn15} Wisnioski, E., F{\"o}rster Schreiber, N.~M., Wuyts, S., et al.\ 2015, \apj, 799, 209 

\bibitem[Wolf et al.(2007)]{wolf07} Wolf, M.~J., Drory, N., Gebhardt, K., \& Hill, G.~J.\ 2007, \apj, 655, 179

\end{thebibliography}
\end{document}